\documentclass[a4paper,12pt]{article}
\usepackage{amsmath, amssymb, graphicx, physics, bm}
\usepackage[a4paper, margin=1in]{geometry}
\usepackage{tikz}
\usetikzlibrary{arrows.meta,calc}
\usepackage{jheppub, appendix, mathrsfs, placeins}

\title{\boldmath On the Universality of Probe Complexity in $\mathcal{N}=4$ SYM}

\author{Eric L. Graef$^{a}$, Jeff Murugan$^{b}$, Horatiu Nastase$^{a}$ and Hendrik J.R. Van Zyl$^{b}$}
\affiliation{$^{a}$Instituto de F\'{i}sica Te\'{o}rica,\\ UNESP-Universidade Estadual Paulista\\
R. Dr. Bento T. Ferraz 271, Bl. II, Sao Paulo 01140-070, SP, Brazil\\
$^{b}$The Laboratory for Quantum Gravity \& Strings,\\
Department of Mathematics and Applied Mathematics,\\
University of Cape Town, Private Bag, Rondebosch, 7701,\\
South Africa}

\emailAdd{eric.graef@unesp.br}
\emailAdd{jeff.murugan@uct.ac.za}
\emailAdd{horatiu.nastase@unesp.br}
\emailAdd{hjrvanzyl@gmail.com}

\abstract{We investigate Krylov complexity for single-trace operators dual to open strings attached to giant gravitons in planar $\mathcal{N}=4$ super Yang–Mills theory. We show that in protected and few-body sectors, Krylov dynamics is governed by orthogonal polynomial theory associated to the seed spectral measure, leading to bounded Lanczos coefficients determined solely by spectral support. In particular, for fixed magnon number $M$ and open-string length $L\to\infty$, we derive $a_n=2Mg$ and $b_n\to Mg$, demonstrating integrable, band-limited dynamics. This establishes that such sectors are insufficient to test recently proposed gravity-side universality of operator complexity growth. We therefore formulate a finite-density program in which magnons scale with system size, and propose a concrete universality test: whether the leading Krylov growth depends only on coarse thermodynamic data $(\rho,\varepsilon)$ and not on microscopic probe structure. This provides a precise boundary-field-theory framework for testing gravitational universality conjectures.
}

\begin{document}
\maketitle
\flushbottom

\section{Introduction}
\label{sec:intro}
\noindent
Recently it has been suggested that the complexity growth of infalling probes in asymptotically AdS spacetimes exhibits a remarkable degree of universality \cite{Nastase:2026lhz}. Understanding this universality from the dual field theory perspective requires identifying the appropriate boundary states corresponding to different classes of probes and determining which features of their complexity are universal and which retain information about the microscopic nature of the probe.\\

\noindent
A natural arena in which to address this question is planar $\mathcal{N}=4$ super Yang–Mills theory, where the AdS/CFT correspondence \cite{Maldacena:1997re, Gubser:1998bc, Witten:1998qj} provides a dual gravitational description, while the field theory itself admits a rich set of analytically tractable sectors. In particular, operator dynamics in many sectors of $\mathcal{N}=4 $ SYM can be mapped to integrable spin chains \cite{Minahan:2002ve, Beisert:2003ys, Beisert:2005tm, Beisert:2010jr, deMelloKoch:2016nxq, deMelloKoch:2018tlb}, allowing for surprisingly precise control over both spectral and dynamical properties \cite{Gromov:2013pga, Levkovich-Maslyuk:2019awk}. In this work, we will focus on Krylov complexity \cite{Parker:2018yvk, Nandy:2024evd, Baiguera:2025dkc,Rabinovici:2025otw} as a diagnostic of operator growth. Given a Hamiltonian $H$ and an initial operator or state $|K_0\rangle$, the Krylov construction generates an orthonormal basis $\{|K_n\rangle\}$ via the Lanczos algorithm \cite{Lanczos1950AnIM,viswanath2008recursion}, 
\begin{eqnarray}
    H|K_n\rangle = b_{n+1}|K_{n+1}\rangle + a_n |K_n\rangle + b_n |K_{n-1}\rangle\,,
\end{eqnarray}
with coefficients $\{a_n,b_n\}$ encoding the dynamics. These coefficients define an effective one-dimensional tight-binding chain, whose structure captures the spreading of the initial operator in Krylov space. Importantly, the Lanczos coefficients are determined entirely by the spectral measure of the initial state \cite{Muck:2022xfc,Adhikari:2025vdl}, providing a direct link between operator growth and spectral properties of the theory. This framework offers a particularly sharp way to formulate the universality question. Rather than asking whether different probes exhibit identical dynamics in full detail, we ask instead whether they share the same leading Krylov growth, as encoded in the asymptotic behaviour of the Lanczos coefficients or the resulting Krylov complexity
\begin{eqnarray}
    K(t)=\sum_{n\ge0} n\,|\phi_n(t)|^2.
\end{eqnarray}
Our goal is to test this idea within $\mathcal{N}=4$ SYM, focusing on operators dual to open strings attached to giant gravitons \cite{deMelloKoch:2007rqf, deMelloKoch:2007nbd, Bekker:2007ea}. This sector is especially well-suited for our purposes for several reasons. These include the facts that it admits a precise spin-chain description at one loop, interpolates between protected (BPS) and highly excited states,
and provides a natural field-theoretic realization of extended probes in the dual gravitational picture.\\

\noindent
The universality of the complexity of infalling probes, in the case
of ${\cal N}=4$ SYM, as conjectured  in 
\cite{Nastase:2026lhz}, had two aspects to it. The late time behaviour 
was the same for all probes, and equal to the one for the particle,  not just for the $K(t)\propto t^2$ behaviour, but also the proportionality coefficient\footnote{Only in the case of the baryon vertex 
was the coefficient renormalized, but that case depends on the 
explicit interaction Hamiltonian of ${\cal N}=4$ SYM, as well as on the introduction of external ``quark" fieds}. Differences between probes arose
in the subleading $1/t$ behaviour. However, a second, weaker, universality was observed in the short time behaviour where $K(t)\propto  t^2$ for probes (both point-like and extended) without R-charge, and 
$K(t)\propto t$  for probes with R-charge (with motion on the 
$S^5$), while now both the proportionality coefficient and the subleading in $t$ behaviour depended on the probe. Our aim in this article is to try test these ideas.\\

\noindent
A key structural insight underlying our analysis is that Krylov dynamics is governed by the spectral measure of the initial state \cite{Muck:2022xfc,Adhikari:2025vdl}. In particular, when the spectral measure is supported on a single interval, standard results from the theory of orthogonal polynomials imply that the Lanczos coefficients asymptote to constants determined solely by the band edges. This provides a powerful analytic handle on operator growth in integrable and few-body regimes. Using this then, we show that few-body sectors, such as those with a fixed number of magnons on an open spin chain, are analytically tractable and exhibit nontrivial Krylov dynamics. However, their spectral measures are band-limited and absolutely continuous, leading to bounded Lanczos coefficients and hence integrable, non-universal behaviour at leading order. Since the asymptotic Lanczos coefficients depend only on the spectral support, they also cannot encode probe-dependent structure in a way compatible with the conjectured universality.\\ 

\noindent
It appears then that few-body sectors are structurally incapable of exhibiting the conjectured universality. Instead, we will argue that the correct setting in which to test universality is a finite-density many-body regime, in which the number of excitations scales with system size,
$L\to\infty$, $M\to\infty$, with $\rho=\frac{M}{L}$ fixed. In this latter regime, the system becomes genuinely many-body, and we can meaningfully ask whether the leading Krylov dynamics depends only on coarse thermodynamic data, such as energy density and filling fraction, rather than on microscopic details of the initial state. We therefore propose a concrete field-theoretic test of universality: At fixed energy density and filling fraction, do different classes of initial states exhibit the same leading Krylov growth?

\section{Krylov Complexity and Spectral Measures}
\label{sec:Spectral-Measures}
A key challenge in the study of Krylov complexity is to relate the dynamical quantities that characterize operator growth, such as the Lanczos coefficients $\{a_n,b_n\}$, to intrinsic properties of the underlying quantum system. While the Lanczos algorithm provides a constructive definition of these coefficients, it does not by itself reveal their origin or asymptotic behavior.\\

\noindent
An alternative, powerful and conceptually clarifying perspective is obtained by recasting the Krylov construction in terms of the \textit{spectral measure} associated to the initial state \cite{Muck:2022xfc, Muck:2024fpb, Balasubramanian:2025xkj, Alishahiha:2026fnu}. This viewpoint establishes an exact equivalence between Krylov dynamics and the theory of orthogonal polynomials in that the Lanczos coefficients are precisely the Jacobi parameters of the measure. In this setting, the Krylov basis corresponds to orthonormal polynomials in the Hamiltonian. This reformulation has two important consequences. First, it identifies the spectral measure as the fundamental object controlling operator growth, allowing one to reduce a dynamical problem to a spectral one. Second, it provides access to a large body of mathematical results, particularly concerning the asymptotics of orthogonal polynomials, that can be used to extract universal features of Krylov dynamics without computing the Lanczos coefficients explicitly. We will now develop this correspondence in detail, with the aim of making it directly applicable to the analysis of operator growth in $\mathcal{N}=4$ SYM.\\

\subsection{Spectral measure of the seed}
\noindent
Let $H$ be a self-adjoint Hamiltonian acting on a Hilbert space $\mathscr H$, and let $|K_0\rangle \equiv |\psi_0\rangle$ be a normalized seed state. The Krylov subspace generated by $|K_0\rangle$ is
\begin{eqnarray}
    \mathcal K(H,|K_0\rangle) = \mathrm{span}\big\{|K_0\rangle,\ H|K_0\rangle,\ H^2|K_0\rangle,\dots\big\}\,.
\end{eqnarray}
Applying the Lanczos algorithm to the sequence $\{|K_0\rangle,H|K_0\rangle,H^2|K_0\rangle,\dots\}$ gives an orthonormal basis $\{|K_n\rangle\}_{n\ge0}$ satisfying
\begin{eqnarray}
    H|K_n\rangle = b_{n+1}|K_{n+1}\rangle + a_n|K_n\rangle + b_n|K_{n-1}\rangle, \qquad b_0:=0,
\end{eqnarray}
with $a_n\in\mathbb R$ and $b_n>0$. The coefficients $\{a_n,b_n\}$ define the effective one-dimensional Krylov chain and therefore encode the operator growth of the seed. The key point here is that this construction is equivalent to the theory of orthogonal polynomials associated to the seed’s spectral measure.\\

\noindent
To see why, note that, by the spectral theorem, the pair $(H,|K_0\rangle)$ determines a unique positive measure $\mu$ on $\mathbb R$, defined by
\begin{eqnarray}
    d\mu(E)=\langle K_0|\delta(E-H)|K_0\rangle\,dE\,.
\end{eqnarray}
Equivalently, for any bounded Borel function $f$,
\begin{eqnarray}
    \langle K_0|f(H)|K_0\rangle = \int f(E)\, d\mu(E)\,.
\end{eqnarray}
In particular, the moments of the measure are 
\begin{eqnarray}
    \mu_n := \int E^n\,d\mu(E) = \langle K_0|H^n|K_0\rangle\,,
\end{eqnarray}
and the spectral measure contains exactly the information about the Hamiltonian that is accessible from the seed $|K_0\rangle$.
From the point of view of Krylov complexity, the measure $\mu$ is the fundamental object in that it encodes the spectral ``fingerprint" of the seed. Each vector in the Krylov subspace can be written as a polynomial in $H$ acting on $|K_0\rangle$. In particular, the Lanczos basis vectors take the form
$|K_n\rangle = p_n(H)\,|K_0\rangle$,
where $p_n(E)$ is a polynomial of degree $n$. The first few examples illustrate the pattern,
\begin{eqnarray}
    |K_0\rangle &=& p_0(H)|K_0\rangle,\qquad p_0(E)=1,\nonumber\\
    \\
    |K_1\rangle &\propto& (H-a_0)|K_0\rangle \quad \Rightarrow \quad p_1(E)\propto E-a_0,\nonumber
\end{eqnarray}
and so on. The point is that the Lanczos algorithm is nothing but Gram–Schmidt orthogonalization of the monomials $1,\ E,\ E^2,\ E^3,\dots$ with respect to the inner product induced by $d\mu$.
Indeed, using $|K_n\rangle = p_n(H)|K_0\rangle$, we find
\begin{eqnarray}
    \langle K_n|K_m\rangle = \langle K_0|p_n(H)p_m(H)|K_0\rangle = \int p_n(E)p_m(E)\,d\mu(E)\,.
\end{eqnarray}
Since the Lanczos basis is orthonormal, the polynomials satisfy
\begin{eqnarray}
    \int p_n(E)p_m(E)\,d\mu(E)=\delta_{nm}\,.
\end{eqnarray}
The $\{p_n(E)\}$ are precisely the orthonormal polynomials with respect to the seed measure $\mu$. The next step is to show that the Lanczos recursion is exactly the three-term recurrence relation for orthonormal polynomials.\\

\noindent
To this end, consider the polynomial $p_n(E)$. Since it has degree $n$, the product $E\,p_n(E)$ must have degree $n+1$, so it lies in 
$\mathrm{span}\{p_0(E),p_1(E),\dots,p_{n+1}(E)\}$, and we can write
\begin{eqnarray}
    E\,p_n(E)=\sum_{k=0}^{n+1} c_{n,k}\,p_k(E)\,.
\end{eqnarray}
The orthogonality of the $p_k$ forces most of these coefficients to vanish. For $k\le n-2$,
\begin{eqnarray}
    \int p_k(E)\,E\,p_n(E)\,d\mu(E) = \int E\,p_k(E)\,p_n(E)\,d\mu(E)\,,
\end{eqnarray}
but since $E\,p_k(E)$ has degree at most $k+1\le n-1$, it is a linear combination of $p_0,\dots,p_{n-1}$, each of which is orthogonal to $p_n$. Hence
\begin{eqnarray}
    \int p_k(E)\,E\,p_n(E)\,d\mu(E)=0 \qquad (k\le n-2)\,.
\end{eqnarray}
Therefore only the components along $p_{n+1},p_n,p_{n-1}$ survive, and we find
\begin{eqnarray}
    E\,p_n(E)=b_{n+1}p_{n+1}(E)+a_n p_n(E)+b_n p_{n-1}(E)\,.
\end{eqnarray}
This is the standard three-term recurrence for orthonormal polynomials with the coefficients  given explicitly by
\begin{eqnarray}
    a_n &=& \int E\,p_n(E)^2\,d\mu(E)\,,\nonumber\\
    \\
    b_{n+1} &=& \int E\,p_n(E)\,p_{n+1}(E)\,d\mu(E)\,,\nonumber
\end{eqnarray}
with $b_{n+1}>0$ after fixing the sign convention for $p_{n+1}$.
Now, acting with this identity on $|K_0\rangle$, we get
\begin{eqnarray}
    H\,p_n(H)|K_0\rangle = b_{n+1}p_{n+1}(H)|K_0\rangle + a_n p_n(H)|K_0\rangle + b_n p_{n-1}(H)|K_0\rangle\,,
\end{eqnarray}
or, using $|K_n\rangle = p_n(H)|K_0\rangle$, we can write this as
\begin{eqnarray}
    H|K_n\rangle = b_{n+1}|K_{n+1}\rangle + a_n|K_n\rangle + b_n|K_{n-1}\rangle,
\end{eqnarray}
which is exactly the Lanczos recursion. So, in this sense, the Krylov chain is simply the \textit{Jacobi matrix} of the orthogonal polynomial problem defined by the spectral measure $\mu$.

\subsection{Jacobi matrix and Jacobi parameters}
The recurrence coefficients $\{a_n,b_n\}$ are known in the orthogonal polynomial literature as the Jacobi parameters of the measure $\mu$ and define the tridiagonal Jacobi matrix
\begin{eqnarray}
    J_\mu =
\begin{pmatrix}
a_0 & b_1 & 0 & 0 & \cdots \\
b_1 & a_1 & b_2 & 0 & \cdots \\
0 & b_2 & a_2 & b_3 & \cdots \\
0 & 0 & b_3 & a_3 & \cdots \\
\vdots & \vdots & \vdots & \vdots & \ddots
\end{pmatrix}\,.
\end{eqnarray}
This matrix represents multiplication by $E$ in the orthonormal polynomial basis. In other words, the Krylov Hamiltonian is the Jacobi matrix of the seed spectral measure.
This observation is extremely useful, because it converts the dynamical operator-growth problem into a spectral problem for a measure.\\

\noindent
In many cases, it is much easier to compute the moments of the spectral measure,
\begin{eqnarray}
    \mu_n = \int E^n\, d\mu(E) = \langle K_0|H^n|K_0\rangle,
\end{eqnarray}
than to explicitly construct the orthonormal polynomials $p_n(E)$. It is therefore useful that the Lanczos coefficients $\{a_n,b_n\}$ can be reconstructed directly from the moments.

To see how, we define the Hankel determinants
\begin{eqnarray}
    \Delta_n = \det\big[\mu_{i+j}\big]_{i,j=0}^{n-1}, \qquad \Delta_0 := 1.
\end{eqnarray}
These are simply Gram determinants of the monomial basis $\{1,E,\dots,E^{n-1}\}$ with respect to the measure $\mu$. A standard result in the theory of orthogonal polynomials is that the off-diagonal recurrence coefficients are given by \cite{viswanath2008recursion}
\begin{eqnarray}
    b_n^2 = \frac{\Delta_{n+1}\Delta_{n-1}}{\Delta_n^2}, \qquad n\ge 1\,,
\end{eqnarray}
which follows from expressing the squared norms of the monic orthogonal polynomials in terms of ratios of Hankel determinants and using the three-term recurrence relation.\\

\noindent
The diagonal coefficients $a_n$ admit a similar expression. Specifically, defining the shifted Hankel determinants
\begin{eqnarray}
    \sigma_n := \det \begin{pmatrix} \mu_0 & \mu_1 & \cdots & \mu_{n-2} & \mu_n\\ \mu_1 & \mu_2 & \cdots & \mu_{n-1} & \mu_{n+1}\\ \vdots & \vdots & & \vdots & \vdots\\ \mu_{n-1} & \mu_n & \cdots & \mu_{2n-3} & \mu_{2n-1} \end{pmatrix}, \qquad \sigma_0 := 0,
\end{eqnarray}
it follows that 
\begin{eqnarray}
    a_n = \frac{\sigma_{n+1}}{\Delta_{n+1}} - \frac{\sigma_n}{\Delta_n}\,.
\end{eqnarray}
These expressions provide an explicit route from spectral data to Krylov dynamics: $\{\mu_n\}\ \longrightarrow\ \{\Delta_n\}\ \longrightarrow\ \{a_n,b_n\}$.
In particular, they will allow us to extract the Lanczos coefficients without performing the Gram–Schmidt orthogonalization explicitly. As a sanity-check, let's compute the first few coefficients:
\begin{itemize}
    \item $\bm{b_{1}}:$ Here, we know that 
    \begin{eqnarray}
        \Delta_0=1,\qquad \Delta_1=\mu_0,\qquad \Delta_2=\mu_0\mu_2-\mu_1^2\,,
    \end{eqnarray}
    So
    \begin{eqnarray}
      b_1^2=\frac{\Delta_2\Delta_0}{\Delta_1^2} = \frac{\mu_0\mu_2-\mu_1^2}{\mu_0^2}\,.
    \end{eqnarray}
    In particular, if $\mu_0=1$ as it would for a probability measure, then $b_1^2=\mu_2-\mu_1^2$, the variance, exactly as expected.
    \item $\bm{a_0}:$ Since $\sigma_0=0$ and $\sigma_1=\mu_1$, it follows that
    \begin{eqnarray}
        a_0=\frac{\sigma_1}{\Delta_1}-\frac{\sigma_0}{\Delta_0} = \frac{\mu_1}{\mu_0}\,,
    \end{eqnarray}
    or, if $\mu_0=1$, then $a_0=\mu_1$. Again exactly as expected.
    \item $\bm{a_1}:$ For this, we use
    \begin{eqnarray}
        \sigma_2= \det \begin{pmatrix} 
        \mu_0 & \mu_2\\ \mu_1 & \mu_3 
        \end{pmatrix} = \mu_0\mu_3-\mu_1\mu_2\,,
    \end{eqnarray}
    so that
    \begin{eqnarray}
        a_1=\frac{\sigma_2}{\Delta_2}-\frac{\sigma_1} {\Delta_1} = \frac{\mu_3-\mu_1\mu_2}{\mu_2-\mu_1^2}-\mu_1\,,
    \end{eqnarray}
    with $\mu_0=1$. This agrees with the direct Gram–Schmidt computation.
\end{itemize}
This representation of Krylov dynamics in terms of the spectral measure $\mu$ is particularly useful for extracting the large-$n$ behavior of the Lanczos coefficients. In favorable cases, this behavior can be determined entirely by coarse properties of $\mu$, without requiring detailed knowledge of its full form. In fact, a central result from the theory of orthogonal polynomials is the following. Suppose the spectral measure is supported on a single compact interval
$\mathrm{supp}(\mu) = [A,B]$,
and has an absolutely continuous density that is positive almost everywhere on this interval. Then the associated Jacobi parameters converge to constants \cite{Muck:2022xfc,Griffel1979AnIT},
\begin{eqnarray}
    a_n \;\longrightarrow\; \frac{A+B}{2}, \qquad b_n \;\longrightarrow\; \frac{B-A}{4}\,,
\end{eqnarray}
as $n\to\infty$. Physically, this statement has a simple interpretation. The Krylov recursion defines an effective one-dimensional tight-binding chain with site energies $a_n$ and hoppings $b_n$. If these coefficients approach constants $(a_\infty,b_\infty)$, the asymptotic chain becomes translationally invariant, with dispersion relation
\begin{eqnarray}
    E(k) = a_\infty + 2b_\infty \cos k, \qquad k\in[0,\pi].
\end{eqnarray}
The corresponding spectrum is the interval $[a_\infty - 2b_\infty,\; a_\infty + 2b_\infty]$. Consistency with the spectral support [A,B] then forces
\begin{eqnarray}
    a_\infty = \frac{A+B}{2}, \qquad b_\infty = \frac{B-A}{4}.
\end{eqnarray}
Consequently, in the single-band case, the asymptotic Krylov chain is completely fixed by the band edges seen by the initial state. In particular, the leading large-$n$ behavior of $\{a_n,b_n\}$ depends only on the support of  $\mu$.\\

\noindent
More refined information about the spectral measure enters only at subleading order. For example, if the density behaves as
$\rho(E)\sim (E-A)^\alpha (B-E)^\beta$ near the endpoints,
then the approach to the asymptotic values is governed by these exponents, typically producing $1/n^2$-type corrections to $b_n$. Such details will be important for distinguishing different probe classes, but do not affect the leading asymptotic behavior. Finally, note that this simple picture can break down if the spectral support is not a single interval. In particular, if $\mu$ is supported on multiple disjoint intervals, the Lanczos coefficients generically exhibit quasiperiodic rather than convergent behavior. This also provides a useful diagnostic for identifying qualitatively different regimes of operator growth.

\section{Protected and Near-Protected Sectors in $\mathcal{N} = 4$ SYM}
Before turning to genuinely many-body regimes, it is instructive to analyze Krylov complexity in sectors of $\mathcal{N}=4$ SYM that are either protected or close to protected. These sectors provide analytically controlled examples in which the Krylov construction can be carried out explicitly, allowing us to isolate the mechanisms that generate nontrivial operator growth. At the same time, they will serve as a diagnostic for the universality conjecture in \cite{Nastase:2026lhz}. As we will see, while these sectors exhibit increasingly rich Krylov dynamics as we move away from exact BPS configurations, they remain too constrained to realize any notion of universal leading growth. Instead, they reveal a hierarchy from trivial to purely spectral to integrable few-body dynamics. This progression will make clear why a finite-density many-body regime is required to meaningfully test any notion of universality.

\subsection{The 1/2-BPS sector}

Before analyzing Krylov dynamics in protected states, it is useful to briefly recall the structure of the 1/2-BPS sector of $\mathcal{N}=4$ super Yang–Mills theory, which provides the simplest class of exactly solvable operators. The field content of $\mathcal{N}=4$ SYM includes six real scalar fields $\phi^I$,  $I=1,\dots,6,$ transforming in the adjoint representation of the gauge group and in the vector representation of the $SO(6)_R$ R-symmetry. It is convenient to combine these into three complex scalars,
\begin{eqnarray}
    Z = \phi^1 + i \phi^2\,,\qquad Y = \phi^3 + i \phi^4\,,\qquad X = \phi^5 + i \phi^6\,.
\end{eqnarray}
The 1/2-BPS sector is defined by holomorphic operators constructed entirely from a single complex scalar, say $Z$, without derivatives or mixing with other fields. These operators preserve half of the supersymmetry and are protected from quantum corrections. A basis of such operators is given by gauge-invariant polynomials in $Z$, for example
\begin{eqnarray}
    \mathrm{Tr}(Z^J),\qquad \mathrm{Tr}(Z^{J_1})\mathrm{Tr}(Z^{J_2}),\qquad \text{etc.}
\end{eqnarray}
More generally, it is convenient to organize these sums of 
multi-trace operators in terms of Schur polynomials labeled by 
Young diagrams $R$,\footnote{To understand the notation, consider for example the 
element $\sigma\in S_n$ exchanging 1 and 2, in the cycle notation 
$\sigma =(12)$, then $\Tr(\sigma Z^{\otimes n})\equiv \Tr(Z^2)
\Tr(Z^{n-2})$. The character $\chi_R(\sigma)$ is the trace of the 
permutation $\sigma $ acting on the vector space of the Young 
diagram $R$.}
\begin{eqnarray}
    \chi_R(Z^{\otimes n}) = \frac{1}{n!} \sum_{\sigma\in S_n} \chi_R(\sigma)\, \mathrm{Tr}(\sigma Z^{\otimes n}),
\end{eqnarray}
which provide an orthogonal basis of 1/2-BPS operators at finite $N$ \cite{Corley:2001zk}.
These operators are eigenstates of the dilatation operator,
\begin{eqnarray}
    H\,|\chi_R\rangle = E_R\,|\chi_R\rangle\,,
\end{eqnarray}
with scaling dimensions determined entirely by their classical dimension,
$E_R = \Delta_R = n$,
where $n$ is the total number of $Z$-fields. In particular, they do not acquire anomalous dimensions and do not mix under renormalization.\\

\noindent
From the perspective of spin-chain dynamics, the 1/2-BPS sector is trivial. Since only a single field is present, there are no nontrivial nearest-neighbor interactions, and the dilatation operator acts diagonally. Equivalently, the sector corresponds to a free matrix harmonic oscillator, which is exactly solvable. This simplicity makes the 1/2-BPS sector an ideal starting point for analyzing Krylov complexity. As we now show, the absence of mixing and the trivial spectral structure imply that individual BPS operators exhibit completely trivial Krylov dynamics.\\

\noindent
Let's now consider a protected (BPS) operator in the $SU(2)$ sector, $\mathcal O_J \sim \mathrm{Tr}(Z^J)$, which is an eigenstate of the dilatation operator, $H|\mathcal O_J\rangle = E_J |\mathcal O_J\rangle$. At planar one loop, and in fact to all orders in the protected sector, such operators do not mix and their anomalous dimension is fixed. The Krylov construction in this case is immediate. Acting repeatedly with the Hamiltonian gives
\begin{eqnarray}
    H^n|\mathcal O_J\rangle = E_J^n |\mathcal O_J\rangle\,,
\end{eqnarray}
so no new directions in Hilbert space are generated. The Krylov subspace therefore collapses to
\begin{eqnarray}
    \mathcal K(H,|\mathcal O_J\rangle) = \mathrm{span}\{|\mathcal O_J\rangle\}\,.
\end{eqnarray}
From the spectral perspective, the seed resolves only a single energy eigenvalue. The associated spectral measure is
\begin{eqnarray}
    d\mu(E)=\langle \mathcal O_J|\delta(E-H)|\mathcal O_J\rangle\,dE = \delta(E-E_J)\,dE\,,
\end{eqnarray}
so all moments 
\begin{eqnarray}
    \mu_n = \int E^n d\mu(E) = E_J^n\,,
\end{eqnarray}
are trivial. Applying the Lanczos algorithm, we find $a_0 = \langle \mathcal O_J|H|\mathcal O_J\rangle = E_J$, while for all higher steps the orthogonalization procedure terminates immediately, giving
$b_1 = 0$ and $b_n = 0$ for all $n\ge 1$.\\

\noindent
Equivalently, the Krylov Hamiltonian reduces to a $1\times 1$ matrix, $H_{\text{Krylov}} = (E_J)$,
and the associated orthogonal polynomial problem is trivial since the measure consists of a single point mass, so only the zeroth polynomial is nonvanishing.
However you look at it though, exact BPS states have trivial Krylov dynamics. Physically, this reflects the complete absence of operator growth. Since the operator is protected, time evolution only generates a phase,
\begin{eqnarray}
    |\psi(t)\rangle = e^{-iE_J t}|\mathcal O_J\rangle\,,
\end{eqnarray}
and the state never spreads in the Krylov basis. Correspondingly, the Krylov complexity,
\begin{eqnarray}
    K(t)=\sum_{n\ge 0} n\,|\phi_n(t)|^2\,,
\end{eqnarray}
remains identically zero for all times. In this sense, exact BPS states provide a baseline; they saturate the notion of ``no growth", with any nontrivial Krylov structure arising only once we consider superpositions or deformations away from protected configurations.

\subsection{Superpositions of BPS states}
The simplest way to obtain nontrivial Krylov dynamics within the 1/2-BPS sector appears to be via a superpositions of eigenstates. Specifically, let
\begin{eqnarray}
    |\psi_0\rangle = \sum_{J} c_J\,|\mathcal O_J\rangle,
    \qquad \sum_J |c_J|^2 = 1,
\end{eqnarray}
where $|\mathcal O_J\rangle$ are orthonormal BPS eigenstates of the dilatation operator. Since the Hamiltonian is diagonal in this basis, the time evolution is trivial at the level of each component,
\begin{eqnarray}
    |\psi(t)\rangle = \sum_J c_J\, e^{-iE_J t}\,|\mathcal O_J\rangle\,,
\end{eqnarray}
but the superposition structure leads to nontrivial Krylov dynamics. The spectral measure associated to $|\psi_0\rangle$ is then
\begin{eqnarray}
    d\mu(E)=\langle \psi_0|\delta(E-H)|\psi_0\rangle\,dE = \sum_J |c_J|^2\,\delta(E-E_J)\,dE\,,
\end{eqnarray}
which is a purely discrete measure supported on the set of BPS energies. The moments are therefore given by
\begin{eqnarray}
    \mu_n = \int E^n\, d\mu(E) = \sum_J |c_J|^2\, E_J^n\,,
\end{eqnarray}
and all Krylov data is encoded in the distribution $|c_J|^2$ over eigenvalues. This is sufficient to write down the first few Lanczos coefficients directly in terms of these moments as
\begin{eqnarray}
    a_0 &=& \mu_1 = \sum_J |c_J|^2 E_J,\nonumber\\
    b_1^2 &=& \mu_2 - \mu_1^2
    = \sum_J |c_J|^2 E_J^2 - \left(\sum_J |c_J|^2 E_J\right)^2,
\end{eqnarray}
i.e. the variance of the energy distribution. Higher coefficients depend on higher moments in an analogous way, for example
\begin{eqnarray}
    a_1 = \frac{\mu_3 - \mu_1 \mu_2}{\mu_2 - \mu_1^2} - \mu_1,
\end{eqnarray}
and so on. In this sense, the Lanczos sequence is entirely determined by the statistics of the discrete spectrum sampled by the seed. If the superposition involves only a finite number of distinct eigenvalues,
\begin{eqnarray}
    |\psi_0\rangle = \sum_{i=1}^{M} c_i |E_i\rangle\,,
\end{eqnarray}
then the Krylov subspace has dimension at most $M$. This follows immediately from the fact that all powers of $H$ act diagonally,
\begin{eqnarray}
    H^n|\psi_0\rangle = \sum_{i=1}^{M} c_i E_i^n |E_i\rangle\,,
\end{eqnarray}
so the vectors $\{|\psi_0\rangle, H|\psi_0\rangle, \dots\}$ span at most the $M$-dimensional space generated by the $|E_i\rangle$.
Equivalently, the orthogonal polynomial problem reduces to a finite discrete measure with $M$ support points, for which the recurrence terminates,
$b_M = 0$ and $b_n=0$ for $n\ge M$.
From the point of view of operator growth, this regime is qualitatively distinct from interacting sectors. In particular, there is no mixing between different $J$ under the Hamiltonian, and the only source of nontriviality is the initial superposition. In other words, the dynamics is effectively that of a classical random variable $E$ with distribution $|c_J|^2$ and Krylov complexity reduces to a problem in spectral statistics rather than many-body dynamics.

\subsection{Local primary wavepackets and pointlike probes}
\label{subsec:primary-wave}

To make contact with the gravitational description of infalling probes in \cite{Nastase:2026lhz}, we need to first identify the appropriate class of operators in the boundary theory. In the bulk analysis, pointlike particles and extended objects are treated on equal footing as semiclassical probes whose dynamics is governed by their motion in the emergent radial direction. A central claim in \cite{Nastase:2026lhz} is that, despite their very different microscopic structure, these probes exhibit universal leading behavior in their complexity growth (both at small times and 
large times), with differences appearing only at subleading order.\\

\noindent
The simplest candidate operator dual to a pointlike particle is a local primary $\mathcal O_\Delta$. Under the state-operator correspondence, this defines a highest-weight state
$|\mathcal O_\Delta\rangle = \mathcal O_\Delta(0)|0\rangle$,
with definite scaling dimension
$D|\mathcal O_\Delta\rangle = \Delta |\mathcal O_\Delta\rangle$.
As such, it is an energy eigenstate of the dilatation operator and therefore exhibits trivial Krylov dynamics analogous to the exact BPS states discussed in Section 3.1, where the absence of any spectral width precludes any nontrivial complexity. Moreover, a falling bulk particle is a localized, dynamical object, while a single primary state is completely delocalized in the radial direction and stationary under time evolution. Consequently, the boundary dual of a localized bulk excitation cannot be a single eigenstate, but must instead be a wavepacket. In the conformal field theory, the natural building blocks for such a wavepacket are the descendants of a given primary, generated by repeated action of the translation operators $P_\mu$. These descendants span the conformal tower $\mathcal H_{\mathcal O}
=
\mathrm{span}\big\{|\mathcal O_\Delta\rangle,\; P_\mu|\mathcal O_\Delta\rangle,\; P_\mu P_\nu|\mathcal O_\Delta\rangle,\ldots\big\}$,
with scaling dimensions $E_n=\Delta+n$ (or $E_n=\Delta+2n$ in the scalar sector). A semiclassical state corresponding to a localized pointlike probe is then obtained by forming a wavepacket over this tower. The simplest such construction is a Gaussian packet in the descendant level,
\begin{eqnarray}
    |\Psi_0\rangle = \mathcal N \sum_{n=0}^\infty \exp\!\left[-\frac{(n-n_0)^2}{4\sigma^2}\right] e^{iqn} \,|\mathcal O_\Delta; n\rangle\,,
    \label{eq:pp-wavepacket}
\end{eqnarray}
which localizes the state in the emergent radial direction with the descendant level $n$ playing the role of a discrete radial coordinate, the spread $\sigma$ controlling the degree of localization and the scalar descendant basis defined as
\begin{eqnarray}
    |\mathcal O_\Delta; n\rangle = \frac{1}{\sqrt{n!\,(2\Delta)_n}}\,(P^2)^n |\mathcal O_\Delta\rangle\,,
\end{eqnarray}
with Pochhammer symbol, $(\Delta)_n$. The wavepacket is normalized by choosing
\begin{eqnarray}
    |\mathcal N|^{-2} = \sum_{n=0}^\infty \exp\!\left[-\frac{(n-n_0)^2}{2\sigma^2}\right]\,.
\end{eqnarray}
We can get a better idea of what this normalization factor is by taking $1 \ll \sigma \ll n_0$,
so that we may approximate the sum by a Gaussian integral,
\begin{eqnarray}
    \sum_{n=0}^\infty e^{-(n-n_0)^2/(2\sigma^2)} \;\approx\; \int_{-\infty}^{\infty} dn\, e^{-(n-n_0)^2/(2\sigma^2)} = \sqrt{2\pi}\,\sigma,
\end{eqnarray}
which gives $\mathcal N \approx (2\pi\sigma^2)^{-1/4}$.\\

\noindent
Since $D\,|\mathcal O_\Delta; n\rangle = (\Delta+2n)|\mathcal O_\Delta; n\rangle$, following our arguments above, the spectral measure is 
\begin{eqnarray}
    d\mu(E)=\sum_{n\ge0} w_n\,\delta(E-(\Delta+2n))\,dE\,,
    \label{eq:gaussian-spectral-measure}
\end{eqnarray}
where the weights
\begin{eqnarray}
    w_n := |\mathcal N|^2 |c_n|^2 =|\mathcal N|^2 \exp\!\left[-\frac{(n-n_0)^2}{2\sigma^2}\right]\,,
    \label{eq:gaussian-weights}
\end{eqnarray}    
satisfy
\begin{eqnarray}
    \sum_{n\ge0} w_n = 1\,,
\end{eqnarray}
with our choice of normalization. Consequently though, all Krylov data are determined by the moments
\begin{eqnarray}
    \mu_k = \sum_{n\ge0} w_n (\Delta+2n)^k\,.
\end{eqnarray}
To compute the coefficients, define
\begin{eqnarray}
    \langle n\rangle := \sum_n w_n n\,, \qquad \mathrm{Var}(n) := \sum_n w_n (n-\langle n\rangle)^2\,.
\end{eqnarray}
Then, for a Gaussian packet away from the boundary, $\langle n\rangle \approx n_0$, $\mathrm{Var}(n)\approx \sigma^2$.
With this, the first moment
\begin{eqnarray}
    \mu_1 = \sum_n w_n (\Delta+2n) = \Delta + 2\langle n\rangle \;\approx\; \Delta + 2n_0.
\end{eqnarray}
Similarly, the second moment
\begin{eqnarray}
    \mu_2 &=& \sum_n w_n (\Delta+2n)^2 = \Delta^2 + 4\Delta\langle n\rangle + 4\langle n^2\rangle\,,\nonumber\\
    &\approx& \Delta^2 + 4\Delta n_0 + 4(n_0^2+\sigma^2)\,,
\end{eqnarray}
where we have used the fact that $\langle n^2\rangle = n_0^2 + \sigma^2$ in the last step. This suffices to compute the first few Lanczos coefficients as
\begin{eqnarray}
    a_0 &=& \mu_1 \approx \Delta + 2n_0\nonumber\\
    b_1 &=& \sqrt{b_1^2} 
    = \sqrt{\mu_2 - \mu_1^2}\nonumber\\
    &=& \sqrt{\left[\Delta^2 + 4\Delta n_0 + 4(n_0^2+\sigma^2)\right] - \left[\Delta + 2n_0\right]^2}= \sqrt{4\sigma^2} \approx 2\sigma\nonumber\\
    a_1 &=& \frac{\mu_3 - 2\mu_1\mu_2 + \mu_1^3}{\mu_2-\mu_1^2} \approx \Delta + 2n_0\,,
\end{eqnarray}
and so on. Consequently, the earliest Krylov growth is controlled entirely by the width of the packet in descendant space, and the Krylov complexity
\begin{eqnarray}
    K(t) &=& \sum_n n|\phi_n(t)|^2 = b_1^2 t^2 + O(t^4)\nonumber\\ 
    &\approx& 4\sigma^2 t^2 + O(t^4).
    \label{eq:pp-complexity}
\end{eqnarray}
Evidently, the early-time dynamics depends only on the variance of the spectral distribution and is insensitive to higher details of the state.\\

\noindent
It is quite instructive to connect this directly to our earlier discussion on orthogonal polynomials. In order to do so, recall that the spectral measure for the descendant wavepacket \eqref{eq:pp-wavepacket}, is given by \eqref{eq:gaussian-spectral-measure} with associated weights \eqref{eq:gaussian-weights}. Passing to the integer variable
$x := n$, with energy $E=\Delta+2x$, then latter become
\begin{eqnarray}
    d\mu(E)=\sum_{x\ge0} w_x\,\delta(E-(\Delta+2x))\,dE, \qquad w_x \propto e^{-(x-n_0)^2/(2\sigma^2)}\,,
\end{eqnarray}
so that, up to the affine map $E=\Delta+2x$, the problem is essentially that of orthogonal polynomials on a discrete Gaussian weight supported on (a truncation of) $\mathbb Z_{\ge 0}$. Now let $\{p_k(x)\}$ be the orthonormal polynomials with respect to the weights $w_x$ in the sense that
\begin{eqnarray}
    \sum_{x\ge0} w_x\, p_k(x)\,p_\ell(x)=\delta_{k\ell}\,.
\end{eqnarray}
They satisfy the three-term recurrence
\begin{eqnarray}
    x\,p_k(x)=\beta_{k+1}\,p_{k+1}(x)+\alpha_k\,p_k(x)+\beta_k\,p_{k-1}(x)\,,
\end{eqnarray}
with Jacobi parameters $\{\alpha_k,\beta_k\}$. From our discussion in Section \ref{sec:Spectral-Measures}, the Lanczos coefficients for the energy variable can be read off as
\begin{eqnarray}
    a_k = \Delta + 2\,\alpha_k\,,\qquad b_k = 2\,\beta_k\,.
\end{eqnarray}
As a quick consistency check, using the moment formulas already derived above,
$\alpha_0=\langle x\rangle \approx n_0$, and $\beta_1^2=\mathrm{Var}(x)\approx \sigma^2$,
we recover
$a_0 \approx \Delta+2n_0$, and  $b_1 \approx 2\sigma$,
in agreement with the direct computation. Now suppose that the packet is centered deep in the bulk, far from the lower edge and is sufficiently broad, $1\ll \sigma \ll n_0$, then
the discreteness of $\mathbb Z$ becomes negligible at the scale of the weight, and we can pass to the continuum variable $y := (x-n_0)/\sigma$, for which
$w_x \ \longrightarrow\ w(y)\,dy \propto e^{-y^2/2}\,dy$. In this regime the orthogonal polynomials approach the (orthonormal) Hermite polynomials $H_k(y)$, which obey the recurrence relation
\begin{eqnarray}
    y\,H_k(y)=\sqrt{k+1}\,H_{k+1}(y)+\sqrt{k}\,H_{k-1}(y)\,,
\end{eqnarray}
which, written in terms of the original variable $x = n_0 + \sigma y$, is
\begin{eqnarray}
    x\,p_k(x) = \sigma\sqrt{k+1}\,p_{k+1}(x) + n_0\,p_k(x) + \sigma\sqrt{k}\,p_{k-1}(x)\,,
\end{eqnarray}
so that, in the bulk, where $k\ll \sigma^2$,
\begin{eqnarray}
    \alpha_k \approx n_0\,,\qquad \beta_k \approx \sigma\sqrt{k}\,.
\end{eqnarray}
Mapping back to energy,
\begin{eqnarray}
    a_k \approx \Delta + 2 n_0,\qquad
b_k \approx 2\sigma \sqrt{k}\,.
\end{eqnarray}
In other words, for a tridiagonal chain with $b_k \propto \sqrt{k}$, the evolution is that of a (discrete) harmonic oscillator, and quadratic-in-time growth of Krylov complexity over a parametrically large window,
\begin{eqnarray}
    K(t)\ \sim\ (2\sigma)^2\, t^2 \quad \text{for } \ \sigma^{-1}\ll t \ll 1\,,
\end{eqnarray}
up to order-one constants depending on normalization conventions. Already at the spectral level, a broad Gaussian packet produces a ballistic growth regime before saturation effects eventually set in. These can be estimated as well. For the exact discrete measure, the support in $x$ is effectively confined to
$x \in [\,n_0 - O(\sigma),\ n_0 + O(\sigma)\,]$, and hence the energy support lies in the interval
$E \in [A,B]$, where
\begin{eqnarray}
    A \approx \Delta + 2(n_0 - c\sigma),\quad B \approx \Delta + 2(n_0 + c\sigma),
\end{eqnarray}
for some $O(1)$ constant $c$. Under the standard single-interval, absolutely-continuous approximation, the Jacobi asymptotics of Section \ref{sec:Spectral-Measures} give
\begin{eqnarray}
    a_k \longrightarrow \frac{A+B}{2} \approx \Delta + 2n_0,\qquad b_k \longrightarrow \frac{B-A}{4} \approx c\,\sigma\,.
\end{eqnarray}

\subsection{Coherent states and exact Krylov dynamics}

The Gaussian wavepackets considered above provide a convenient semiclassical description of localized probes, but their Krylov dynamics was obtained only approximately.  It is therefore natural to ask whether there exists a class of states for which the dynamics can be computed exactly. This is in fact the case when the conformal descendants organize into an $su(1,1)$ algebra. In this setting, certain wavepackets can be identified with $SU(1,1)$ coherent states, for which the time evolution is exactly solvable \cite{Caputa:2021sib,Balasubramanian:2022tpr,Chattopadhyay:2023fob,Balasubramanian:2025xkj}.  As we will see, this construction reproduces the characteristic $\sin^2 t$ behavior anticipated from equally spaced spectra, and provides an exact counterpart to the approximate Gaussian analysis above. Toward this end, we identify an $su(1,1)$ subalgebra as
\begin{eqnarray}
\left[ u \cdot K, u \cdot P  \right] & = & 2 D\,,    \nonumber \\
\left[ D, u \cdot K \right] & = & - u \cdot K\,, \\
\left[ D, u\cdot P  \right] & = & u\cdot P\,,    \nonumber
\end{eqnarray}
where $u \cdot u = 1$.  If, in addition we introduce a vector $v$ such that $v \cdot v = 1$ and $u \cdot v = 0$ then
\begin{eqnarray}
\left[  u^\mu v^\nu M_{\mu\nu}, u \cdot P \right] & = & v \cdot P\,,    \nonumber \\
\left[  u^\mu v^\nu M_{\mu\nu}, u \cdot K \right] & = & v \cdot K\,,     \\
\left[  u \cdot K, v \cdot P \right] & = & -2 u^\mu v^\nu M_{\mu\nu}\,,   \nonumber
\end{eqnarray}
packaging the full conformal algebra.  The simplest states to consider involve only the $su(1,1)$ generators. With this in mind, we choose 
\begin{eqnarray}
    |z\rangle = N e^{ z (u \cdot P)}| O_{\Delta}\rangle\,.
\end{eqnarray}
These states are precisely $SU(1,1)$ coherent states \cite{Perelomov} built from the highest-weight representation defined by $|O_\Delta\rangle$.  In particular, they correspond to coherent superpositions of conformal descendants generated along a fixed direction $u^\mu$.  The complexity for an arbitrary $SU(1,1)$ target state $|z_t\rangle$, starting from an $SU(1,1)$ reference state $|z_r\rangle$ is known \cite{Chattopadhyay:2023fob} to be\footnote{Just for this subsection, to avoid confusion with the conformal generator, we will denote the Krylov complexity by $C(t)$ rather than $K(t)$.}
\begin{equation}
C = \frac{\Delta |z_t - z_r|^2 }{(1 - |z_t|^2)(1 - |z_r|^2) }\,.
\end{equation}
For the purposes of computing complexity, our target state is the time-evolved reference state
\begin{eqnarray}
    |z_t\rangle = e^{-i t D} |z_r\rangle =  |e^{-i t} z_r\rangle\,.
\end{eqnarray}
This simple-looking phase rotation reflects the equally spaced spectrum of the dilatation operator within this subsector, and is the algebraic origin of the periodic behavior observed below.
The complexity for this state is given by
\begin{eqnarray}
    C(t) = \Delta \frac{ |z_0|^2  }{1 - |z_0|^2} \sin^2\left(\frac{t}{2}\right)\,.
\end{eqnarray}
It is instructive to compare this result with the Gaussian wavepacket analysis of the previous subsection.  For small times, 
\begin{eqnarray}
    C(t) \sim \Delta \frac{|z_0|^2}{1 - |z_0|^2} \frac{t^2}{4},
\end{eqnarray}
which matches the quadratic growth $K(t)\sim \sigma^2 t^2$ obtained earlier, if we identify the effective variance of the packet with $\sigma^2 \sim \Delta \frac{|z_0|^2}{1 - |z_0|^2}$. The full $\sin^2(t/2)$ dependence, however, is a genuinely nonperturbative effect that arises from the exact $SU(1,1)$ structure and the perfectly equally spaced spectrum of the dilatation operator in this sector. This provides a concrete realization of the general picture developed above; Gaussian wavepackets capture the universal short-time behavior, while coherent-state constructions expose the exact periodic dynamics associated with an affine spectral ladder.

\subsection{The BMN sector: integrable few-body Krylov dynamics}
The Berenstein–Maldacena–Nastase (BMN) sector of $\mathcal{N}=4$ super Yang–Mills theory \cite{Berenstein:2002jq} is a kinematic regime of operators carrying a large charge under a chosen $U(1)\subset SO(6)_R$. Typically, one selects a complex scalar, say Z, and considers single-trace operators built predominantly from $Z$, with a small number of insertions of other fields (impurities), such as another complex scalar $X$, fermions, or covariant derivatives.
In the simplest $SU(2)$ subsector, operators are composed of just the two complex scalars $\{Z,X\}$. A general operator with $M$ impurities then takes the form
\begin{eqnarray}
    \mathcal O \sim \sum_{\{\ell_i\}} \psi(\{\ell_i\})\, \mathrm{Tr}\Big(Z^{\ell_1} X Z^{\ell_2-\ell_1-1} X \cdots Z^{J-\ell_M-1}\Big),
\end{eqnarray}
where $J$ is the total number of $Z$-fields and the impurity number $M$ is held fixed in the limit $J\to\infty$. To be clear, the BMN limit is not just a large-$J$ limit; it is a \textit{double-scaling limit} in which the ’t Hooft coupling $\lambda = g_{\text{YM}}^2 N$ is taken large together with the R-charge $J$, while keeping the effective coupling $\lambda' \equiv \lambda/J^2$ fixed. More precisely, we require
\begin{eqnarray}
    J \to \infty\,,\qquad \lambda \to \infty\,,\qquad \lambda'=\frac{\lambda}{J^2}\ \text{fixed}\,,\qquad M\ \text{fixed}\,.
\end{eqnarray}
In the planar theory, the one-loop dilatation operator acting on these operators reduces to the Hamiltonian of an integrable nearest-neighbor spin chain (the $XXX_{1/2}$ Heisenberg model), with the $Z$-fields forming the vacuum and the impurities behaving as magnons propagating along the chain. The spectrum is therefore exactly solvable via Bethe ansatz, and the dynamics corresponds to a few-body problem on a long background. Physically, this sector describes small excitations around a BPS state of large R-charge. In the dual string theory, it corresponds to string excitations in the pp-wave (Penrose) limit of $\text{AdS}_5\times S^5$, where the large charge $J$ sets the light-cone momentum and the impurities map to string oscillators.\\

\noindent
In the $SU(2)$ spin chain, a single impurity state is
(formally, ignoring the cyclicity of the trace)
$|\ell\rangle = \mathrm{Tr}(Z^\ell X Z^{J-\ell-1})$, and the Hamiltonian acts as a nearest-neighbor hopping operator \cite{Minahan:2002ve},
$H|\ell\rangle = 2g|\ell\rangle - g(|\ell+1\rangle + |\ell-1\rangle)$. This is just a tight-binding particle on a ring which can be diagonalized by
$|k\rangle = \frac{1}{\sqrt{J}} \sum_{\ell} e^{ik\ell} |\ell\rangle$, with dispersion $E(k) = 2g(1-\cos k)$. We have two options for a seed $|\psi_0\rangle$. If $|\psi_0\rangle = |k\rangle$ is a momentum eigenstate, then
$H|\psi_0\rangle = E(k)|\psi_0\rangle$, and the Krylov dynamics is trivial, exactly like BPS. If, on the other hand, $|\psi_0\rangle = |\ell=0\rangle$ is a localized state, then the spectral measure $d\mu(E) = \rho(E)\,dE,$ is continuous (for large $J$) with
$E \in [0,4g]$, so the Krylov dynamics is nontrivial and, in fact, exactly solvable. The spectral density is known explicitly to be $\rho(E) \sim 1/\sqrt{E(4g - E)}$ and corresponds to a constant Jacobi matrix, $a_n = 2g$, $b_n = g$.
In other words, the Krylov chain is exactly translationally invariant at all $n$ and 
the Krylov data is essentially fixed and universal but in a trivial way. We note however that the cyclicity of the trace, requires us to insert another impurity $X$ at the 
origin of the trace, or to consider two general impurities.
\\

\noindent
Consequently, we will work in the $SU(2)$ subsector, with fields $\{Z,X\}$, and consider operators of the form
\begin{eqnarray}
    \mathcal O \sim \sum_{\ell_1<\ell_2} \psi(\ell_1,\ell_2)\,\mathrm{Tr}\big(Z^{\ell_1} X Z^{\ell_2-\ell_1-1} X Z^{J-\ell_2-1}\big)\,,
\end{eqnarray}
with two impurities X propagating on a background of Z-fields. At planar one loop, the dilatation operator reduces to the Heisenberg XXX$_{1/2}$ spin chain with Hamiltonian \cite{Minahan:2002ve}
\begin{eqnarray}
    H = g \sum_{\ell=1}^{J}(1 - P_{\ell,\ell+1}), \qquad g = \frac{\lambda}{8\pi^2}.
    \label{xxx-chain}
\end{eqnarray}
This system is famously integrable and diagonalized by Bethe ansatz. The eigenstates are labeled by magnon momenta $\{k_1,k_2\}$, with energies \cite{Beisert:2005tm}
\begin{eqnarray}
    E(k_1,k_2) = \varepsilon(k_1) + \varepsilon(k_2), \qquad \varepsilon(k) = 2g(1 - \cos k)\,.
\end{eqnarray}
For the seed, we take $|\psi_0\rangle$ to be a generic two-impurity state. Expanding in the energy eigenbasis,
\begin{eqnarray}
    |\psi_0\rangle = \sum_{k_1,k_2} c_{k_1,k_2} |k_1,k_2\rangle\,,
\end{eqnarray}
the spectral measure is
\begin{eqnarray}
    d\mu(E) = \sum_{k_1,k_2} |c_{k_1,k_2}|^2 \,\delta\big(E - E(k_1,k_2)\big)\,dE\,.
\end{eqnarray}
In the large-$J$ limit, the momenta become continuous variables and the spectrum fills a compact interval, $E \in [0, 8g]$. To see why, note that when $J \to \infty$ with fixed number of magnons, the spacing between allowed momenta scales as
$\Delta k \sim \frac{2\pi}{J}$, so the discrete set of allowed $k$’s becomes dense in $[0,2\pi]$ and effectively a continuous variable. At one loop, the magnon dispersion is $\varepsilon(k) = 2g(1 - \cos k)$. Since $k \in [0,2\pi]$, we have $\cos k \in [-1,1]$,
so $\varepsilon(k) \in [0,4g]$. For two impurities (neglecting binding effects, which do not change the support), $E \in [0,4g] + [0,4g] = [0,8g]$ and because $k_1,k_2$ are continuous variables in the large-$J$ limit, the map
$(k_1,k_2) \mapsto E(k_1,k_2)$
sweeps out all values in $[0,8g]$.\\

\noindent
Using the general results of Section 2, the single-interval support immediately implies that
\begin{eqnarray}
    a_n \;\longrightarrow\; \frac{A+B}{2} = 4g, \qquad b_n \;\longrightarrow\; \frac{B-A}{4} = 2g.
\end{eqnarray}
Again, the asymptotic Krylov chain becomes translationally invariant,
\begin{eqnarray}
    H_{\text{Krylov}} \sim \sum_n \left( 2g\,|n\rangle\langle n+1| + \text{h.c.} + 4g\,|n\rangle\langle n| \right),
\end{eqnarray}
corresponding to a tight-binding chain with bandwidth $8g$. The lesson here is that \textit{the BMN sector exhibits bounded, band-limited Krylov dynamics}. In particular, note that the Lanczos coefficients do not grow with $n$, in contrast with chaotic systems where $b_n\sim n$. The asymptotic behaviour of the complexity then 
depends on the initial state. However, simple choices of initial states do not seem to lead to the conjectured $K(t)\propto t^2$ behaviour.\\

\noindent
While the asymptotic behavior of the Lanczos coefficients is fixed by spectral support, the early Lanczos coefficients retain detailed dependence on the initial state. To see this explicitly,  expand the seed in Hamiltonian eigenstates as
\begin{eqnarray}
    |\psi_0\rangle=\sum_\alpha c_\alpha |E_\alpha\rangle, \qquad H|E_\alpha\rangle=E_\alpha |E_\alpha\rangle, \qquad \sum_\alpha |c_\alpha|^2=1.
\end{eqnarray}
In the BMN two-impurity sector, $\alpha$ may be taken to stand for the pair of magnon quantum numbers $(k_1,k_2)$, so we can also write
$|\psi_0\rangle=\sum_{k_1,k_2} c_{k_1,k_2}\,|k_1,k_2\rangle$.
Next, define the spectral weights
$w_\alpha:=|c_\alpha|^2$,
with
$\sum_\alpha w_\alpha=1$.
Then the spectral measure is
$d\mu(E)=\sum_\alpha w_\alpha\,\delta(E-E_\alpha)\,dE$. Let's use this to compute the first few Lanczos coefficients:
\begin{itemize}
    \item \textbf{First Lanczos coefficient} $\bm{a_0}$: By definition, 
    \begin{eqnarray}
        a_0 &=& \langle \psi_0|H|\psi_0\rangle\nonumber\\
        &=& \sum_{\alpha,\beta} c_\alpha^* c_\beta \langle E_\alpha|H|E_\beta\rangle\,.
    \end{eqnarray}
    Since $H|E_\beta\rangle=E_\beta|E_\beta\rangle$ and the eigenstates are orthonormal,
    \begin{eqnarray}
        \langle E_\alpha|H|E_\beta\rangle = E_\beta\,\delta_{\alpha\beta}.
    \end{eqnarray}
   Therefore
   \begin{eqnarray}
       a_0 = \sum_\alpha |c_\alpha|^2 E_\alpha = \sum_\alpha w_\alpha E_\alpha \to \int E\,d\mu(E). \,.
   \end{eqnarray}
   So $a_0$ is just the mean energy of the seed with respect to the spectral distribution $w_\alpha$. In the BMN notation this is
   \begin{eqnarray}
       a_0=\sum_{k_1,k_2} |c_{k_1,k_2}|^2\,E(k_1,k_2)\,.
   \end{eqnarray}
   \item \textbf{Second moment and} $\bm{b_1^2}$: The first residual vector in the Lanczos algorithm is $|f_1\rangle=(H-a_0)|\psi_0\rangle$, and by definition $b_1^2=\langle f_1|f_1\rangle$. Putting this gives
   \begin{eqnarray}
       b_1^2 = \langle \psi_0|(H-a_0)^2|\psi_0\rangle = \langle \psi_0|H^2|\psi_0\rangle-a_0^2\,.
   \end{eqnarray}
   Now
   \begin{eqnarray}
       \langle \psi_0|H^2|\psi_0\rangle &=& \sum_{\alpha,\beta} c_\alpha^* c_\beta \langle E_\alpha|H^2|E_\beta\rangle\nonumber\\ &=& \sum_\alpha |c_\alpha|^2 E_\alpha^2 = \sum_\alpha w_\alpha E_\alpha^2\,,
   \end{eqnarray}
   So
   \begin{eqnarray}
       b_1^2 = \sum_\alpha w_\alpha E_\alpha^2 - \left(\sum_\alpha w_\alpha E_\alpha\right)^2.
   \end{eqnarray}
   Equivalently,
   \begin{eqnarray}
       b_1^2=\int E^2\,d\mu(E)-\left(\int E\,d\mu(E)\right)^2\,.
   \end{eqnarray}
   In this form, it is particularly clear that $b_1^2$ is simply the variance of the seed energy distribution. In BMN notation,
   \begin{eqnarray}
       b_1^2 = \sum_{k_1,k_2}|c_{k_1,k_2}|^2 E(k_1,k_2)^2 - \left(\sum_{k_1,k_2}|c_{k_1,k_2}|^2 E(k_1,k_2)\right)^2.
   \end{eqnarray}
\end{itemize}
The point is that the coefficients $c_{k_1,k_2}=\langle k_1,k_2|\psi_0\rangle$ are the overlaps of the seed with the BMN eigenbasis, and depend on the precise microscopic form of the operator. Let's illustrate with a two-impurity seed peaked around $(k_1^{(0)},k_2^{(0)})$.\\

\noindent
In the $SU(2)$ BMN sector, a two-impurity operator is
\begin{eqnarray}
    |\ell_1,\ell_2\rangle \;\equiv\; \mathrm{Tr}\!\big(Z^{\ell_1} X Z^{\ell_2-\ell_1-1} X Z^{J-\ell_2-1}\big), \qquad \ell_1<\ell_2.
\end{eqnarray}
These form a natural position-space basis for the two-magnon sector. The corresponding energy eigenstates are the Fourier modes
\begin{eqnarray}
    |k_1,k_2\rangle = \frac{1}{J}\sum_{\ell_1<\ell_2} e^{i(k_1 \ell_1 + k_2 \ell_2)} \,|\ell_1,\ell_2\rangle\,,
\end{eqnarray}
up to symmetrization and Bethe scattering phases (which are subleading for our purposes).
These states diagonalize the Hamiltonian in the sense that
$H|k_1,k_2\rangle = E(k_1,k_2)|k_1,k_2\rangle$. Now we can define the seed as a superposition of momentum eigenstates,
\begin{eqnarray}
    |\psi_0\rangle = \mathcal N \sum_{k_1,k_2} \exp\!\left[ -\frac{(k_1-k_1^{(0)})^2}{4\sigma_1^2} -\frac{(k_2-k_2^{(0)})^2}{4\sigma_2^2} \right] |k_1,k_2\rangle\,,
\end{eqnarray}
where $(k_1^{(0)},k_2^{(0)})$ is the center of the packet, $\sigma_{1,2}$ are the widths, and $\mathcal N$ is a normalization constant. Fourier transforming, in position space
\begin{eqnarray}
    |\psi_0\rangle = \sum_{\ell_1<\ell_2} \Psi(\ell_1,\ell_2)\,|\ell_1,\ell_2\rangle\,,
\end{eqnarray}
where
\begin{eqnarray}
    \Psi(\ell_1,\ell_2) = \mathcal N \sum_{k_1,k_2} e^{i(k_1 \ell_1 + k_2 \ell_2)} \exp\!\left[-\frac{(k_1-k_1^{(0)})^2}{4\sigma_1^2} -\frac{(k_2-k_2^{(0)})^2}{4\sigma_2^2}\right].
\end{eqnarray}
In the large-$J$ limit, sums are replaced by integrals and each factor is a standard Fourier transform of a Gaussian, giving
$\Psi(\ell_1,\ell_2)
\propto
e^{i(k_1^{(0)}\ell_1 + k_2^{(0)}\ell_2)}
\,
e^{-\sigma_1^2 \ell_1^2}
\,
e^{-\sigma_2^2 \ell_2^2}$. In other words, in position space the impurities are localized wavepackets with spatial widths $\sim 1/\sigma_a$, and modulated by plane-wave phases. Putting it all together, the seed operator is 
\begin{eqnarray}
    \mathcal O_{\psi} = \sum_{\ell_1<\ell_2} e^{i(k_1^{(0)}\ell_1 + k_2^{(0)}\ell_2)} \,e^{-\sigma_1^2 \ell_1^2 - \sigma_2^2 \ell_2^2}\, \mathrm{Tr}\Big(Z^{\ell_1} X Z^{\ell_2-\ell_1-1} X Z^{J-\ell_2-1}\Big)\,.
\end{eqnarray}
This is a wavepacket of impurities propagating on a long $Z$-background. Computing the overlap for this seed then,
\begin{eqnarray}
    |c_{k_1,k_2}|^2 \propto \exp\!\left[-\frac{(k_1-k_1^{(0)})^2}{2\sigma_1^2} -\frac{(k_2-k_2^{(0)})^2}{2\sigma_2^2}\right]\,.
\end{eqnarray}
For simplicity, we will assume that $J$ is large enough that $k_1,k_2$ may be treated as continuous, the packet is narrow and correlations between $k_1$ and $k_2$ are negligible. Then we can expand the energy around the packet center by writing $k_a = k_a^{(0)} + \delta k_a$, for  $a=1,2$, and
\begin{eqnarray}
    E(k_1,k_2) = E_0 + \sum_{a=1}^2 \left.\frac{\partial E}{\partial k_a}\right|_0 \delta k_a + \frac12\sum_{a,b}\left.\frac{\partial^2 E}{\partial k_a\partial k_b}\right|_0 \delta k_a\delta k_b +\cdots,
\end{eqnarray}
where $E_0 = E(k_1^{(0)},k_2^{(0)})$. Since the dispersion is additive,
\begin{eqnarray}
    \frac{\partial E}{\partial k_1}=2g\sin k_1,\qquad \frac{\partial E}{\partial k_2}=2g\sin k_2,
\end{eqnarray}
and
\begin{eqnarray}
    \frac{\partial^2 E}{\partial k_1^2}=2g\cos k_1,\qquad \frac{\partial^2 E}{\partial k_2^2}=2g\cos k_2,\qquad \frac{\partial^2 E}{\partial k_1\partial k_2}=0.
\end{eqnarray}
To linear order then,
\begin{eqnarray}
    E \approx E_0 + 2g\sin k_1^{(0)}\,\delta k_1 + 2g\sin k_2^{(0)}\,\delta k_2\,.
\end{eqnarray}
Since the Gaussian is centered at the origin in $\delta k_a$, the average $\langle \delta k_a\rangle = 0$. So at leading order,
$a_0 \approx E_0 = 4g - 2g\Big(\cos k_1^{(0)}+\cos k_2^{(0)}\Big)$. Physically, this tells us that the packet mean energy is just the BMN energy at its center. To compute $b_1^2$ for this seed note that, at leading order, the variance comes entirely from the linearized fluctuation
$\delta E \approx 2g\sin k_1^{(0)}\,\delta k_1 + 2g\sin k_2^{(0)}\,\delta k_2$. Therefore
$b_1^2 = \langle (\delta E)^2\rangle$. Assuming the packet factorizes and $\langle \delta k_1 \delta k_2\rangle=0$,
\begin{eqnarray}
    b_1^2 &\approx& (2g)^2\left(\sin^2 k_1^{(0)}\,\langle \delta k_1^2\rangle +\sin^2 k_2^{(0)}\,\langle \delta k_2^2\rangle\right)\nonumber\\
    &=& 4g^2\left(\sin^2 k_1^{(0)}\,\sigma_1^2+\sin^2 k_2^{(0)}\,\sigma_2^2\right)\,,
\end{eqnarray}
since $\langle \delta k_a^2\rangle = \sigma_a^2$. Taking the square root,
\begin{eqnarray}
    b_1 \approx 2g\sqrt{\sin^2 k_1^{(0)}\,\sigma_1^2+\sin^2 k_2^{(0)}\,\sigma_2^2}.
\end{eqnarray}
To make the physics a little more transparent, choose both widths to be the same, $\sigma_1=\sigma_2=\sigma$,
so that
\begin{eqnarray}
    b_1 \approx 2g\,\sigma\,\sqrt{\sin^2 k_1^{(0)}+\sin^2 k_2^{(0)}}\,.
\end{eqnarray}
Now we can distinguish between three separate cases:
\begin{itemize}
    \item \textbf{Exact eigenstate:} As $\sigma\to 0$,
    $b_1\to 0$.

    \item \textbf{Packet centered near band edges:} If $k_a^{(0)}\approx 0$ or $\pi$, then $\sin k_a^{(0)}\approx 0$, so even a moderately broad packet gives small $b_1$.

    \item \textbf{Packet centered near the middle of the band:} If $k_1^{(0)}\approx k_2^{(0)}\approx \pi/2$, then $\sin k_a^{(0)}\approx 1$, and
    $b_1 \approx 2\sqrt2\, g\,\sigma$.
\end{itemize}
\noindent
In other words, the packet has the largest early-time Krylov growth when it is centered where the dispersion has the largest slope.  In appendix \ref{BMNNumerics} we also provide numerical checks for these claims.  

\subsection{Comments on previous proposals for complexity of 
BMN-type states}

\noindent
In \cite{Fatemiabhari:2025cyy}, where the holographic Krylov complexity of a generic infalling particle was computed, a proposal was put forward for calculating the Krylov complexity of a restricted class of BMN-type operators in $\mathcal N=4$ SYM. The idea was to consider not the familiar $SU(2)$ sector, but rather the $SL(2)$ sector, whose operators are built from the complex scalar $Z$ and covariant derivatives $D_+ = D_1 + iD_2$. These operators may be viewed as living on either the cylinder coordinate $w$ or the complex plane coordinate $z$, restricted to a two-dimensional subspace corresponding to the boundary of an $AdS_3 \subset AdS_5$ geometry. As first studied in \cite{Beisert:2003jj}, the $SL(2)$ sector has the important advantage that its underlying $SL(2)$ symmetry matches the isometry group of $AdS_3$. This correspondence allows one to make more robust comparisons with holographic predictions, including for example \cite{Caputa:2024sux}.\\

\noindent
The key observation of \cite{Fatemiabhari:2025cyy} is that operators containing $D_+$ insertions acting on a background of $Z$ fields admit a spin-chain description. The $J$ scalar fields $Z$ define a chain of length $J$, while the \(S\) derivative insertions \(D_+\) play the role of excitations propagating along the chain. At fixed chain length \(J\) and fixed excitation number \(S\), the Krylov basis can be organized by the total derivative number $n$, and labelled by the positions of the insertions,
\begin{eqnarray}
    |l_1,\ldots,l_S\rangle, \qquad
    l_S>\cdots>l_1,
\end{eqnarray}
or equivalently by the total level
\begin{eqnarray}
    n=l_1+\cdots+l_S, \qquad |n\rangle \equiv
    |l_1,\ldots,l_S\rangle.
\end{eqnarray}
At one loop, the dilatation operator in this sector is mapped to the nearest-neighbour Heisenberg XXX$_{1/2}$ spin-chain Hamiltonian. This considerably simplifies the analysis of Krylov dynamics. In particular, as we saw in the previous subsection, the two-impurity $S=2$ sector can be diagonalized explicitly, allowing the associated Krylov chain to be constructed analytically. Unlike the $SU(2)$ sector, the $SL(2)$ sector shares the same noncompact symmetry group that underlies the AdS$_3$/CFT$_2$ correspondence. While both sectors are integrable at one loop, the \(SL(2)\) spin chain furnishes representations of a noncompact symmetry algebra and is therefore expected to capture features of \(AdS_3\) dynamics more faithfully. This makes it a natural arena in which to search for the quadratic Krylov growth,
$K(t)\propto t^2$, predicted holographically for probes in a CFT$_2$ defined on an infinite spatial line.\\

\noindent
There are, however, two important caveats. First, a meaningful comparison with semiclassical gravity requires the strong-coupling regime of the AdS/CFT correspondence, namely \(\lambda\to\infty\). From this perspective, restricting attention to the one-loop dilatation operator is insufficient, since it captures only the weak-coupling dynamics of the gauge theory. Second, an asymptotically growing Krylov complexity requires an infinite Krylov chain. At fixed impurity number \(S\), such as the two-impurity sector considered in the previous subsection, this in turn requires taking the spin-chain length \(L\) to infinity, thereby generating an infinite set of Krylov states \(|n\rangle\).\\

\noindent
The standard BMN limit, in which \(\lambda/J^2\) is held fixed while the number of impurities remains finite, does not satisfy the physical requirements of the holographic problem. The corresponding states are dual to strings localized near the centre of \(AdS_5\) and rotating on the \(S^5\), rather than to probes falling in from the boundary of \(AdS_5\). Moreover, although the interacting SYM Hamiltonian,
$H_{\rm int,SYM}\propto \lambda,$
continues to act through nearest-neighbour interactions on states with a finite number of impurities, this structure alone is not sufficient to reproduce the dynamics expected of infalling holographic probes. To make contact with the gravity picture, one must therefore consider both a more general class of states and a more appropriate strong-coupling limit.

\subsection{BMN wavepackets and the LLM droplet}

\noindent
In order to bridge to the conjectured universality in \cite{Nastase:2026lhz}, it will be useful to translate the BMN wavepacket construction into the dual geometric language. In the 1/2-BPS sector, states built from the complex scalar $Z$ admit an exact description in terms of the Lin–Lunin–Maldacena (LLM) geometry \cite{Lin:2004nb}, in which the gauge theory state is mapped to a droplet of free fermions in a two-dimensional phase space. The ground state corresponds to a circular droplet, whose boundary supports chiral edge excitations. Small fluctuations of this boundary can be expanded in Fourier modes,
\begin{eqnarray}
    \delta r(\theta,t)=\sum_{n\ge 1}\left(a_n e^{in\theta}+a_n^\dagger e^{-in\theta}\right)\,,
\end{eqnarray}
with mode number $n$ identified with angular momentum along the droplet edge.\\

\noindent
In this language, the BMN momentum quantum number is naturally identified with the Fourier mode number of the boundary excitation. A momentum eigenstate therefore corresponds to a single harmonic deformation of the droplet boundary, delocalized along the angular direction. By contrast, a superposition of nearby momenta, such as the Gaussian wavepacket constructed above, corresponds to a localized disturbance along the droplet edge. In particular, a wavepacket centered at $(k_1^{(0)},k_2^{(0)})$ defines a two-impurity excitation whose profile is localized in the angular coordinate and propagates along the boundary. This in turn gives us a clear geometric interpretation of the field-theoretic construction; BMN wavepackets correspond to localized edge excitations propagating on the LLM droplet.\\

\noindent
The distinction between eigenstates and wavepackets is now also transparent. A pure momentum eigenstate produces a rigid, delocalized deformation, while a wavepacket describes a localized excitation whose motion reflects the underlying dispersion relation. In the strict 1/2-BPS limit, these edge modes propagate without distortion, corresponding to the absence of interactions. The inclusion of impurities in the BMN sector introduces nontrivial dispersion and mixing, deforming this picture into that of a weakly interacting excitation propagating on a large-$J$ background. This perspective also clarifies the contrast with subsection 3.5. Superpositions of BPS states correspond to quantum superpositions of distinct droplet geometries, with no notion of propagation within a single background. In the BMN sector however, the background geometry is fixed, and the dynamics arises from the motion of localized excitations on that background.\\

\noindent
This is important to the interpretation of Krylov dynamics in this sector. The early-time Lanczos coefficients, which depend sensitively on the spectral distribution of the seed, now have a geometric meaning in that they probe how a localized excitation samples the dispersion relation of the background geometry. In particular, the dependence of $b_1$ on the width and location of the wavepacket reflects how sharply the excitation is localized and where it sits relative to the band structure.
However, despite this appealing geometric picture, the BMN sector remains a few-body, integrable system. The resulting Krylov dynamics, while nontrivial, is still governed by a compact spectral band and exhibits bounded Lanczos coefficients. As such, it captures probe-dependent early-time structure but does not realize the universal leading growth expected in genuinely many-body or strongly chaotic regimes.\\

\noindent
This motivates the transition to the next class of operators, in which extended probes, such as strings attached to giant gravitons, hopefully provide a richer setting for testing universality.

\section{Giant Gravitons and Giant–Open String Operators}
\label{sec:giants}
As we have seen, the BMN sector provides perhaps the simplest realization of nontrivial Krylov dynamics in $\mathcal{N}=4$ SYM, corresponding to a small number of excitations propagating on a large-$J$ BPS background. However, it remains a few-body, integrable system, and therefore does not capture the structural features expected of universal operator growth. A natural next step is to consider extended probes in the dual geometry. In the gauge theory, these are described by giant graviton operators and their excitations \cite{McGreevy:2000cw, Corley:2001zk, Balasubramanian:2001nh}. These sectors retain analytic control while introducing qualitatively new ingredients such as nontrivial representation-theoretic structure, open-string degrees of freedom and boundary effects in the effective spin chain. We would therefore expect that they provide a more refined testing ground for the interplay between probe structure and Krylov dynamics.

\subsection{Giant graviton operators in $\mathcal{N}=4$ SYM}
Giant gravitons are described by 1/2-BPS operators built from the complex scalar $Z$, organized in terms of Schur polynomials \cite{Corley:2001zk, deMelloKoch:2007uu} labeled by Young diagrams $R$,
\begin{eqnarray}
    \chi_R(Z^{\otimes n}) = \frac{1}{n!}\sum_{\sigma\in S_n} \chi_R(\sigma)\,\mathrm{Tr}(\sigma Z^{\otimes n})\,,
    \label{eq:schur}
\end{eqnarray}
where $n$ is the number of fields and $\chi_R(\sigma)$ is the character of the permutation $\sigma$ in representation $R$.
At large $N$, these operators have a direct geometric interpretation. Specifically, Young diagrams with a large number of rows correspond to sphere giant gravitons, while those with a large number of columns correspond to AdS giant gravitons. In both cases, the operator describes a coherent, semiclassical object in the bulk, rather than a small fluctuation around the vacuum. Now, take a Schur operator $\chi_R(Z)$,
with $R$ a Young diagram of size $|R|=p=\mathcal O(N)$ for a giant graviton.
Since $\chi_R(Z)$ is 1/2-BPS,
$[D,\chi_R]=p\,\chi_R$. Consequently, following our arguments above, $\mu_1=p$, $\mu_2=p^2$ and $b_1^2=\mu_2-\mu_1^2=0$, so exact giant graviton operators are Krylov-trivial. This is exactly parallel to the exact BPS and exact BMN-eigenstate story. To get anything nontrivial, we need to use a non-eigen seed. 

\subsection{Wavepackets of Giants}
\label{subsec:schur_wavepacket}
From the gravitational perspective, (sphere) giant-gravitons are semiclassical D3-brane configurations that wrap an $S^3\subset S^5$, carry angular momentum, and are allowed to move in the AdS radial direction $r(t)$ as well as in internal coordinates $\theta(t)$, and $\phi(t)$. Indeed, in \cite{Nastase:2026lhz} they are considered to be pointlike, but structured probes. A falling giant-graviton should therefore be dual to a state or operator obtained by taking a static 1/2-BPS giant-graviton and forming a time-dependent semiclassical packet in the relevant collective coordinates, so that it is localized in $r,\theta,\phi$ rather than being a stationary eigenstate. Schematically, an infalling giant-graviton should take the form
\begin{eqnarray}
    |\Psi_{\text{falling giant}}\rangle \sim \sum_{R,\;J,\;\text{collective modes}} c_{R,J,\ldots}\,|R,J,\ldots\rangle\,,
\end{eqnarray}
where $|R,J,\ldots\rangle$ are giant-graviton states labeled by Young-diagram data and angular-momentum/collective-coordinate quantum numbers, and the coefficients $c_{R,J,\ldots}$ are chosen so that the state has a semiclassical profile corresponding to a brane localized in the bulk and then falling inward.\\

\noindent
Having motivated above for the Schur polynomial \eqref{eq:schur} as the general giant-graviton operator in the field theory, it will suffice to make our argument to focus on one particular representation and so, without loss of generality, we take $R$ to be the totally antisymmetric representation; a single column of length $p$, say corresponding to a sphere giant. More precisely, for $p=0,1,\dots,N$, let
$|A_p\rangle$ denote the state dual to the totally antisymmetric Schur polynomial with one column of length $p$, $\chi_{A_p}(Z)$. These are 1/2-BPS states, so in radial quantization they are eigenstates of the dilatation operator,
\begin{eqnarray}
    H\,|A_p\rangle = E_p\,|A_p\rangle\,.
\end{eqnarray}
In the strict BPS sector, $E_p = p$
up to a choice of units. We will use that normalization from now on.
The states are orthonormal, with 
$\langle A_p|A_{p'}\rangle=\delta_{pp'}$ and so furnish a perfectly good discrete basis for giant-graviton states. Next we define the seed
\begin{eqnarray}
    |\Psi_0\rangle = \mathcal N \sum_{p=0}^{N} \exp\!\left[-\frac{(p-p_0)^2}{4\Sigma^2}\right] e^{iqp}\, |A_p\rangle\,,
\end{eqnarray}
where $p_0$ is the mean giant size,
$\Sigma$ is the width of the packet in Young-diagram size,
$q$ is a phase, and $\mathcal N$ is fixed by normalization. Since the states are orthonormal,
\begin{eqnarray}
    1=\langle \Psi_0|\Psi_0\rangle = |\mathcal N|^2\sum_{p=0}^{N} \exp\!\left[-\frac{(p-p_0)^2}{2\Sigma^2}\right]\,.
\end{eqnarray}
setting
\begin{eqnarray}
    \mathcal N^{-2} = \sum_{p=0}^{N} \exp\!\left[-\frac{(p-p_0)^2}{2\Sigma^2}\right]\,.
\end{eqnarray}
In order to mimic the computation in section \ref{subsec:primary-wave}, it will be convenient to define the probability weights
\begin{eqnarray}
    w_p:=|\langle A_p|\Psi_0\rangle|^2 = |\mathcal N|^2 \exp\!\left[-\frac{(p-p_0)^2}{2\Sigma^2}\right]\,.
\end{eqnarray}
Notice that the phase $e^{iqp}$ drops out of all spectral moments in the exact BPS basis, so in the strict Schur problem, the Krylov data will depend only on $w_p$, not on $q$.\\

\noindent
Adapting our discussion above, for a normalized seed
\begin{eqnarray}
    |\Psi_0\rangle=\sum_{p=0}^N c_p |A_p\rangle, \qquad w_p=|c_p|^2\,,
\end{eqnarray}
the spectral measure is defined by
$d\mu(E)=\langle \Psi_0|\delta(E-H)|\Psi_0\rangle\,dE$. Since
$H|A_p\rangle=p|A_p\rangle$,
the operator-valued delta function acts diagonally as
\begin{eqnarray}
    \delta(E-H)|A_p\rangle=\delta(E-p)|A_p\rangle\,.
\end{eqnarray}
Consequently, 
\begin{eqnarray}
    \langle \Psi_0|\delta(E-H)|\Psi_0\rangle &=& \sum_{p,p'} c_{p'}^*c_p \langle A_{p'}|\delta(E-H)|A_p\rangle\\ &=& \sum_{p,p'} c_{p'}^*c_p \delta(E-p)\langle A_{p'}|A_p\rangle\nonumber\\
    &=& \sum_{p=0}^N |c_p|^2\,\delta(E-p) = \sum_{p=0}^N w_p\,\delta(E-p)\,.
\end{eqnarray}
Hence, the spectral measure is given by
\begin{eqnarray}
   d\mu(E)=\sum_{p=0}^N w_p\,\delta(E-p)\,dE\,.
\end{eqnarray}
Another way of stating this is that the spectral measure is just the probability distribution of energy eigenvalues seen by the seed in the sense that, for any function $f$,
\begin{eqnarray}
    \langle \Psi_0|f(H)|\Psi_0\rangle = \sum_p w_p f(p) = \int f(E)\,d\mu(E)\,.
\end{eqnarray}
In any event, it is a purely discrete measure supported on the integer set $\{0,1,\dots,N\}$, weighted by the Gaussian packet and all Krylov data is determined by its moments
\begin{eqnarray}
    \mu_n=\langle \Psi_0|H^n|\Psi_0\rangle = \sum_{p=0}^{N} w_p\, p^n.
\end{eqnarray}
From the general formulae discussed above,
\begin{eqnarray}
    a_0=\mu_1, \qquad b_1^2=\mu_2-\mu_1^2,
\end{eqnarray}
and
\begin{eqnarray}
    a_1=\frac{\mu_3-2\mu_1\mu_2+\mu_1^3}{\mu_2-\mu_1^2}\,,
\end{eqnarray}
etc. Consequently,
\begin{eqnarray}
    a_0&=&\sum_{p=0}^{N} w_p\,p,\nonumber\\
    b_1^2&=&\sum_{p=0}^{N} w_p\,p^2-\left(\sum_{p=0}^{N} w_p\,p\right)^2,\nonumber\\
    a_1 &=& \frac{\sum_{p=0}^{N} w_p\,p^3 -2\left(\sum_{p=0}^{N} w_p\,p\right)\left(\sum_{p=0}^{N} w_p\,p^2\right) +\left(\sum_{p=0}^{N} w_p\,p\right)^3}{ \sum_{p=0}^{N} w_p\,p^2-\left(\sum_{p=0}^{N} w_p\,p\right)^2}\,.
    \label{eq:gg-packet-coefficients}
\end{eqnarray}
This is already an explicit, if unwieldy, closed-form construction.
As a first pass at extracting the physics of this probe, let's assume the packet is well inside the allowed range, $1\ll \Sigma \ll N$, with $p_0 \gg \Sigma$,
and  $N-p_0\gg \Sigma$,
so that truncation at $p=0$ and $p=N$ can be neglected. Then, the discrete Gaussian may be approximated by a continuous Gaussian on $\mathbb R$, and we recover the standard moments,
\begin{eqnarray}
    \mu_1 &\approx& p_0\,,\nonumber\\ 
    \mu_2 &\approx& p_0^2+\Sigma^2\,,\\ 
    \mu_3 &\approx& p_0^3+3p_0\Sigma^2\,.\nonumber
\end{eqnarray}
Substituting into \eqref{eq:gg-packet-coefficients} gives the first few coefficients 
\begin{eqnarray}
    a_0 &\approx& p_0\,,\nonumber\\
    b_1^2 &\approx& (p_0^2+\Sigma^2)-p_0^2 = \Sigma^2 \Rightarrow b_1 \approx \Sigma\,,\\
    a_1 &\approx& \frac{(p_0^3+3p_0\Sigma^2)-2p_0(p_0^2+\Sigma^2)+p_0^3}{\Sigma^2} = p_0\,.
\end{eqnarray}
In words, for a Gaussian wavepacket well away from the boundaries, the first two diagonal Lanczos coefficients are both approximately the packet center, while the first off-diagonal coefficient is the packet width.
Consequently, the early Krylov growth is controlled simply by how broad the Schur packet is. In particular,
\begin{eqnarray}
    K(t)=b_1^2 t^2 + O(t^4) \approx \Sigma^2 t^2 + O(t^4)\,.
    \label{eq:gg-complexity}
\end{eqnarray}
In this approximation at least,    it is clear that a narrow packet in $p$ has very small early-time complexity growth, while a broad packet has larger early-time growth.\\

\noindent
If, on the other hand, the packet is close to $p=0$ or $p=N$, the Gaussian is truncated, the continuum approximation breaks down and the moments shift. This is important physically, because packets near $p=0$ correspond to very small giants, while packets near $p=N$ correspond to giants approaching maximal size. So the finite-$N$ edge of the Schur basis acts like a hard wall in the collective-coordinate space. To uncover possible edge effects, define
\begin{eqnarray}
    Z_0(p_0,\Sigma;N) := \sum_{p=0}^{N} \exp\!\left[-\frac{(p-p_0)^2}{2\Sigma^2}\right]\,,
    \label{eq:Z0}
\end{eqnarray}
in terms of which the weights
\begin{eqnarray}
    w_p=\frac{1}{Z_0}\exp\!\left[-\frac{(p-p_0)^2}{2\Sigma^2}\right]\,.
\end{eqnarray}
We also define the quantities
\begin{eqnarray}
    Z_1&:=&\sum_{p=0}^{N} p\,e^{-(p-p_0)^2/(2\Sigma^2)}\,,\nonumber\\ 
    Z_2&:=&\sum_{p=0}^{N} p^2\,e^{-(p-p_0)^2/(2\Sigma^2)}\,,\\
    Z_3&:=&\sum_{p=0}^{N} p^3\,e^{-(p-p_0)^2/(2\Sigma^2)}\,.\nonumber
\end{eqnarray}
Then the first Lanczos coefficients can be expressed as
\begin{eqnarray}
    a_0&=&\frac{Z_1}{Z_0}\,,\nonumber\\
    b_1^2&=&\frac{Z_2}{Z_0}-\left(\frac{Z_1}{Z_0}\right)^2\,,\\
    a_1&=& \frac{\frac{Z_3}{Z_0} -2\frac{Z_1}{Z_0}\frac{Z_2}{Z_0}+\left(\frac{Z_1}{Z_0}\right)^3}{\frac{Z_2}{Z_0}-\left(\frac{Z_1}{Z_0}\right)^2}\,.
\end{eqnarray}
These are exact on the finite interval $0\le p\le N$. Differentiating $Z_0$, we find
\begin{eqnarray}
    \partial_{p_0}\log Z_0 &=& \frac{\langle p\rangle-p_0}{\Sigma^2}\,,\nonumber\\
    \partial_{p_0}^2\log Z_0 &=& \frac{\mathrm{Var}(p)}{\Sigma^4}-\frac{1}{\Sigma^2}\,,\quad etc.
\end{eqnarray}
Consequently, $b_1^2=\mathrm{Var}(p) = \Sigma^2+\Sigma^4\,\partial_{p_0}^2\log Z_0$. Near the upper endpoint $p_0\simeq N$, $Z_0$ decreases in curvature because the positive tail $p>N$ has been removed. In that regime then, $\partial_{p_0}^2\log Z_0<0$, and hence $b_1^2<\Sigma^2$. Physically, the reason is that the upper tail of the Gaussian, which would have contributed large positive fluctuations $p-p_0$, is removed by the finite-$N$ bound $p\le N$. The distribution then becomes narrower and skewed away from the endpoint, so the early Krylov growth $K(t)=b_1^2t^2+\cdots$ is suppressed relative to the infinite-$N$ estimate of $K(t)\simeq \Sigma^2t^2$.

\subsection{Falling point particles vs falling giants}
\label{sec:comparison}
In order to compare Krylov complexities for the falling point-particle and the falling giant graviton, we need one more result from the theory of orthogonal polynomials: suppose a measure $\mu(x)$ has orthonormal-polynomial recurrence
\begin{eqnarray}
    x\,p_n(x)=b_{n+1}p_{n+1}(x)+a_n p_n(x)+b_n p_{n-1}(x)\,.
\end{eqnarray}
If we define a new variable
$E=\alpha+\beta x$, then the new recurrence becomes
\begin{eqnarray}
    E\,p_n(x) = \beta b_{n+1}p_{n+1}(x)+(\alpha+\beta a_n)p_n(x)
    +\beta b_n p_{n-1}(x)\,.
\end{eqnarray}
In other words, under an affine change of spectral variable,
\begin{eqnarray}
    a_n \;\mapsto\; \alpha+\beta a_n, \qquad b_n \;\mapsto\; |\beta|\, b_n\,.
    \label{eq:affine-map}
\end{eqnarray}
In our case, both wavepackets are superpositions of eigenstates on an equally spaced ladder. For the point particle, $E_n^{\rm pp}=\Delta+2n$, while for the giant, $E_p^{\rm gg}=p$, and both spectral measures are discrete Gaussians on a uniform lattice,
\begin{eqnarray}
    d\mu_{\rm pp}(E)=\sum_n w_n\,\delta(E-(\Delta+2n))\,dE,
    \qquad d\mu_{\rm gg}(E)=\sum_p v_p\,\delta(E-p)\,dE\,.
\end{eqnarray}
If we choose the same packet shape in the index variable, $w_n \equiv v_n$, then the two measures are related by the affine map $E_{\rm pp}=\Delta+2E_{\rm gg}$. Applying this to \eqref{eq:affine-map} immediately gives
\begin{eqnarray}
   a_n^{\rm pp}=\Delta+2\,a_n^{\rm gg}\,, \qquad b_n^{\rm pp}=2\,b_n^{\rm gg}\,,
\end{eqnarray}
again, provided the packet weights in the index variable are the same. This means that the two Krylov problems are in the \textit{same affine universality class}. Earlier, in section \ref{subsec:primary-wave} we found that the point-particle descendant packet,
\begin{eqnarray}
    a_0^{\rm pp}\approx \Delta+2n_0\,,\qquad b_1^{\rm pp}\approx 2\sigma\,,\qquad a_1^{\rm pp}\approx \Delta+2n_0\,,
\end{eqnarray}
while for the giant packet, away from the edges,
\begin{eqnarray}
    a_0^{\rm gg}\approx p_0,\qquad b_1^{\rm gg}\approx \Sigma,\qquad a_1^{\rm gg}\approx p_0\,.
\end{eqnarray}
If we identify $p_0=n_0$, and $\Sigma=\sigma$,
then indeed
\begin{eqnarray}
    a_0^{\rm pp}=\Delta+2a_0^{\rm gg}, \qquad b_1^{\rm pp}=2b_1^{\rm gg}, \qquad a_1^{\rm pp}=\Delta+2a_1^{\rm gg}\,,
\end{eqnarray}
and indeed the earliest Krylov data matches exactly under the affine rescaling. To appreciate the consequences this has for the Krylov complexity, recall that for the two packets, we found that $K_{\rm g}(t)\approx \Sigma^2 t^2$, and 
$K_{\rm pp}(t)\approx 4\sigma^2 t^2$. If we would choose
$\Sigma=\sigma$, then
\begin{eqnarray}
    K_{\rm pp}(t)\approx 4\,K_{\rm gg}(t)\,,
\end{eqnarray}
at the same early time $t$, simply because the descendant spectrum has spacing 2 while the Schur spectrum has spacing 1. Equivalently, if we rescale time as
$t_{\rm pp}=\frac12 t_{\rm gg}$,
then the two early-time complexity laws coincide. 
However, the result in \cite{Nastase:2026lhz} suggests that it is 
rather the giant graviton that has a bigger complexity 
(proportionality  constant) than the point particle, so we expect
that, rather, $\Sigma>2\sigma$. This makes sense, since a giant 
graviton should be more complex than a point particle.
At the level of the two wavepackets, the point-particle packet is a packet over descendant levels of one local primary and the giant packet is constructed over giant sizes in the Schur basis. While these are physically very different bases,  mathematically they are both Gaussian packets on a uniform energy ladder. It is not surprising then that their Krylov dynamics is identical up to an affine redefinition of the energy scale.\\

\noindent
There remains an important caveat to this story. Both of these constructions are still purely spectral; the Schur packet is constructed over BPS eigenstates, and the descendant packet over conformal eigenstates. This means that the universality we are seeing here is a feature of spectral packets on equally spaced ladders and not yet the full universality of falling dynamical probes. In particular, neither packet includes interactions, true operator mixing, or any nontrivial radial infall dynamics in the SYM Hamiltonian beyond phase accumulation. As a result, this comparison should be viewed as perhaps a zeroth-order field-theory realization of the conjecture in \cite{Nastase:2026lhz}, with an important caveat that we will now address below.

\section{Finite-density operator growth and asymptotic universality}
\label{sec:finite-density}

The preceding sections focused primarily on protected and few-body sectors of $\mathcal N=4$ SYM. These examples are analytically tractable and provide useful intuition regarding the relation between spectral structure and Krylov dynamics. However, they also exhibit an important limitation in that the number of active excitations remains finite in the thermodynamic limit. Consequently, the associated spectral measures remain effectively few-body and compact, and the resulting Lanczos coefficients either remain bounded or asymptotically constant. Such sectors therefore probe only local or early-time operator growth, rather than the asymptotic Krylov geometry expected in genuinely chaotic many-body dynamics. This observation suggests that a better boundary counterpart of the holographic universality proposed in \cite{Nastase:2026lhz} should be formulated not in a few-body regime, but in a finite-density many-body regime.\\

\noindent
To this end, we now consider the thermodynamic limit
\begin{eqnarray}
    L\to\infty, \qquad M\to\infty, \qquad \rho=\frac{M}{L} \quad\text{fixed},
\end{eqnarray}
where $L$ is the spin-chain length and $M$ is the number of impurities. The quantity $\rho$ is the impurity filling fraction. At this point it is important to emphasize that these states can no longer naturally interpreted as single infalling probes. Rather, they should be viewed as many-body boundary states designed to test whether the proposed holographic universality is fundamentally a thermodynamic statement about asymptotic Krylov geometry. The central question is then whether different classes of probes with the same coarse thermodynamic data such as energy density, filling fraction and conserved charges, exhibit the same asymptotic Lanczos structure. We now explicitly compute the leading Krylov data for several finite-density probe families.

\subsection{Localized impurity product states}
We begin with the simplest finite-density states in the one-loop $SU(2)$ sector. These are the localized impurity configurations that we have already seen are particularly useful because their Krylov data can be computed exactly and transparently. This in turn allows us to isolate the microscopic origin of the leading operator-growth coefficients.\\

\noindent
The one-loop dilatation operator in the $SU(2)$ sector is the ferromagnetic Heisenberg XXX$_{1/2}$ spin chain, \eqref{xxx-chain} where we recall that $P_{\ell,\ell+1}$ exchanges the spins at neighboring sites and periodic boundary conditions are understood unless otherwise stated. The local Hilbert space at each site is two-dimensional,
\begin{eqnarray}
    |Z\rangle = \begin{pmatrix}1\\0\end{pmatrix}, \qquad |X\rangle = \begin{pmatrix}0\\1\end{pmatrix}\,,
\end{eqnarray}
so that the spin chain provides a direct representation of single-trace operators built from the complex scalars $Z$ and $X$.
We consider a basis configuration
\begin{eqnarray}
    |C\rangle = |s_1s_2\cdots s_L\rangle, \qquad s_\ell\in\{Z,X\},
\end{eqnarray}
containing $M=\rho L$
impurities $X$. Such states are completely localized in configuration space and should  be viewed as maximally inhomogeneous finite-density probes. The action of the Hamiltonian is particularly simple in this basis. Define the set of domain walls
\begin{eqnarray}
    {\rm DW}(C) = \{\ell : s_\ell\neq s_{\ell+1}\},
\end{eqnarray}
and let $D(C)=|{\rm DW}(C)|$
be the total number of interfaces between neighboring unlike spins.
The reason domain walls are central is clear from the action of the permutation operator. If two neighboring spins are identical,
\begin{eqnarray}
    P_{\ell,\ell+1}|ZZ\rangle
    =|ZZ\rangle, \qquad P_{\ell,\ell+1}|XX\rangle =|XX\rangle,
\end{eqnarray}
and therefore
$(1-P_{\ell,\ell+1})|ZZ\rangle=0$,
while $(1-P_{\ell,\ell+1})|XX\rangle=0$.
In other words, equal neighboring spins contribute no energy. By contrast, for a domain wall,
\begin{eqnarray}
    P_{\ell,\ell+1}|ZX\rangle
    =|XZ\rangle, \qquad P_{\ell,\ell+1}|XZ\rangle =|ZX\rangle,
\end{eqnarray}
so that $(1-P_{\ell,\ell+1})|ZX\rangle
=
|ZX\rangle-|XZ\rangle$,
and similarly for $|XZ\rangle$.
Consequently, only domain-wall bonds contribute nontrivially to the Hamiltonian. If $|C_\ell\rangle$ denotes the configuration obtained by swapping the spins across the domain wall at bond $\ell$, then
\begin{eqnarray}
    H|C\rangle = gD(C)|C\rangle - g\sum_{\ell\in{\rm DW}(C)} |C_\ell\rangle.
\end{eqnarray}
This also has a simple physical interpretation. The diagonal term
$gD(C)|C\rangle$
counts the energetic cost of interfaces, while the off-diagonal terms describe hopping or diffusion of domain walls along the chain. The Hamiltonian therefore acts locally by rearranging neighboring impurities.\\

\noindent
We can now compute the first Lanczos coefficients exactly.
The initial Lanczos coefficient is simply the expectation value of the Hamiltonian,
$a_0
=
\langle C|H|C\rangle$.
Since the swapped configurations $|C_\ell\rangle$ are orthogonal to the original product state,
$\langle C|C_\ell\rangle=0$,
it immediately follows that 
$a_0=gD(C)$. Consequently, the initial Krylov energy is determined entirely by the number of interfaces between impurities and background fields.\\

\noindent
The next Lanczos coefficient probes the variance of the Hamiltonian,
$b_1^2
=
\|(H-a_0)|C\rangle\|^2$.
Using the explicit action of the Hamiltonian,
\begin{eqnarray}
    (H-a_0)|C\rangle = -g\sum_{\ell\in{\rm DW}(C)} |C_\ell\rangle\,.
\end{eqnarray}
The states $|C_\ell\rangle$ associated with distinct domain walls are orthogonal product configurations. Therefore
\begin{eqnarray}
    \langle C_\ell|C_{\ell'}\rangle = \delta_{\ell\ell'}\,,
\end{eqnarray}
and the norm is
\begin{eqnarray}
    b_1^2 = g^2 \sum_{\ell\in{\rm DW}(C)}1 = g^2D(C).
\end{eqnarray}
Hence $b_1^2 = g^2D(C)$. The quantity $b_1^2$ measures the number of distinct local moves generated by the Hamiltonian acting on the initial state. In these localized impurity configurations, each domain wall contributes precisely one possible local rearrangement. The early Krylov growth is therefore controlled by the mobility of interfaces. Now let's introduce the domain-wall density $\delta
= D(C)/L$, in terms of which
\begin{eqnarray}
    \frac{a_0}{L} = g\delta, \qquad \frac{b_1^2}{L} = g^2\delta.
\end{eqnarray}
Evidently, both the initial energy density and the initial Krylov variance density are controlled entirely by the density of domain walls.
This becomes particularly transparent for random product states. Suppose each site independently contains an impurity with probability $\rho$. Then a domain wall occurs whenever neighboring sites differ
$ZX$ or $XZ$. The probability for $ZX$ is $(1-\rho)\rho$,
and similarly for $XZ$. Therefore the average domain-wall density is
$\delta = 2\rho(1-\rho)$. Substituting into the previous expressions yields
\begin{eqnarray}
    \frac{a_0}{L} = 2g\rho(1-\rho), \qquad \frac{b_1^2}{L} = 2g^2\rho(1-\rho).
\end{eqnarray}
Note the following:
\begin{itemize}
    \item First, both quantities vanish at $\rho=0$ and $\rho=1$, corresponding respectively to the ferromagnetic vacua consisting entirely of $Z$-fields or entirely of $X$-fields. In these cases there are no interfaces and hence no local dynamics.
    \item Second, the domain-wall density is maximal at $\rho=\frac12$, where the chain is maximally disordered and the number of local rearrangements is largest.
\end{itemize}
Most importantly, these results demonstrate explicitly that early-time Krylov growth in finite-density states is strongly probe-dependent in its coefficient, although the $K(t)
\propto t^2$ behaviour still holds. The coefficient
$b_1^2$ is decidedly not universal, but depends sensitively on the microscopic spatial structure of the initial state through the domain-wall density. In particular, the growth is controlled not merely by the impurity density itself, but by how the impurities are arranged spatially. Two states with the same filling fraction $\rho$ can therefore exhibit very different initial Krylov dynamics if their domain-wall structure differs. This observation will play an important role below when comparing localized impurity states with coherent magnon condensates and giant-open-string configurations.

\subsection{Coherent magnon condensates}

Next, let us consider the qualitatively different class of finite-density states that are \textit{coherent magnon condensates}. Unlike the localized impurity configurations above, these states are spatially delocalized and possess long-range phase coherence. They therefore provide a natural probe of how coherent many-body structures modify Krylov dynamics.
We'll start by defining the spin-coherent state
\begin{eqnarray}
    |\Psi_q\rangle
=
\bigotimes_{\ell=1}^{L}
|s_\ell\rangle\,,
\end{eqnarray}
with local spin state
\begin{eqnarray}
    |s_\ell\rangle = \sqrt{1-\rho}\,|Z\rangle 
    + \sqrt{\rho}\,e^{iq\ell}
    |X\rangle\,.
\end{eqnarray}
Here $\rho$ is again the impurity filling fraction, and $q$ is a uniform phase gradient or magnon momentum. This state can also be viewed as a coherent condensate of magnons carrying momentum $q$. In contrast to the localized product states of the previous subsection, notice that the impurities are now spread coherently throughout the chain.
This state also admits a simple geometric interpretation. Each site corresponds to a spin pointing in a common direction on the Bloch sphere, but with an azimuthal angle that winds uniformly along the chain with
$\phi_\ell=q\ell$.
The parameter $q$ therefore measures the twist or phase winding of the condensate.\\

\noindent
The key quantity controlling the local dynamics is the overlap $\alpha = \langle s_\ell|s_{\ell+1}\rangle$ between neighboring spins. Using the explicit form of the coherent state, we can write this as 
$\alpha = (1-\rho)+\rho e^{iq}$ with corresponding overlap probability,
\begin{eqnarray}
    r &=& |\alpha|^2 \nonumber\\
    &=& \left[(1-\rho)+\rho e^{iq}\right] \left[(1-\rho)+\rho e^{-iq}\right]\nonumber\\
    &=& (1-\rho)^2
    +\rho^2+2\rho(1-\rho)\cos q\nonumber\\
    &=& 1-4\rho(1-\rho)\sin^2\frac q2\,.
    \label{eq:overlap}
\end{eqnarray}
Physically, this means that when neighboring spins are perfectly aligned, $r=1$, and the local Hamiltonian vanishes. Misalignment between neighboring sites lowers the overlap and therefore increases the local interaction energy. To compute the expectation value of the Hamiltonian we start by defining the local bond Hamiltonian
\begin{eqnarray}
    h_\ell = g(1-P_{\ell,\ell+1})\,.
\end{eqnarray}
For a product state,
\begin{eqnarray}
    \langle P_{\ell,\ell+1}\rangle
    = |\langle s_\ell|s_{\ell+1}\rangle|^2 = r\,,
\end{eqnarray}
so that $\langle h_\ell\rangle
=
g(1-r)$ and since every bond is equivalent, $\langle H\rangle
=
Lg(1-r)$. Substituting the explicit expression for $r$ from \eqref{eq:overlap}, we obtain the energy density
\begin{eqnarray}
    \varepsilon_q = \frac{\langle H\rangle}{L} = 4g\rho(1-\rho)\sin^2\frac q2\,.
\end{eqnarray}
Several features are immediately apparent. First, $\varepsilon_q=0$
for $q=0$, corresponding to a perfectly aligned ferromagnetic coherent state. In this limit all neighboring spins are identical and the chain is locally BPS-like.
Second, the energy density grows with increasing phase twist. The momentum $q$ therefore controls the local frustration of the condensate. Third, the energy density vanishes again at
$\rho=0$ or $\rho=1$,
where the state becomes a pure $Z$- or $X$-vacuum. This is sufficient to compute the first nontrivial Krylov coefficient,
\begin{eqnarray}
    b_1^2 = \langle H^2\rangle-\langle H\rangle^2\,.
\end{eqnarray}
Unlike the localized impurity states, the calculation is now more subtle because the coherent state is not an eigenstate of the local permutation operators. Writing $H=\sum_\ell h_\ell$,
we have
\begin{eqnarray}
    H^2 = \sum_{\ell,m} h_\ell h_m\,.
\end{eqnarray}
Since the state factorizes, non-overlapping bonds factorize independently and
\begin{eqnarray}
    \langle h_\ell h_m\rangle = \langle h_\ell\rangle\langle h_m\rangle\,, 
    \qquad |m-\ell|>1\,.
\end{eqnarray}
Consequently, disconnected contributions cancel against
$\langle H\rangle^2$.\\

\noindent
In the variance, only coincident and adjacent bonds survive so that
\begin{eqnarray}
    \mathrm{Var}(H) = \sum_\ell \Big(\langle h_\ell^2\rangle-\langle h_\ell\rangle^2 \Big) + 2\sum_\ell \Big(\langle h_\ell h_{\ell+1}\rangle - \langle h_\ell\rangle\langle h_{\ell+1}\rangle \Big)\,.
\end{eqnarray}
The first term measures local bond fluctuations, while the second measures correlations between neighboring bonds. Using
the fact that $P_{\ell,\ell+1}^2=1$, it follows that
\begin{eqnarray}
    h_\ell^2 = g^2(1-P_{\ell,\ell+1})^2 = 2g\,h_\ell,
\end{eqnarray}
and therefore
$\langle h_\ell^2\rangle
=
2g\langle h_\ell\rangle$. The adjacent-bond correlators require three-site overlaps. To facilitate the computation, let us define
\begin{eqnarray}
    \gamma = \langle s_\ell|s_{\ell+2}\rangle = 1-\rho+\rho e^{2iq}\,,
\end{eqnarray}
After simplification, the extensive variance density can be written as
\begin{eqnarray}
    \frac{b_1^2}{L} = 16g^2 \rho(1-\rho)\bigl[1-3\rho(1-\rho)\bigr]\sin^4\frac q2.
\end{eqnarray}
Notice that, unlike the localized impurity configurations, the Krylov variance now depends strongly on momentum $q$. The coherent phase structure therefore modifies the local operator-growth dynamics in an essential way. A second point worth noting is that the variance scales as $\sin^4\frac q2$,
instead of $\sin^2\frac q2$ so that the early Krylov growth is more sensitive to phase gradients than the energy density itself.
This can be put into a more illuminating form by expressing it in terms of the energy density, $\varepsilon_q
=
4g\rho(1-\rho)\sin^2\frac q2$,
in terms of which
\begin{eqnarray}
    \frac{b_1^2}{L} = \frac{1-3\rho(1-\rho)}{\rho(1-\rho)}\,\varepsilon_q^2.
\end{eqnarray}
This is qualitatively different from the corresponding localized impurity result,
$\frac{b_1^2}{L}\propto \varepsilon$,
obtained previously and the difference is physically significant. In localized impurity states, the Krylov variance is controlled by the density of sharp interfaces between neighboring spins. In contrast, in coherent magnon condensates the growth is controlled by smooth phase gradients distributed coherently across the entire chain. Consequently, two states with the same filling fraction and even the same energy density can exhibit parametrically different early-time Krylov dynamics depending on their microscopic coherence structure.\\

\noindent
The important point is that early-time operator growth in finite-density sectors is manifestly probe-dependent. The coefficient $b_1$ is sensitive not only to thermodynamic data such as impurity density, but also to the detailed spatial and phase organization of the state. This observation strongly suggests that any genuine universality, meaning both the same functional form and the 
same coefficient, must emerge only in the asymptotic large-$n$ Lanczos geometry rather than in the first few Krylov coefficients.\\

\subsection{Giant-open-string states}

Finally, we now turn to open strings attached to giant gravitons \cite{deMelloKoch:2007rqf,deMelloKoch:2007nbd,Bekker:2007ea}. At one loop, the bulk dynamics is again governed by the XXX Hamiltonian \eqref{xxx-chain},
supplemented by boundary terms determined by the giant graviton. In the maximal giant case, the giant is described by the determinant operator
\begin{eqnarray}
    \det Z = \epsilon^{i_1\cdots i_N} \epsilon_{j_1\cdots j_N} Z^{j_1}_{i_1}\cdots Z^{j_N}_{i_N}\,,
\end{eqnarray}
and an open-string excitation of the giant is constructed by replacing one of the $Z$’s with a word $W$,
\begin{eqnarray}
    \mathcal O_W \sim \epsilon\epsilon\, Z\cdots Z\, W .
\end{eqnarray}
If the first or last letter of $W$ is itself a $Z$, the operator will factorize as
\begin{eqnarray}
    W=ZW' \qquad\Rightarrow\qquad \mathcal O_W \sim (\det Z)\,\mathrm{Tr}(W').
\end{eqnarray}
Such states therefore do not describe genuine open strings attached to the giant. To avoid such states, the physical Hilbert space must exclude $Z$-fields at the endpoints of the chain. To implement this constraint, we can define the projector onto the $Z$-state,
$Q^Z
=
|Z\rangle\langle Z|$.
The full open-chain Hamiltonian then takes the form
\begin{eqnarray}
    H_{\rm giant-open} = g\sum_{\ell=1}^{L-1}(1-P_{\ell,\ell+1}) + gQ_1^Z + gQ_L^Z\,.
\end{eqnarray}
The endpoint operators
\begin{eqnarray}
    h_{\rm left}=gQ_1^Z\,, \qquad h_{\rm right}=gQ_L^Z\,,
\end{eqnarray}
act only on the first and last sites of the chain. Explicitly,
\begin{eqnarray}
    Q^Z|Z\rangle
    =|Z\rangle\,, \qquad Q^Z|X\rangle=0\,,
\end{eqnarray}
so the boundary Hamiltonian simply penalizes configurations in which the chain begins or ends with a Z-field\footnote{It is important to emphasize that the terminology ``bulk" and ``boundary" here refers to the interior and endpoints of the open spin chain, not to the AdS bulk spacetime.}. The chain interior contributes extensively, $H_{\rm chain}\sim O(L)$,
while the giant-induced endpoint terms remain finite, $h_{\rm left}+h_{\rm right}\sim O(1)$.\\

\noindent
Now let us consider a finite-density state whose excitations occupy the interior of a long open string attached to the giant. The chain contribution scales extensively,
$\langle H_{\rm chain}\rangle
=
L\,\varepsilon_{\rm chain}$, while the endpoint contribution remains finite,
$\langle h_{\rm left}+h_{\rm right} \rangle
=
O(1)$. Consequently, $a_0
=
L\varepsilon_{\rm chain}
+
O(1)$,
and therefore
\begin{eqnarray}
    \frac{a_0}{L} = \varepsilon_{\rm chain} + O(L^{-1})\,.
\end{eqnarray}
Similarly,
$b_1^2
=
\mathrm{Var}(H)$,
and decomposing
$H=H_{\rm chain}+h_{\rm left}+h_{\rm right}$,
gives
\begin{eqnarray}
    \mathrm{Var}(H) = \mathrm{Var}(H_{\rm chain}) + O(1)\,,
\end{eqnarray}
since the endpoint variance and covariance terms remain finite.
Now if we write
$\mathrm{Var}(H_{\rm chain})
=
Lv_{\rm chain}$, then
\begin{eqnarray}
    \frac{b_1^2}{L}
=
v_{\rm chain}
+
O(L^{-1}).
\end{eqnarray}
Physically, this tells us that in a long finite-density open string, the giant graviton modifies the operator dynamics only through subextensive endpoint effects. The leading operator-growth data per site is controlled entirely by the interior many-body dynamics of the chain. The leading extensive contribution forgets all the microscopic details of the boundary probe, while the giant-induced structure survives only in subleading corrections. This analysis also clarifies an important limitation. The suppression of the giant contribution occurs only when the excitation density is concentrated in the chain interior. Boundary-localized excitations, endpoint coherent states, or giant-string bound states may instead be dominated by the boundary terms and could therefore provide a more faithful realization of genuinely infalling giant-graviton dynamics.

\subsection{A Hermite Universality Class for Finite-Density Probes}

The explicit calculations of the previous subsections reveal an important pattern. Although the early-time Krylov growth is strongly probe dependent, the dependence enters only through a small number of extensive thermodynamic quantities. This suggests that the detailed microscopic structure of the probe may become irrelevant in the thermodynamic limit, where the spectral measure is controlled by collective fluctuations rather than individual excitations. To make this idea precise, consider a sequence of finite-density states $|\Psi_L\rangle$ in a spin chain of length $L$, with filling fraction $\rho=M/L$ held fixed as $L\rightarrow\infty$. In addition, let us assume that the state possesses clustering correlations in the sense that connected correlators decay sufficiently rapidly with distance. Such states include localized impurity configurations, coherent magnon condensates and giant-open-string states with finite-density bulk excitations.\\

\noindent
The one-loop dilatation operator may be written as a sum of local terms,
\begin{eqnarray}
    H_L=\sum_{\ell=1}^{L}
    h_\ell\,,
\end{eqnarray}
where each $h_\ell$ acts only on a finite number of neighboring sites. The expectation value of the Hamiltonian is extensive,
$\langle H_L\rangle = L\varepsilon + O(L)$, while the variance scales as
$\mathrm{Var}(H_L) = \langle H_L^2\rangle - \langle H_L\rangle^2 = Lv + O(L)$,
where $v$ is the variance density.
To understand the full spectral measure, it is useful to examine the connected cumulants of $H_L$. Writing
\begin{eqnarray}
    \kappa_n(H_L)
    =\sum_{\ell_1,\ldots,
    \ell_n} \langle
    h_{\ell_1}\cdots h_{\ell_n}
    \rangle_c\,,
\end{eqnarray}
we see that clustering implies that connected correlators become negligible whenever the sites are widely separated. Consequently, only clusters of sites lying within a distance of order the correlation length contribute appreciably. Since there are $O(L)$ such clusters, each cumulant grows only linearly with system size, $\kappa_n(H_L)=O(L)$
for every fixed $n$. Now, if we introduce the centered and normalized Hamiltonian
\begin{eqnarray}
    X_L = \frac{H_L-L\varepsilon}
    {\sqrt{Lv}}\,,
\end{eqnarray}
the cumulants of which are
$\kappa_n(X_L) = (Lv)^{-n/2}
\kappa_n(H_L)$, and therefore satisfy $\kappa_n(X_L) =
O\!\left(L^{1-n/2}\right)$. In particular, $\kappa_2(X_L)\to1$,
while $\kappa_n(X_L)\to0$, for $n\ge3$. The limiting cumulants are therefore exactly those of a standard Gaussian random variable. Equivalently, the spectral measure of $X_L$ converges weakly to
\begin{eqnarray}
    d\mu_L(x) \longrightarrow
    \frac{1}{\sqrt{2\pi}} 
    e^{-x^2/2}\,dx.
\end{eqnarray}
Physically, this is just the quantum analogue of the central limit theorem. The Hamiltonian is the sum of a large number of weakly correlated local contributions, and after subtracting the mean and dividing by $\sqrt L$, the resulting distribution becomes Gaussian.\\

\noindent
At this point an important caveat is necessary. The quantum central limit theorem establishes the limiting spectral measure but does not by itself determine the asymptotic Lanczos coefficients. To do so, we need to additionally assume that the convergence of spectral measures is sufficiently strong for the associated orthogonal-polynomial problem to converge to that of the Gaussian measure. Under this fairly natural assumption, the orthogonal polynomials associated with $d\mu_L$ then approach the Hermite polynomials which satisfy
the three term recursion,
\begin{eqnarray}
    x\,p_n(x) = \sqrt{n+1}
    \,p_{n+1}(x) + \sqrt n
    \,p_{n-1}(x)\,,
\end{eqnarray}
with Jacobi parameters $\alpha_n=0$ and $\beta_n=\sqrt n$. Using the affine relation
\begin{eqnarray}
    H_L = L\varepsilon + 
    \sqrt{Lv}\,X_L\,,
\end{eqnarray}
we identify the corresponding Lanczos coefficients as
$a_n = L\varepsilon + O(L)$, and
$b_n = \sqrt{Lv}\,\sqrt n + O(\sqrt n)$. This suggests the existence of a universal asymptotic Krylov class for finite-density probes. Although the microscopic structure of the probe enters through the variance density $v$, the exponent governing the Lanczos growth is independent of the probe! More precisely, we are led to the conjecture
\begin{eqnarray}
    b_n^{(P)} =
    \sqrt{Lv_P}\,\sqrt n
    + O(\sqrt n)\,,
\end{eqnarray}
for any finite-density probe family $P$ satisfying the clustering assumptions above.
The explicit examples considered in Sections 5.1–5.3 support this picture. For localized impurity configurations, $v_{\rm loc}
= g^2\delta$, while coherent condensates yield
\begin{eqnarray}
    v_{\rm coh} = 
    16g^2\rho(1-\rho) 
    \Bigl[ 1-3\rho(1-\rho) 
    \Bigr] \sin^4\frac q2\,.
\end{eqnarray}
Similarly, giant-open-string states inherit the variance density of the bulk chain up to subleading endpoint corrections. The variance density therefore retains information about the microscopic probe, even though the asymptotic scaling exponent is common to all our examples.\\

\noindent
This in turn motivates the following universality statement:
\textit{Finite-density probes in planar one-loop }$\mathcal N=4$ \ \textit{SYM are expected to belong to a common Hermite Krylov universality class.} In this universality class the asymptotic Lanczos growth is governed by $b_n\propto \sqrt n$, with a probe-dependent amplitude but a universal exponent. It is worth emphasizing that this notion of universality differs fundamentally from the linear Lanczos growth, $b_n\sim n$, associated with chaotic operator dynamics. The latter arises naturally in operator Krylov complexity when the corresponding Liouvillian spectral measure possesses exponential tails. In contrast, the finite-density probe states considered here are governed by ordinary energy spectral measures whose thermodynamic limit is Gaussian. Their asymptotic Krylov geometry is therefore controlled by Hermite rather than exponential orthogonal-polynomial asymptotics.
From this perspective, the appearance of $\sqrt n$ growth should be viewed as a consequence of central-limit physics in many-body systems with clustering correlations.

\section{Conclusions}
\label{sec:conclusion}

\noindent
In this work we investigated Krylov complexity in several analytically controlled sectors of planar $\mathcal N=4$ super Yang–Mills theory with the goal of formulating a precise field-theoretic test of recently proposed universality conjectures for the complexity growth of infalling probes in AdS gravity. Our central observation is that the appropriate language for this problem is the spectral-measure formulation of Krylov dynamics. In this framework, the Lanczos coefficients are identified with the Jacobi parameters of the orthogonal-polynomial problem associated with the seed spectral measure, allowing questions about operator growth to be reformulated as questions about spectral asymptotics.\\

\noindent
This perspective immediately clarifies the behaviour of protected and few-body sectors. Exact BPS states generate trivial Krylov dynamics, while superpositions of BPS states and localized descendant wavepackets produce nontrivial but purely spectral growth controlled by discrete measures. More generally, we showed that fixed-magnon sectors of the open Heisenberg spin chain, including giant-open-string operators, possess compact absolutely continuous spectral support. Standard results from orthogonal-polynomial theory then imply that the corresponding Lanczos coefficients approach constants determined solely by the band edges. In particular, for fixed magnon number $M$ and open-string length $L\to\infty$, we obtained
\begin{eqnarray*}
    a_n \to 2Mg\,,
\qquad
b_n \to Mg\,,
\end{eqnarray*}
demonstrating explicitly that these sectors exhibit integrable, band-limited Krylov dynamics. The leading asymptotic behaviour is therefore controlled only by spectral support and cannot distinguish between different microscopic probe constructions. 
Consequently, the late time behaviour exhibits a kind of universality,  analogous 
to the one in \cite{Nastase:2026lhz}, although it is less clear how to distinguish between probes.\\

\noindent
We also identified a weaker but nevertheless instructive notion of early-time universality already present in protected sectors. Local-primary descendant packets and giant-graviton wavepackets are physically very different states, yet their Krylov dynamics coincide after an affine redefinition of the energy scale. This universality is purely spectral in origin and reflects the fact that both constructions reduce to Gaussian packets on uniformly spaced energy ladders. While this should not be confused with the universality of genuinely dynamical probes, it provides a useful boundary realization of the idea that complexity growth may become insensitive to microscopic probe structure. \\

\noindent
In the case of (BMN) 
operators in the $SU(2)$ sector of ${\cal N}=4$ SYM,
few-body wave-packets of magnons of varying energies 
interacting via the one-loop SYM Hamiltonian, present a similar
behavior, although they do not present a Gaussian distribution over 
uniformly spaced energy ladder. However we note that  small $\lambda$ makes comparison with gravity difficult, while the large ($\lambda,J$) limit of 
BMN does not correspond to an object infalling from the boundary of AdS, but rather to an object (string) at the center of AdS.\\

\noindent
Motivated by these observations, we then argued that the correct arena in which to formulate a better notion of universality is the finite-density thermodynamic regime,
\begin{eqnarray*}
    L\rightarrow\infty\,, \qquad
    M\rightarrow\infty\,, \qquad
    \rho=M/L
    \ \text{fixed}\,,
\end{eqnarray*}
where the number of active degrees of freedom scales extensively with system size. For several distinct classes of finite-density states including localized impurity configurations, coherent magnon condensates and giant-open-string states, we found evidence that the properly rescaled spectral measure approaches a Gaussian distribution. Assuming convergence of the associated orthogonal-polynomial problem, the Krylov dynamics are then governed asymptotically by Hermite polynomials, leading to
$b_n \sim \sqrt{Lv}\,\sqrt n$.
This in turn suggests the existence of a common Hermite Krylov universality class for finite-density probes in planar one-loop $\mathcal N=4$ SYM. In this class, the microscopic structure of the probe survives only through the variance density $v$, while the asymptotic growth exponent is universal. The universality uncovered here differs fundamentally from the linear Lanczos growth, $b_n\sim n$, expected in chaotic operator dynamics. The latter is associated with Liouvillian spectral measures possessing exponential tails, whereas the finite-density states considered in this work are governed by ordinary energy spectral measures whose thermodynamic limit is Gaussian. The resulting $\sqrt n$ behaviour should therefore be viewed as a manifestation of central-limit physics rather than quantum chaos. From this perspective, our analysis identifies a distinct universality class of Krylov growth controlled by many-body thermodynamics.\\

\noindent
Several important questions remain open. First, it would be desirable to establish the Hermite universality conjecture rigorously by proving convergence of the relevant orthogonal-polynomial problems in the thermodynamic limit. Second, it would be interesting to determine how this structure is modified beyond the one-loop $SU(2)$ sector, where integrability-breaking effects and additional interactions become important. Third, and perhaps most importantly, we would like to understand whether a direct connection exists between the Hermite universality identified here and the probe universality observed in gravitational systems. Resolving this issue will require a deeper understanding of the relation between boundary Krylov geometry, thermodynamic coarse-graining and bulk probe dynamics.\\

\noindent
Regardless of the ultimate answer, the picture that emerges from our analysis is conceptually simple. The key distinction is not between point particles, strings and giant gravitons, but between few-body and finite-density states. Few-body probes retain detailed spectral information and exhibit bounded Krylov growth, whereas finite-density states are governed by thermodynamic spectral statistics and appear to flow toward a common asymptotic Krylov geometry. In this sense, universality in complexity growth may be less a property of individual probes than a consequence of many-body collective behavior.

\section*{Acknowledgements}
We would like to thank Pawel Caputa, Rathindra Das and Masataka Watanabe for useful discussions. JM and HJRVZ are supported in part by the ``Quantum Technologies for Sustainable Development" grant
from the National Institute for Theoretical and Computational Sciences of South Africa
(NITheCS).
The work of HN is supported in part by  CNPq grant 
304583/2023-5 and FAPESP grant 2024/15298-0.
HN would also like to thank the ICTP-SAIFR for their support 
through FAPESP grant 2021/14335-0.
ELG is supported by FAPESP grant 2024/13362-2.

\appendices

\section{The 1/2-BPS sector}

\label{NumericsAppendix}

\subsection{Superpositions of descendants with small $\sigma$}

\begin{figure}
	\centering
    \includegraphics[width=0.6
    \textwidth]{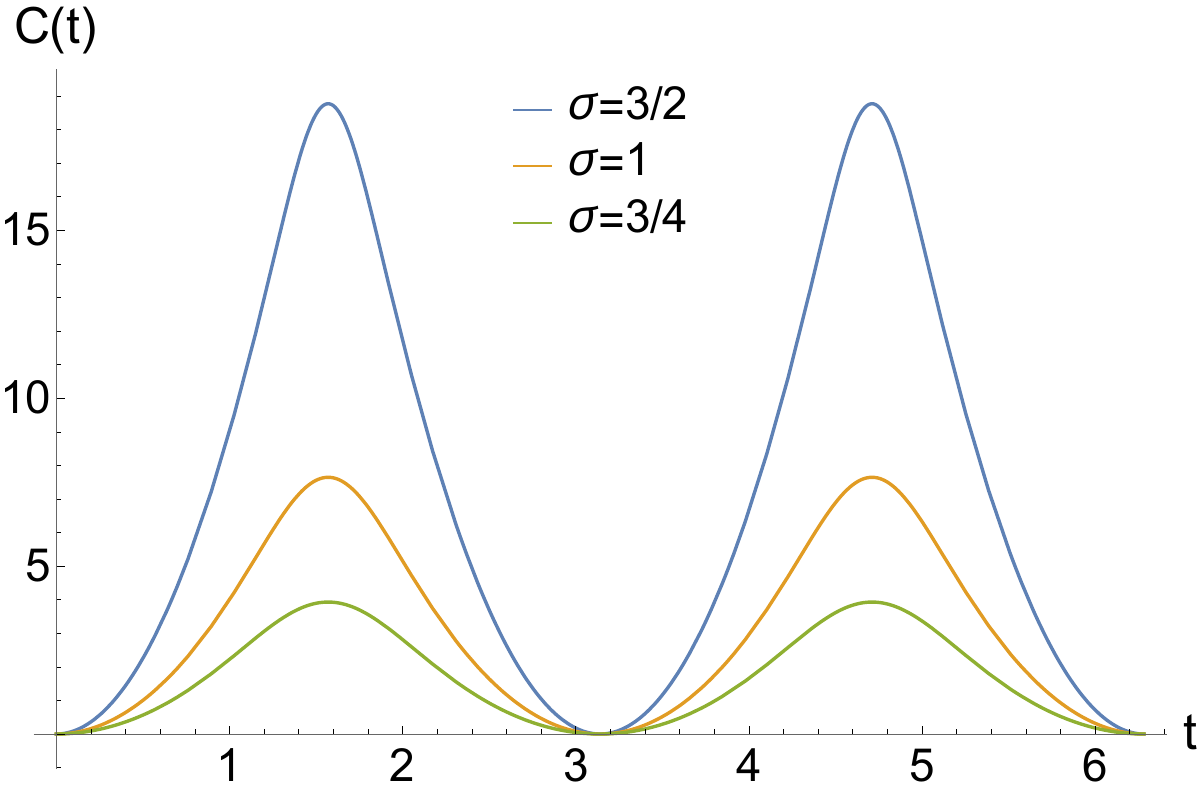}
	\caption{The spread complexity of the Gaussian superposition of descendants.  Here we have taken $n_0= 50$ and the cutoff $\epsilon = 10^{-30}$.  For these values of $\sigma$ the overlap with the $n=0$ state is negligible.  A simple periodic function in time is the result.  As $\sigma$ decreases, the dimension of the Krylov subspace and complexity decreases.  }
	\label{fig:GaussianDescendentsCt}
\end{figure}

\begin{figure}
\begin{minipage}{0.32\textwidth}
		\includegraphics[width=\textwidth]{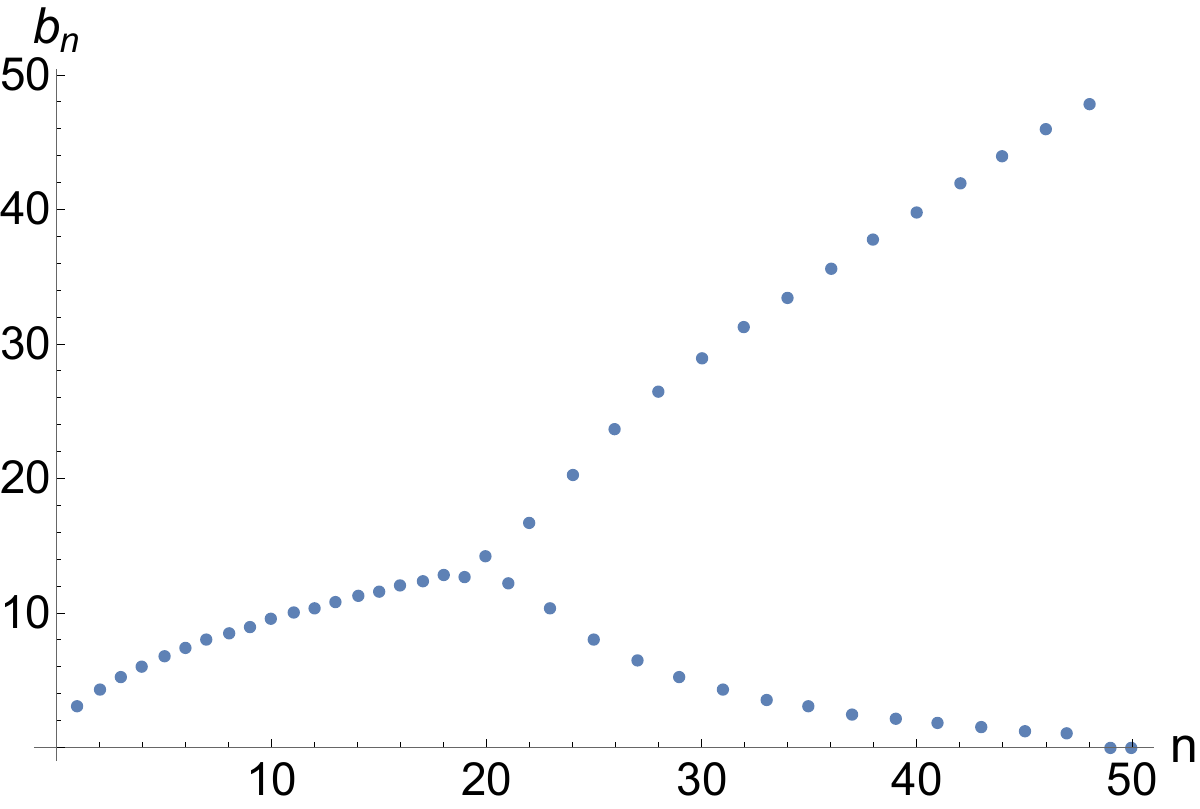}
\end{minipage}
\begin{minipage}{0.32\textwidth}
		\includegraphics[width=1.00\textwidth]{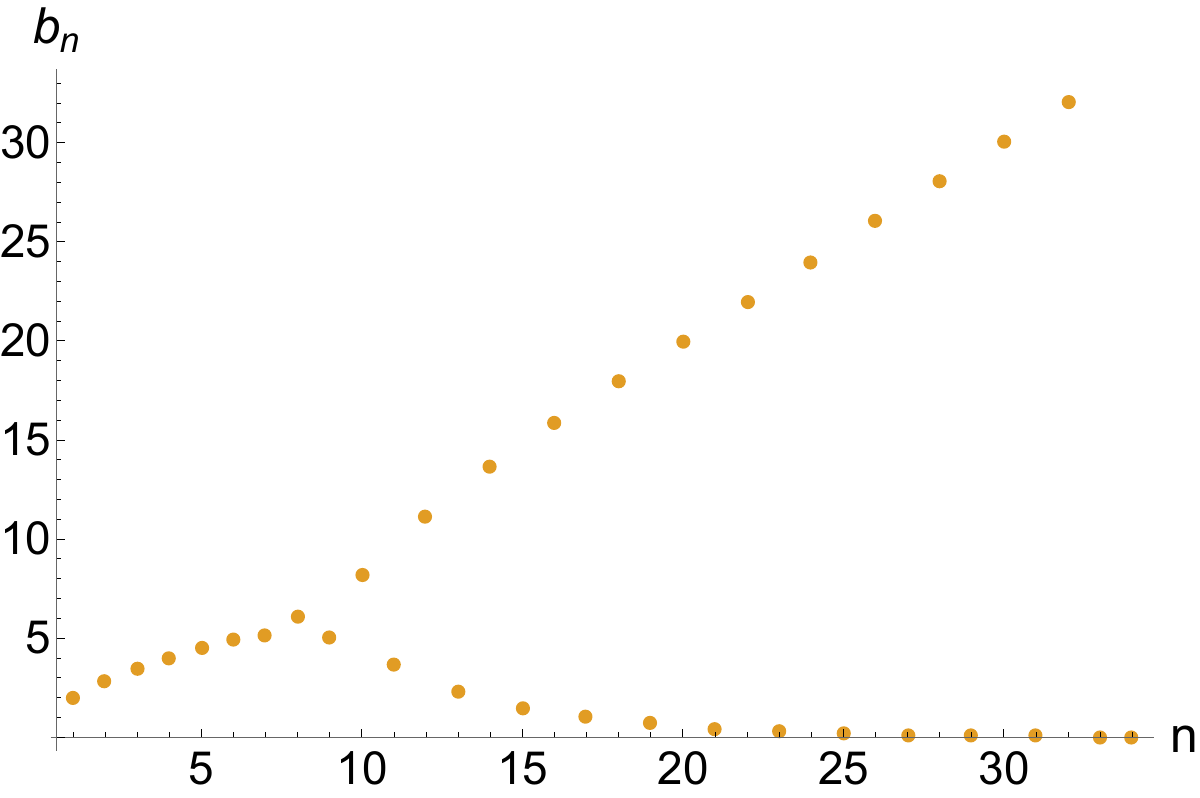}
\end{minipage}
\begin{minipage}{0.32\textwidth}
		\includegraphics[width=1.00\textwidth]{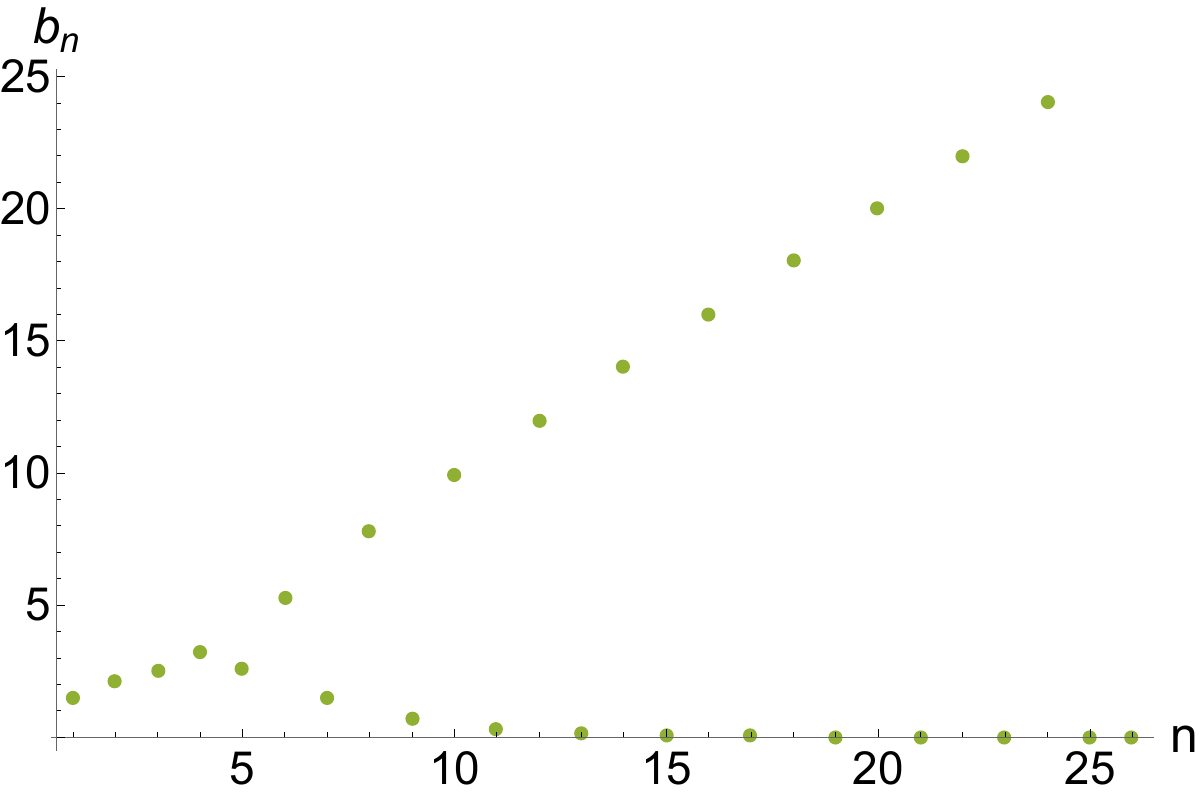}
\end{minipage}
	\caption{The $b_n$ coefficients accompanying the complexity plots in Fig. (\ref{fig:GaussianDescendentsCt}). The $a_n$ coefficients are all constant.    }
	\label{bnsBulkGuassianDescendents}
\end{figure}

The first set of states we will investigate numerically is a Gaussian superposition of descendants
\begin{equation}
    |K_0 \rangle = \sum_{n=0}^\infty e^{-\frac{(n-n_0)^2}{4 \sigma^2}} e^{i q n} |\mathcal{O}_{\Delta}; n\rangle,
\end{equation}
where
\begin{equation}
    H|\mathcal{O}_{\Delta}; n\rangle = 2 n + \Delta.
\end{equation}
Two features of this state are important to note.  First, since the eigenstates don't mix under action of the Hamiltonian, the phase factor $q$ does not contribute.  An alternative way to see this is that the dynamics for Hamiltonians differing by a constant shift are identical - it is only the $a_n$ coefficients that are all shifted by a constant.  The phase factor can be absorbed by shifting $H \rightarrow H - \frac{q}{2}$.  Similarly, the scaling dimension of the primary, $\Delta$, has no effect on the complexity.  It too, only shifts all the $a_n$ coefficients by a constant.  \\ \\
Secondly, due to the linear scaling of the energies, the time-evolved reference state $|K_0(t)\rangle$ is (up to an overall phase) periodic $|K_0(t + \pi)\rangle  = e^{i \phi}|K_0(t)\rangle $.  This periodicity will show up in all the numerical plots.   \\ \\
The complexity is thus only a function of the center $n_0$ and the width $\sigma$.  Numerically we introduce a cut-off $\epsilon$ for the amplitudes, so that if $ |\langle \mathcal{O}_\Delta; n  |K_0\rangle| < \epsilon$ then the term proportional to $|\mathcal{O}_{\Delta}; n\rangle$ is dropped from the superposition.  By doing this we have a reference state effectively composed of a finite number of eigenstates and the dimension of the Krylov subspace is immediately bounded.  Finally, if the overlap with the $n=0$ boundary $ |\langle \mathcal{O}_\Delta; 0  |K_0\rangle| < \epsilon$, then the shift $n_0$ has no effect on the dynamics.  This is again due to the fact that the dynamics are unchanged if the Hamiltonian is shifted by a constant.  \\ \\
\noindent
In Fig. (\ref{fig:GaussianDescendentsCt}) we have plotted the spread complexity as a function of time for such ``bulk" superpositions (that have a negligible overlap with the $n=0$ state) for a selection of $\sigma$ values.  Complexity is always a simple function of time.  As $\sigma$ decreases, the Gaussian superposition becomes more localised and the complexity decreases.  The $b_n$ coefficients accompanying the spread complexity are plotted in Fig. (\ref{bnsBulkGuassianDescendents}).  We find that the coefficients grow slowly up to some point, whereafter there are two branches - one that decays to zero and the other that grows.  In the next subsection we will show that this growth is approximately as $b_n \sim \sqrt{n}$. \\ 

\begin{figure}
	\centering
		\includegraphics[width=0.6\textwidth]{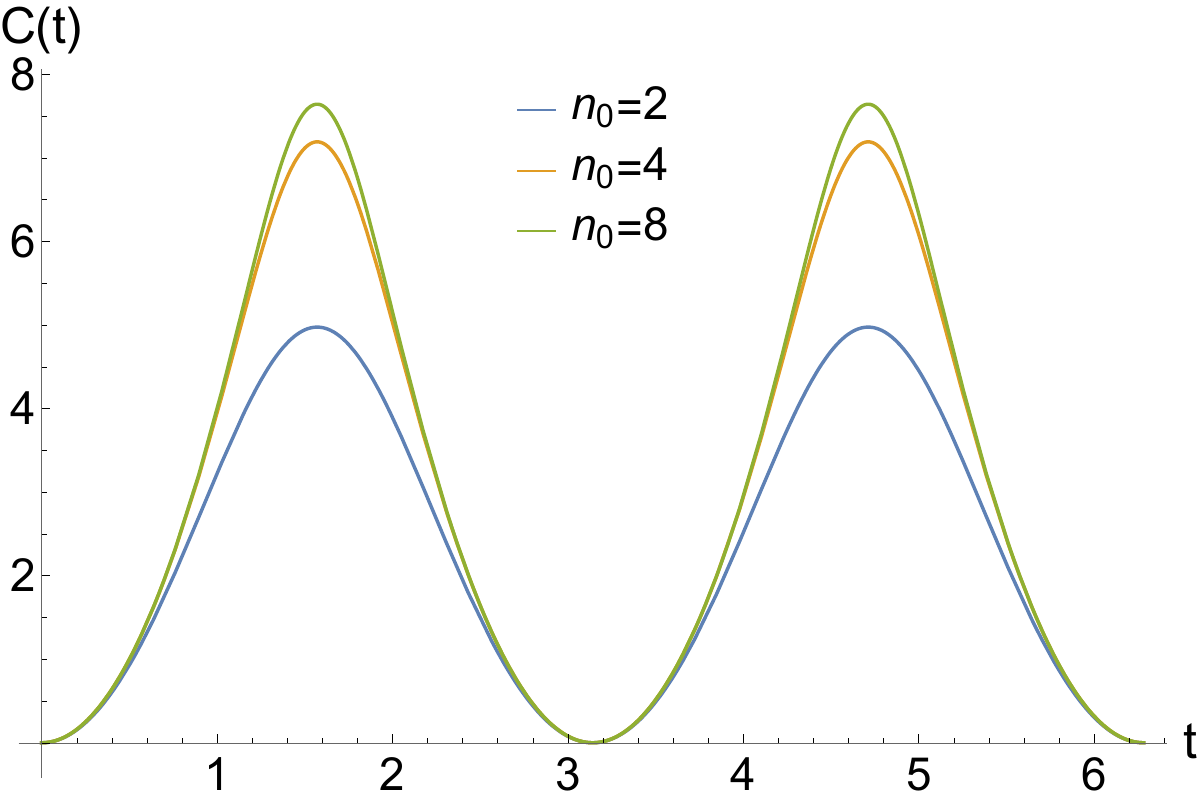}
	\caption{The spread complexity of the Gaussian superposition of descendants.  Here we have taken $\sigma= 1$ and the cutoff $\epsilon = 10^{-30}$.  A simple periodic function in time is the result.  As $n_0$ decreases, the dimension of the Krylov subspace and complexity decreases. }
	\label{fig:GaussianDescendentsCtdiffn0}
\end{figure}

\begin{figure}
\begin{minipage}{0.49\textwidth}
		\includegraphics[width=1.00\textwidth]{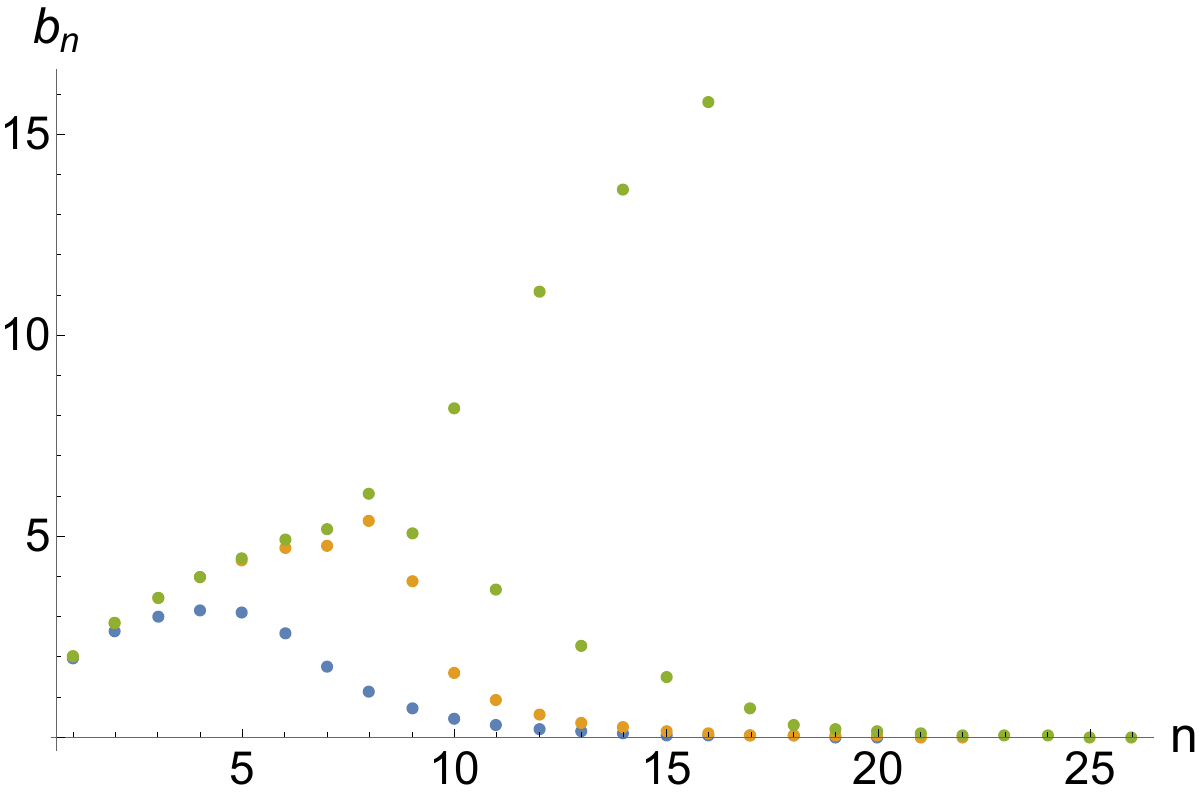}
\end{minipage}
\begin{minipage}{0.49\textwidth}
		\includegraphics[width=1.00\textwidth]{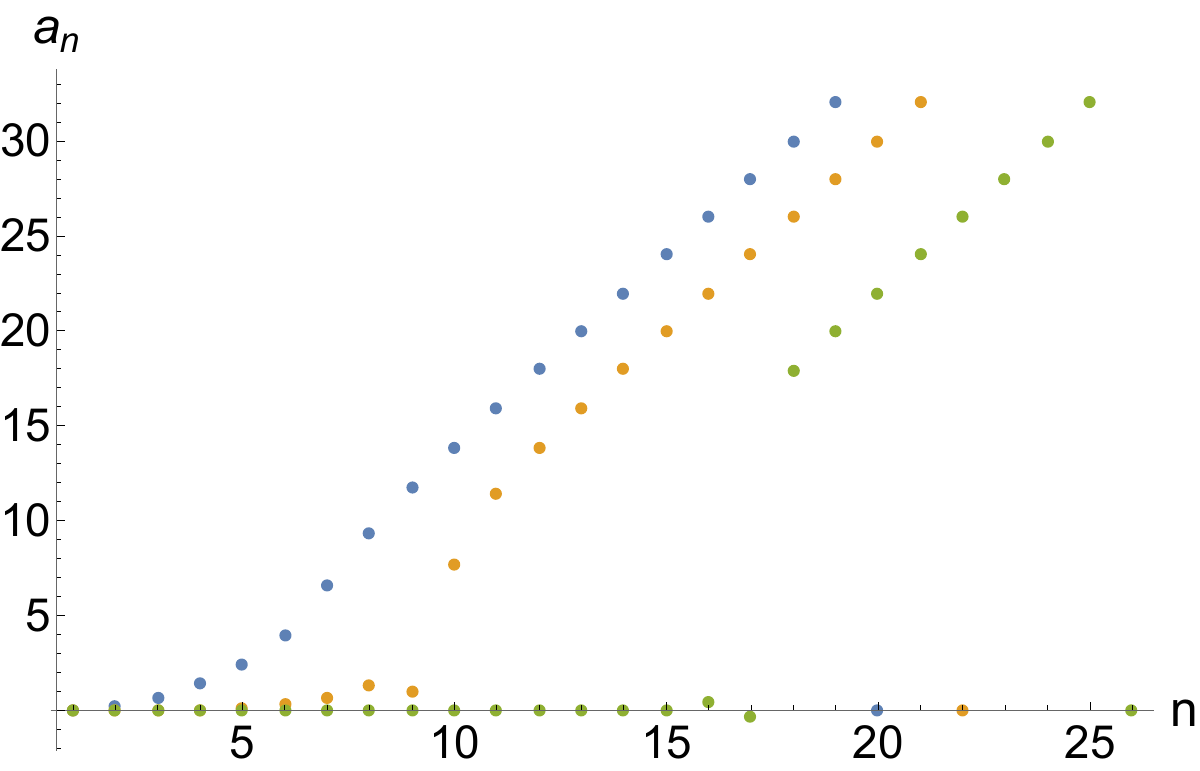}
\end{minipage}
	\caption{The $b_n$ coefficients accompanying the complexity plots in Fig. (\ref{fig:GaussianDescendentsCtdiffn0}).   }
	\label{Lanczosdiffn0}
\end{figure}

\noindent
In Fig. (\ref{fig:GaussianDescendentsCtdiffn0}) we investigate the dependence of complexity on $n_0$.  Generally, we find that complexity decreases as $n_0$ decreases.  The accompanying Lanczos coefficients are plotted in Fig. (\ref{Lanczosdiffn0}).  The $a_n$'s are no longer constant.  They are close to $0$ up to some value for $n$, after which they appear to increase linearly. Regardless of the choice of $\sigma$ and $n_0$, however, spread complexity is always a simple periodic function.  \\ \\

\FloatBarrier
\subsection{Superpositions of descendants with large $\sigma$}

With large enough $\sigma$, we can verify the predictions for the continuous limit found using the spectral measure and orthogonal polynomials in section \ref{subsec:primary-wave}. These are
\begin{equation}
    a_n \approx \Delta + 2 n_0,\qquad b_n \approx 2\sigma \sqrt{n},
\end{equation}
when $n$ is small enough for the coefficients to be free of effects induced by the spectral boundaries, and
\begin{eqnarray}
    a_n \approx \frac{A+B}{2} \approx \Delta + 2n_0,\qquad b_n \approx \frac{B-A}{4} \approx c\,\sigma,
\end{eqnarray}
for large enough $n$ ($A$ and $B$ are, respectively, the lower and upper spectral boundaries and $c$ is a constant). Of course, all of these predictions are valid when $1\ll \sigma \ll n_0$. In the following analysis, we set $\Delta=1$ and we maintain the cutoff $\epsilon=10^{-30}$, guaranteeing a finite Krylov basis.\\

\begin{figure}[tbh]
	\centering
    \includegraphics[width=0.45\textwidth]{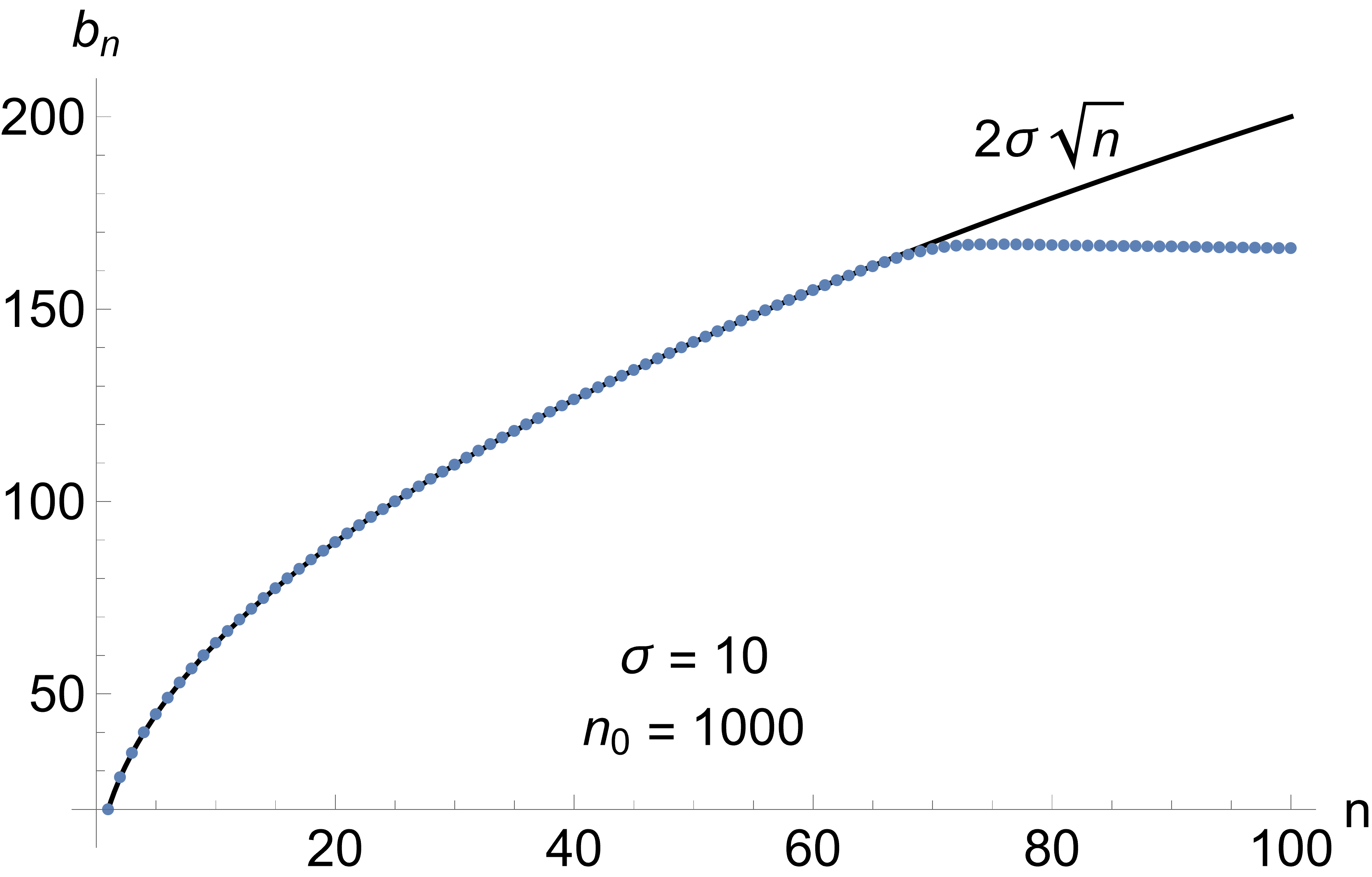}
    \includegraphics[width=0.45\textwidth]{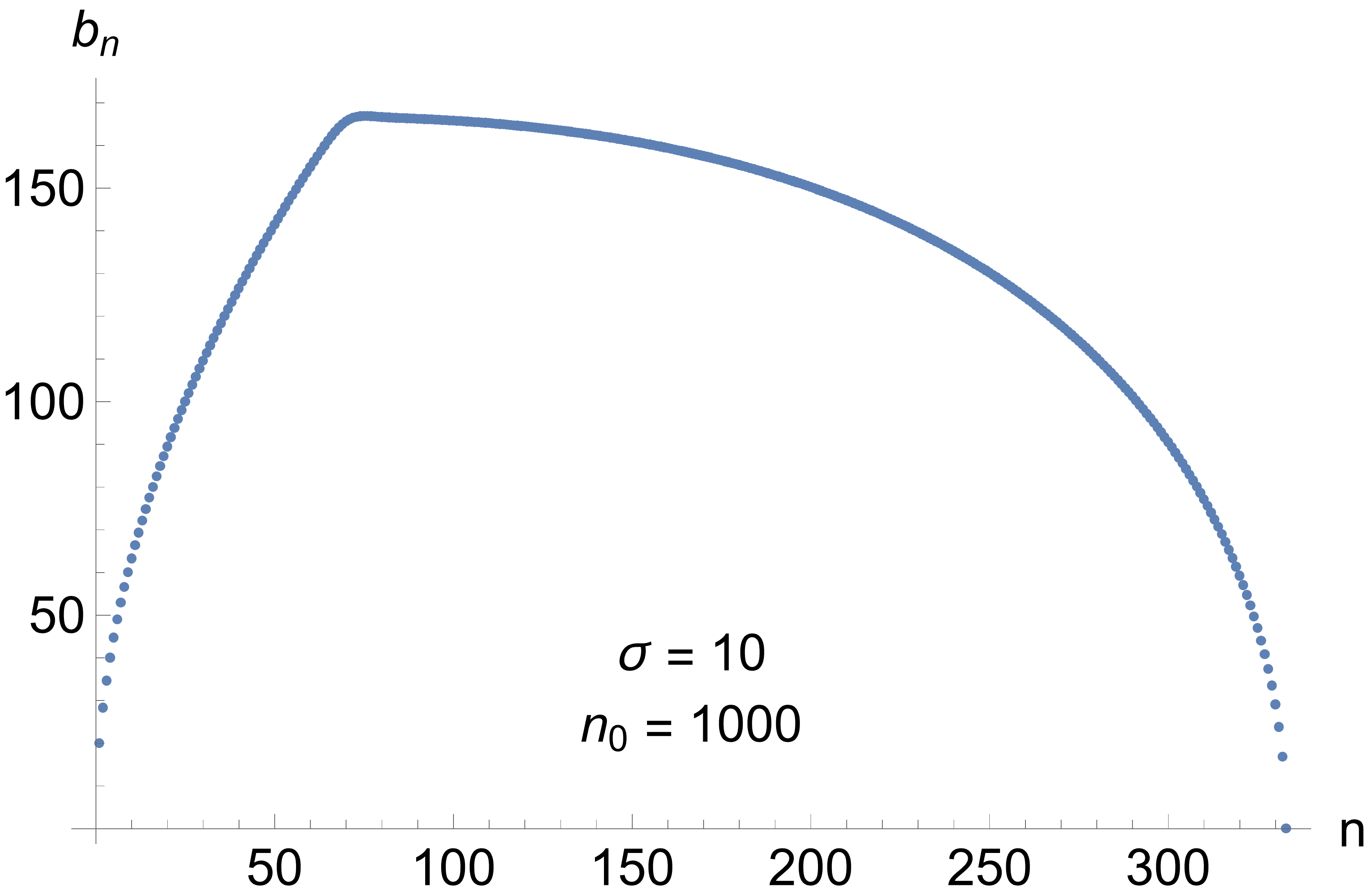}
	\caption{The $b_n$ coefficients of the gaussian superposition of descendants with $\sigma=10,\,n_0=1000,\,\Delta=1$. The left plot displays the coefficients up to $n=100$ and the right plot contains the full set. The corresponding $a_n$ coefficients are all equal to $2001$.}
	\label{fig:descendants bn s10 n1000}
\end{figure}

\begin{figure}[tbh]
	\centering
    \includegraphics[width=0.6\textwidth]{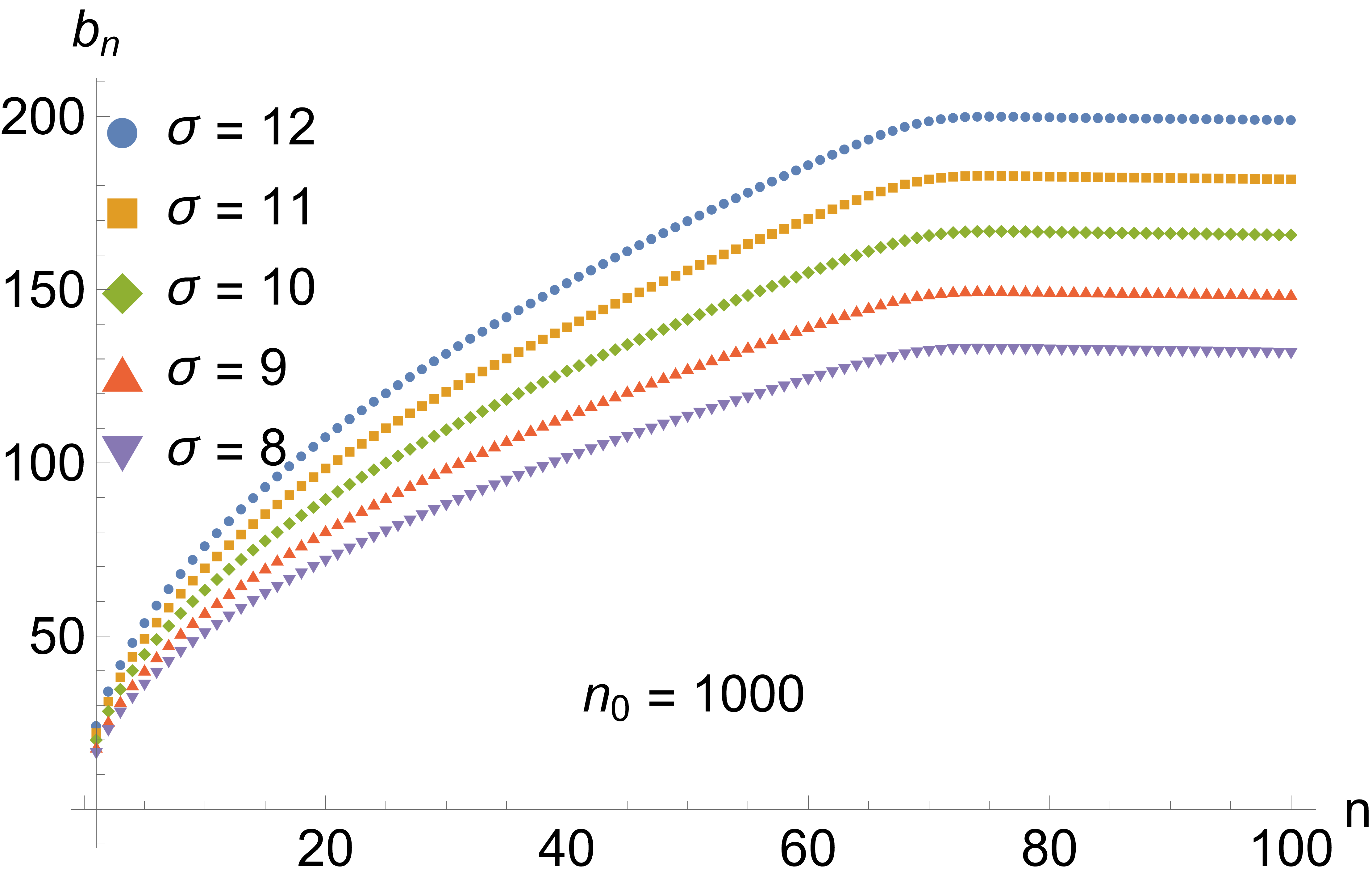}
	\caption{The $b_n$ coefficients up to $n=100$ of gaussian superpositions of descendants with $\,n_0=1000,\,\Delta=1$ and several $\sigma$. The numerical values for the highest coefficients shown here are given in Table \ref{tab:bns varying s}.}
	\label{fig:descendants bn n1000 many s}
\end{figure}

\noindent
We start with the case defined by the parameters $\{\sigma=10,\,n_0=1000,\,A=1669,\,B=2333\}$, in which all the predictions can be verified. The corresponding data, presented in Fig. \ref{fig:descendants bn s10 n1000}, shows that $b_n\approx2\sigma\sqrt{n}$ is valid for all $b_n$'s up to some value of $n$, after which the coefficients become sensitive to the spectral boundaries. At this point, the $b_n$'s start a slow decay towards $0$, an inevitable consequence of the finiteness of the Krylov space. The $a_n$'s are not shown, but are given by $a_n = (A+B)/2 = \Delta + 2 n_0 = 2001$, for all $n$. The maximum value of the $b_n$'s is found to be $\approx166.85$, very close to $(B-A)/4=166$, demonstrating that the slowly decaying plateau points to the asymptotic regime predicted for the continuous case. The asymptotic prediction is further confirmed by the data presented in Fig. \ref{fig:descendants bn n1000 many s} and Table \ref{tab:bns varying s}, corresponding to $n_0=1000$ and varying $\sigma$.
\begin{table}[tbh]
    \centering
    \begin{tabular}{c|c|c|c}
         $\sigma$ & Max$(b_n)$ & $(B-A)/4$ & $\text{Max}(b_n)/\sigma$ \\ \hline
         8 & 132.77 & 132 & 16.596 \\
         9 & 149.81 & 149 & 16.646 \\
         10 &  166.85  &  166  &  16.685 \\
         11 & 182.89 & 182 & 16.626 \\
         12 & 199.93 & 199 & 16.661
    \end{tabular}
    \caption{Comparison between the asymptotic predictions and the maximum values of the $b_n$ found from gaussian superpositions of descendants with $\,n_0=1000,\,\Delta=1$ and several $\sigma$.}
    \label{tab:bns varying s}
\end{table}\\

\begin{figure}
	\centering
    \includegraphics[width=0.45\textwidth]{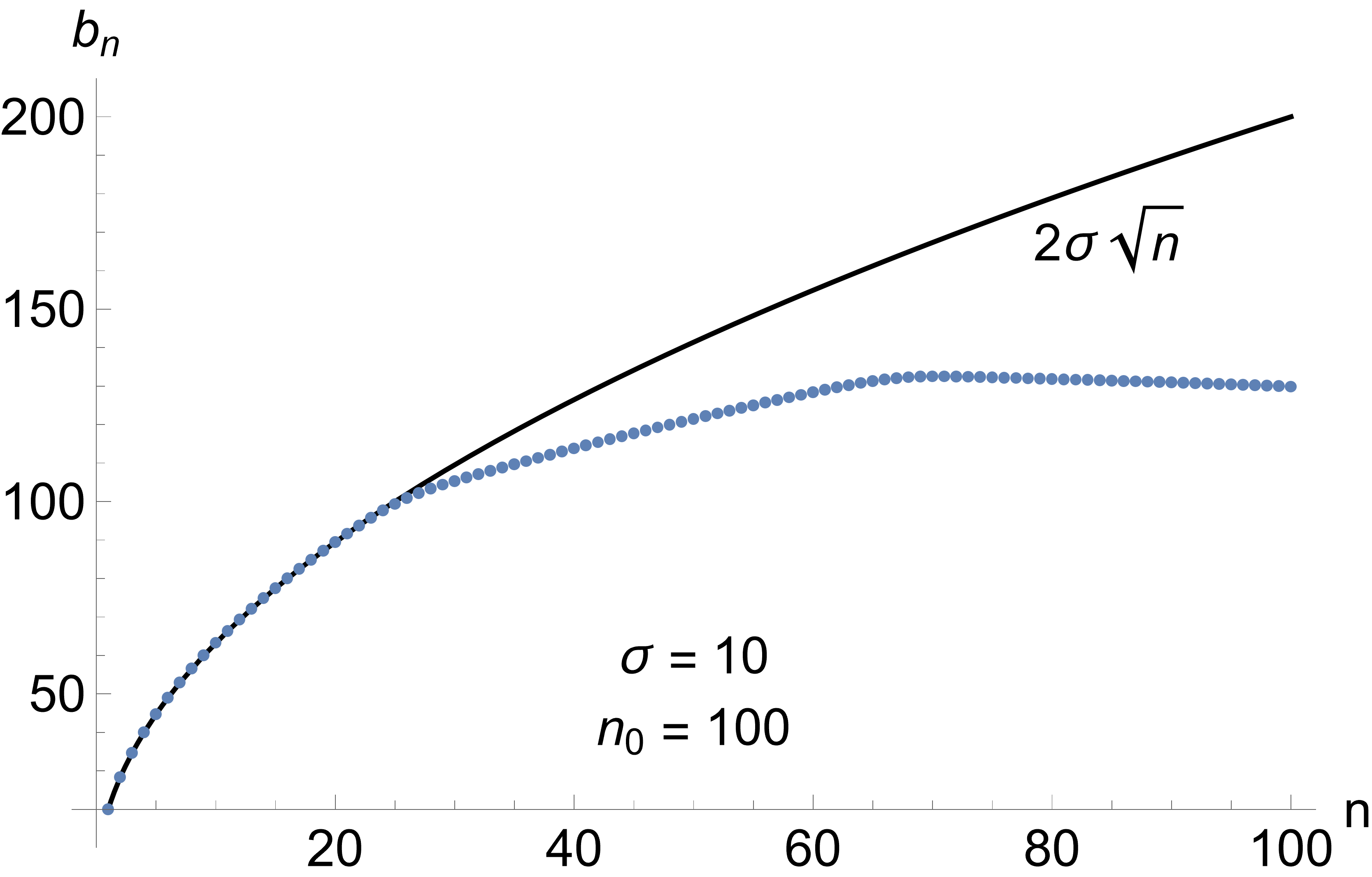}
    \includegraphics[width=0.45\textwidth]{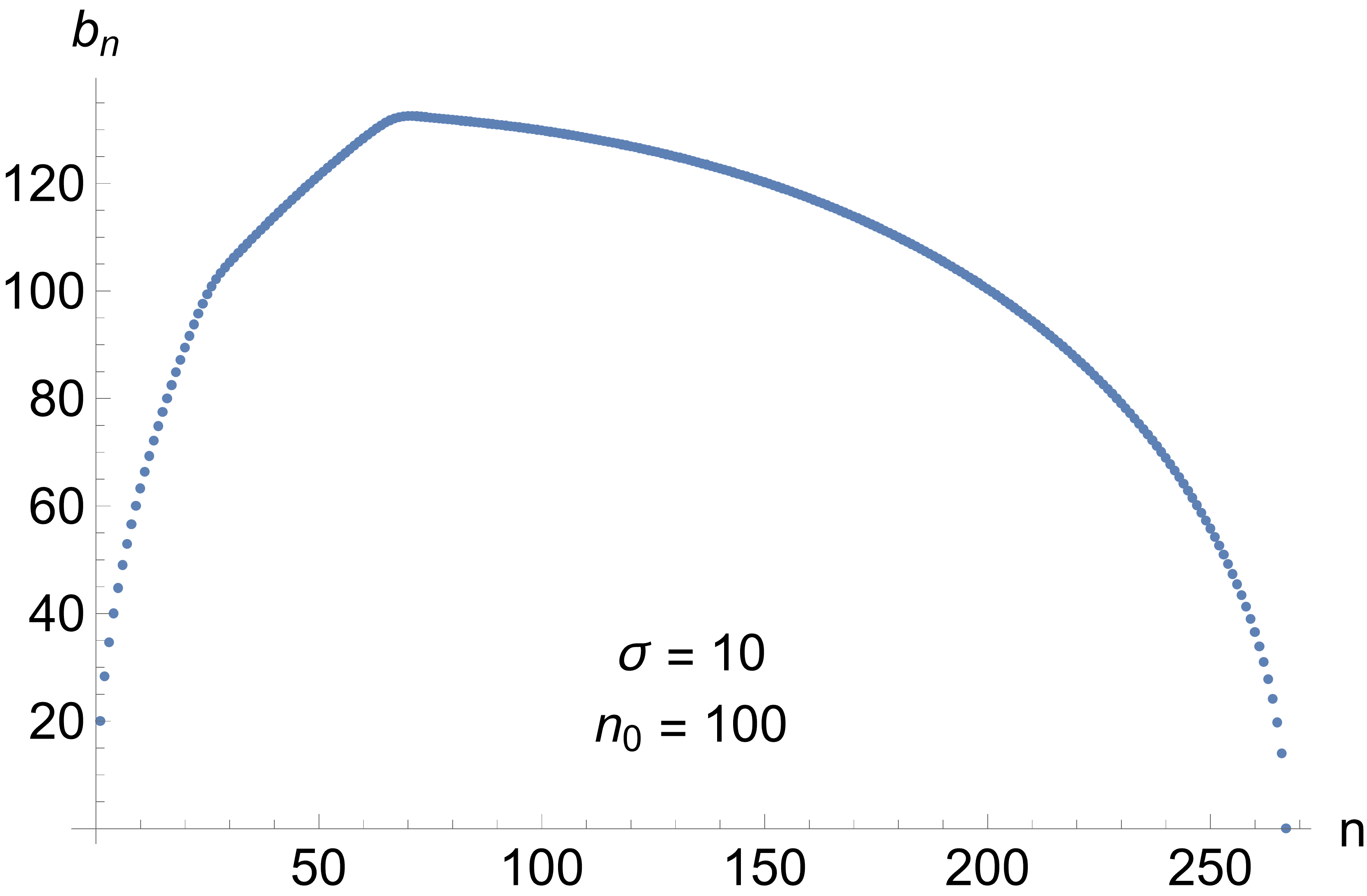}
    \includegraphics[width=0.45\textwidth]{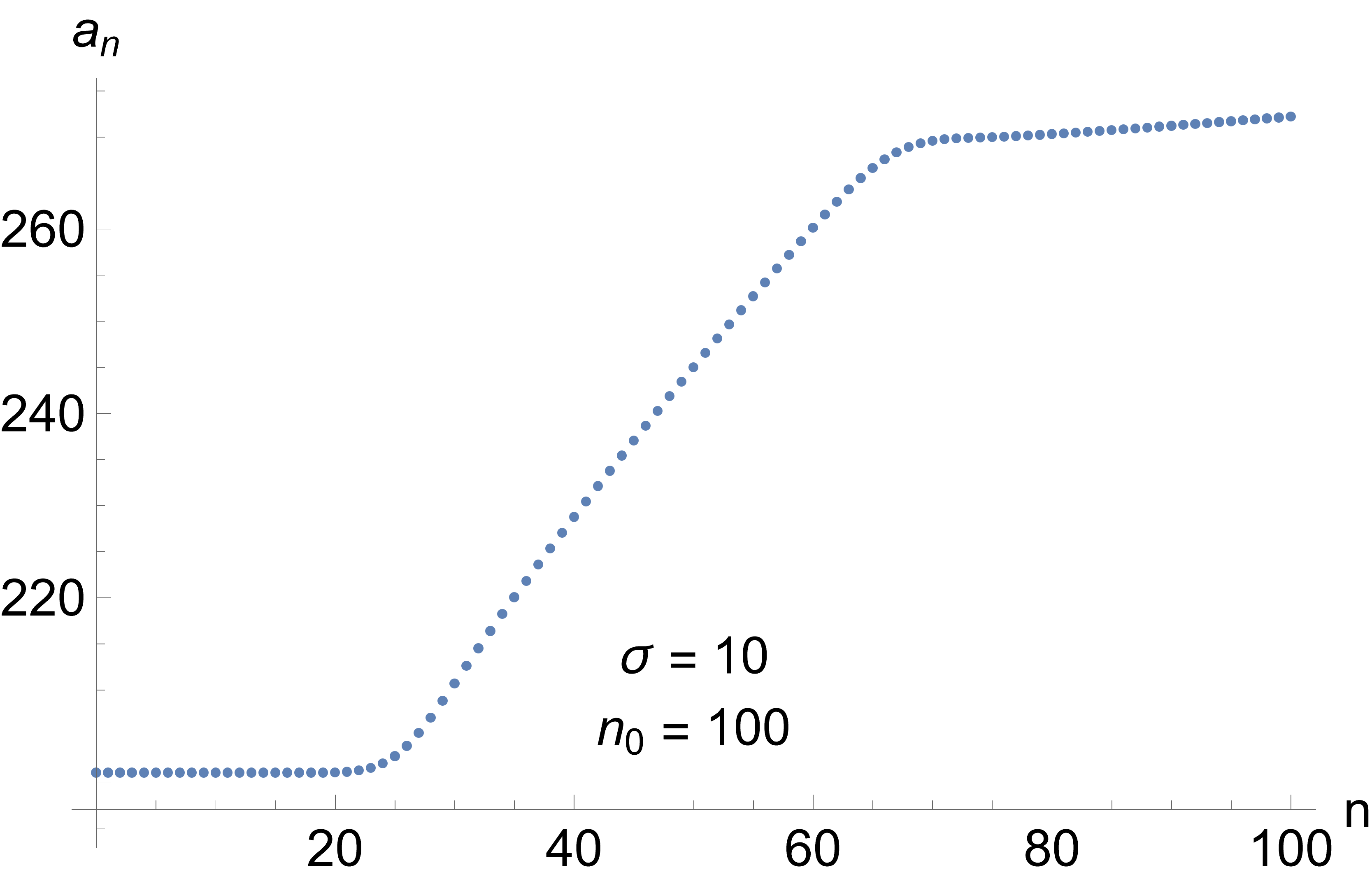}
    \includegraphics[width=0.45\textwidth]{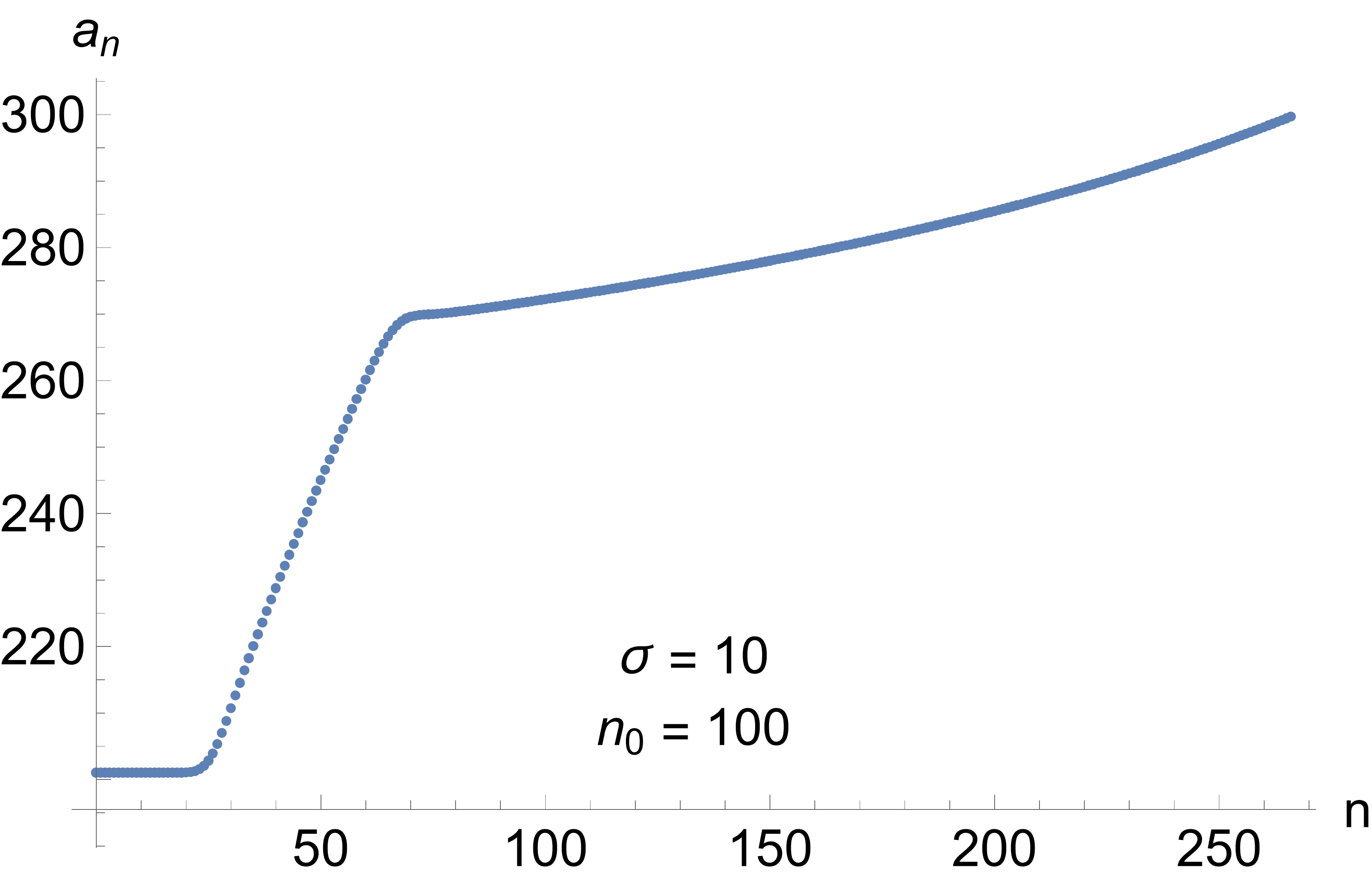}
	\caption{The Lanczos coefficients of the gaussian superposition of descendants with $\sigma=10,\,n_0=100,\,\Delta=1$. The $b_n$ coefficients are given in the top row and the $a_n$'s on the bottom. The left plots displays the coefficients up to $n=100$ and the right plots contain the full sets.}
	\label{fig:descendants s10 n100}
\end{figure}

\noindent
New behaviors can be observed when the center of the distribution is relatively close to the ground state. Fig. \ref{fig:descendants s10 n100} refers to the case $\{\sigma=10,\,n_0=100\}$, and shows that the initial $b_n\propto\sqrt{n}$ regime is succeeded by a different one with slower growth. Further on, this gives way to a third regime, in which the $b_n$'s decay. The $a_n$'s also behave differently in each of these three phases. We can infer that the first regime change occurs as the coefficients become sensitive to the lower spectral boundary (corresponding, in this case, to the ground state), while the second one occurs when they become sensitive to the higher spectral boundary, which is further away from $n_0$. Although the $n_0=100$ case is quite different from the ones with $n_0=1000$, we find that Max$(b_n)$ still satisfies the asymptotic predictions, with Max$(b_n)\approx132.5\approx(B-A)/4=133$.
\\

\begin{figure}[tbh]
	\centering
    \includegraphics[width=0.45\textwidth]{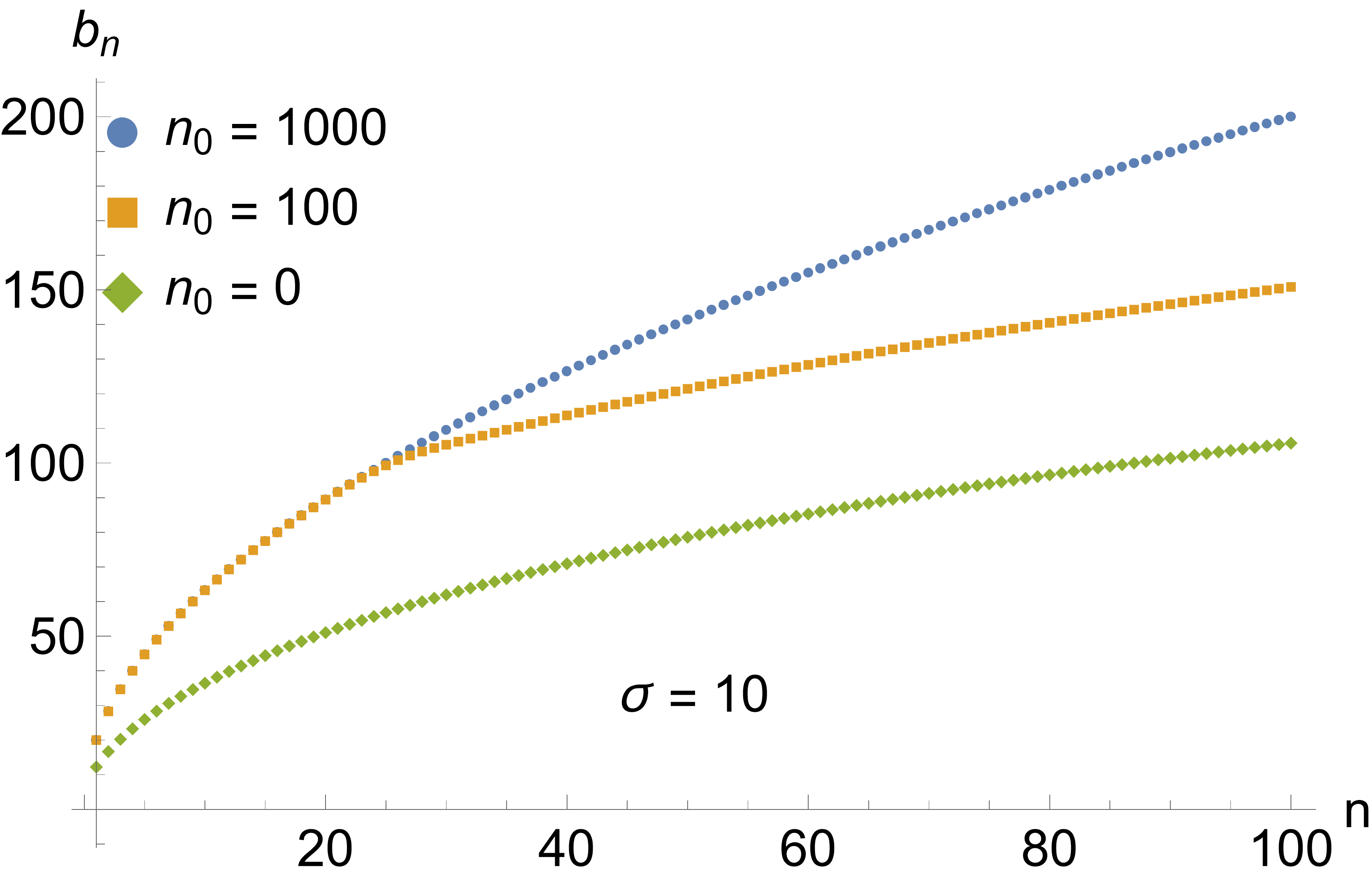}
    \includegraphics[width=0.45\textwidth]{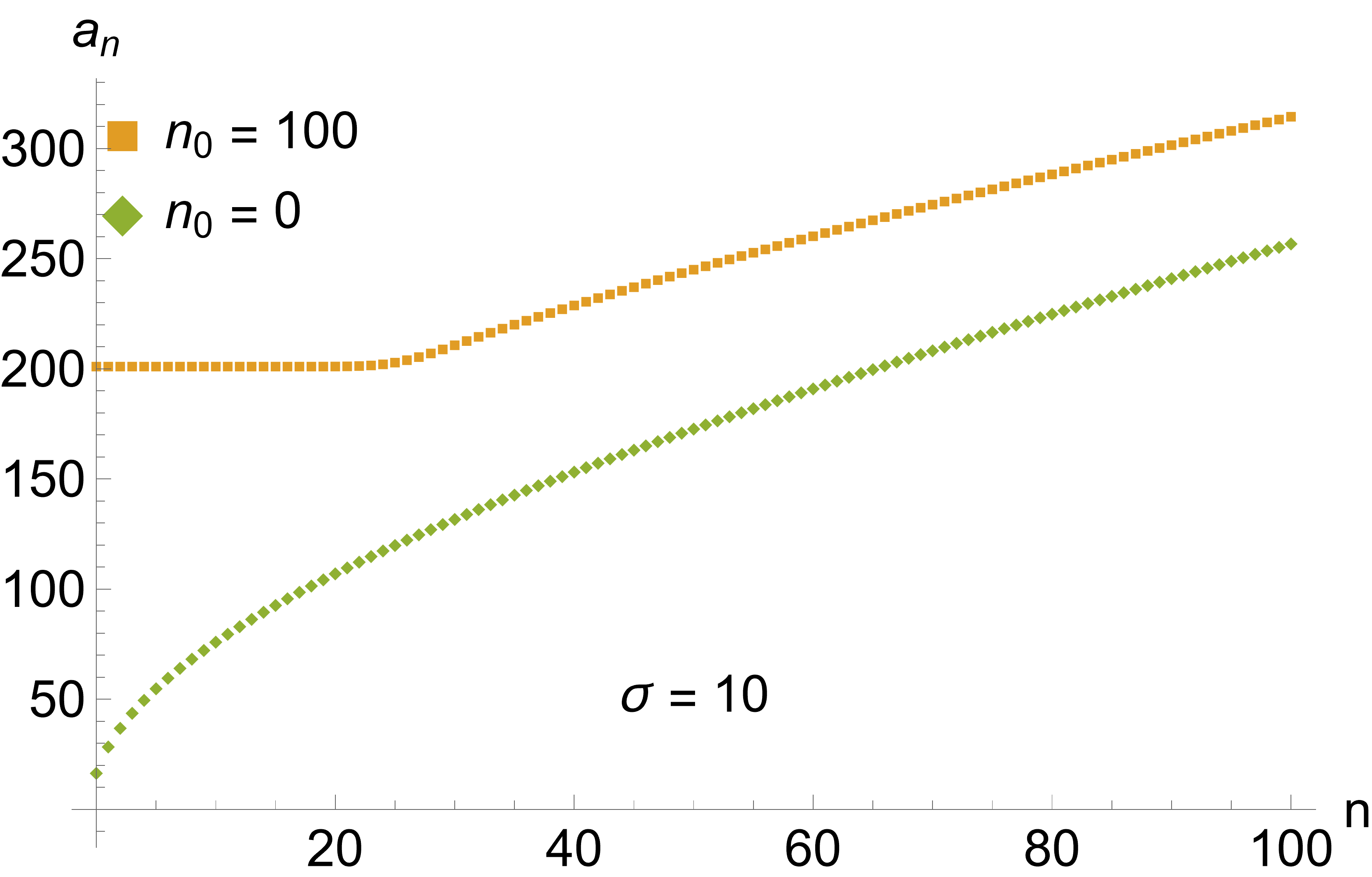}
	\caption{The first Lanczos coefficients of the gaussian superpositions of descendants without a cutoff $\epsilon$ and with $\sigma=10,\,\Delta=1$ and several $n_0$. Here the Krylov basis is infinite and the spectral measure is bound only from below.}
	\label{fig:descendants infty}
\end{figure}

\noindent
We can get rid of the upper boundary by repeating the calculations with a different formulation: now we don't truncate the initial superposition with the cutoff $\epsilon$, and instead approximate all infinite sums, so that we no longer have a limitation for the dimensionality of the Krylov basis and the spectral measure is bounded only from below. The results are shown in Fig. \ref{fig:descendants infty}.
\\

\noindent
The $n_0=0$ case exhibits no regime change, because the effect of the boundary is felt from the start. The $n_0=100$ case exhibits only the first regime change observed in Fig. \ref{fig:descendants s10 n100}, and the second regime here is expected to be eternal, given that there's no upper spectral boundary. The $n_0=1000$ case exhibits no regime change in the range studied, but one is expected, since the coefficients have to sense the spectral boundary at some point. After that, however, the second regime must be eternal. The $a_n$'s for this last case are not shown and are all equal to $2001$ in the range studied.

\subsection{Other probability distributions}

\begin{figure}[tbh]
	\centering
    \includegraphics[width=0.45\textwidth]{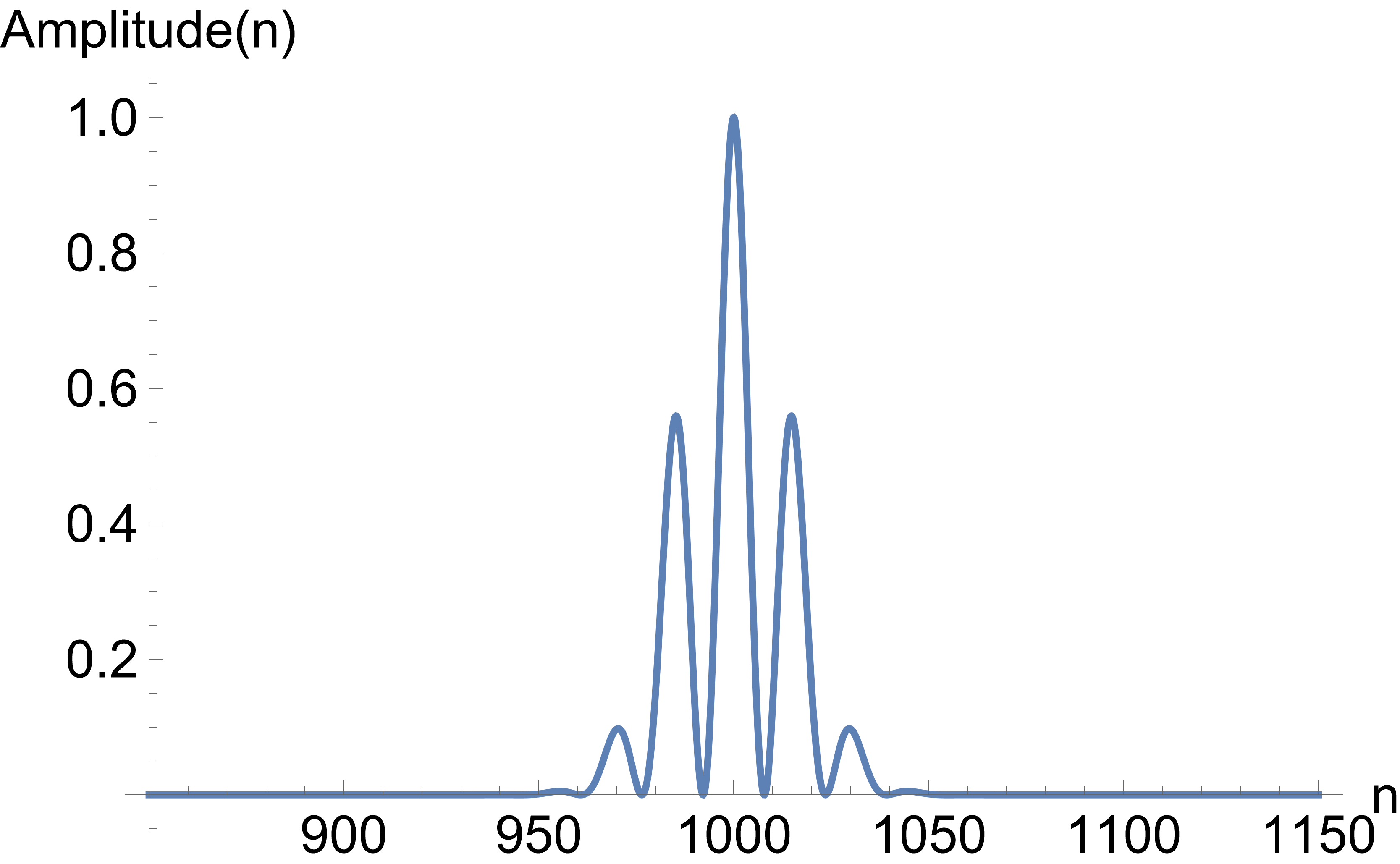}
    \includegraphics[width=0.45\textwidth]{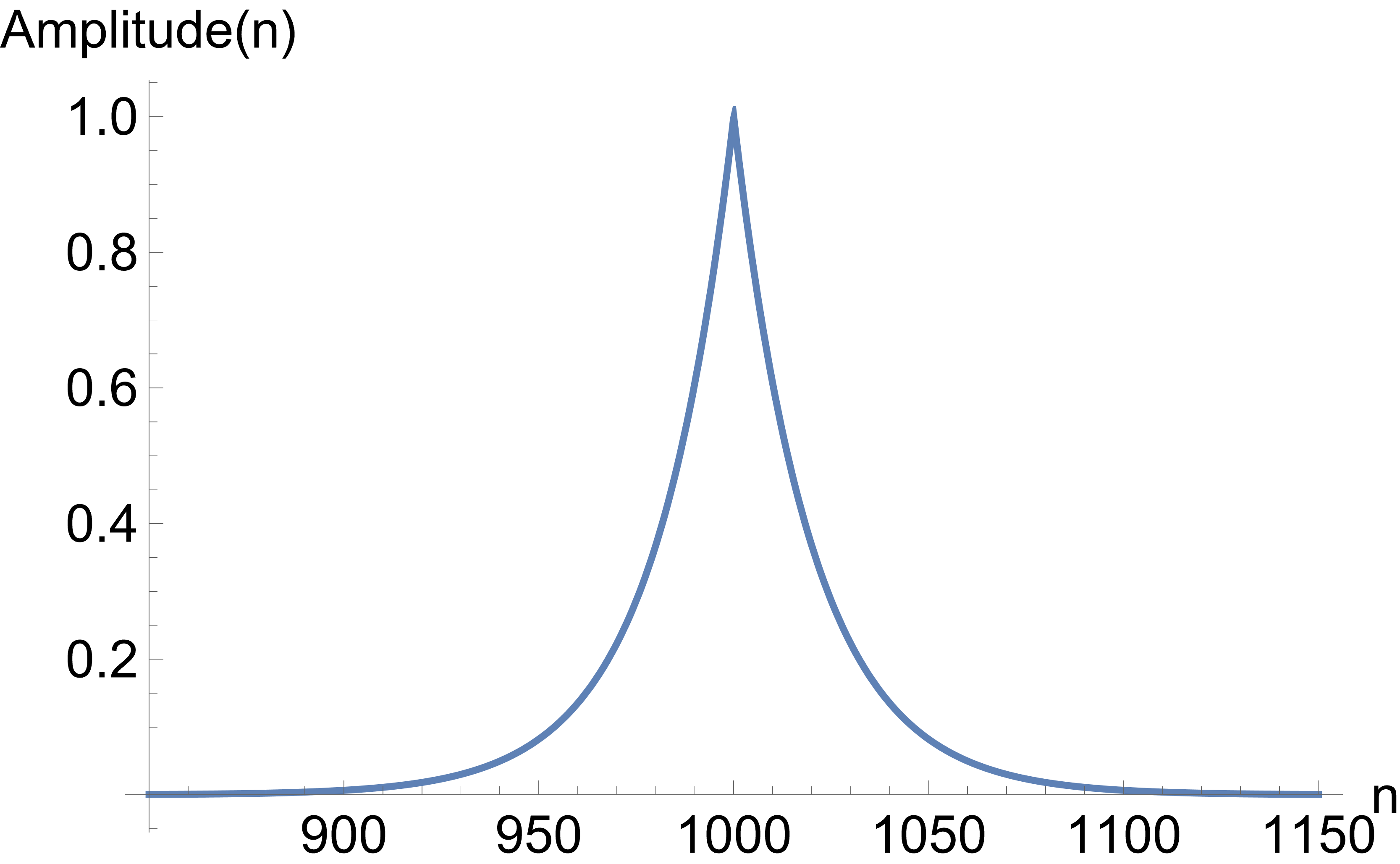}
	\caption{The non-Gaussian probability distributions given in Eqs. \ref{eq:non gaussian cos} (left) and \ref{eq:non gaussian exp} (right).}
	\label{fig:non gaussian dist}
\end{figure}

As a final demonstration for this section, we obtain Lanczos coefficients for two seeds defined by the non-Gaussian distributions
\begin{equation}\label{eq:non gaussian cos}
    |K_0 \rangle = \sum_n e^{-\frac{(n-n_0)^2}{4 \sigma^2}}\cos^2\left( \frac{2(n-n_0)}{\sigma} \right) |\mathcal{O}_{\Delta}; n\rangle,
\end{equation}
and
\begin{equation}\label{eq:non gaussian exp}
    |K_0 \rangle = \sum_n e^{-\frac{|n-n_0|}{2 \sigma}} |\mathcal{O}_{\Delta}; n\rangle,
\end{equation}
up to a normalization constant. We pick $n_0=1000$ and $\sigma=10$, and adjust the limits of the sum over $n$ so that the spectral measure is bounded by $A=1669$ and $B=2333$, as in the case studied in the previous subsection and given in Fig. \ref{fig:descendants bn s10 n1000}. The unnormalized distributions are plotted in Fig. \ref{fig:non gaussian dist} for reference.
\\

\begin{figure}[tbh]
	\centering
    \includegraphics[width=0.45\textwidth]{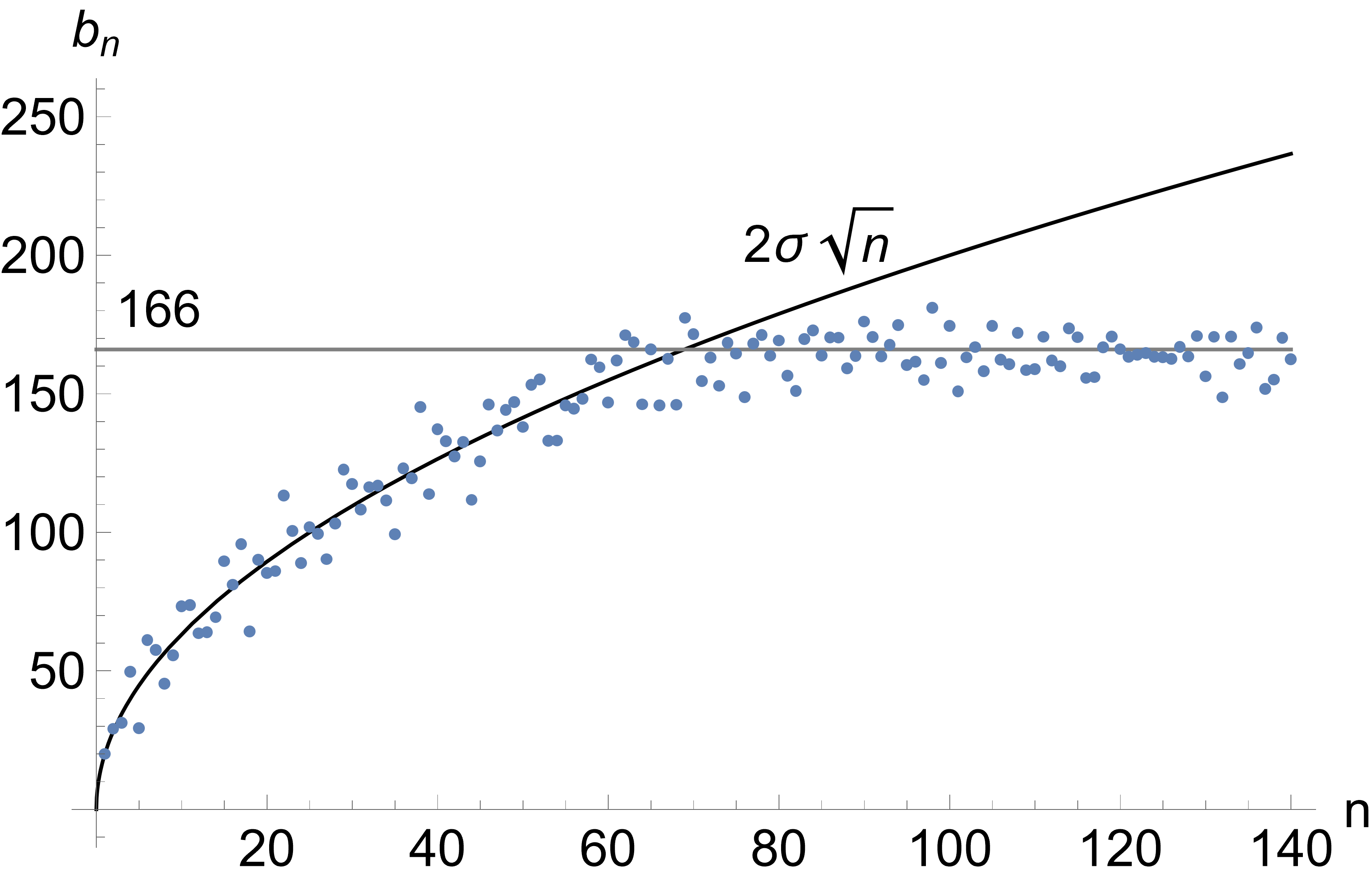}
    \includegraphics[width=0.45\textwidth]{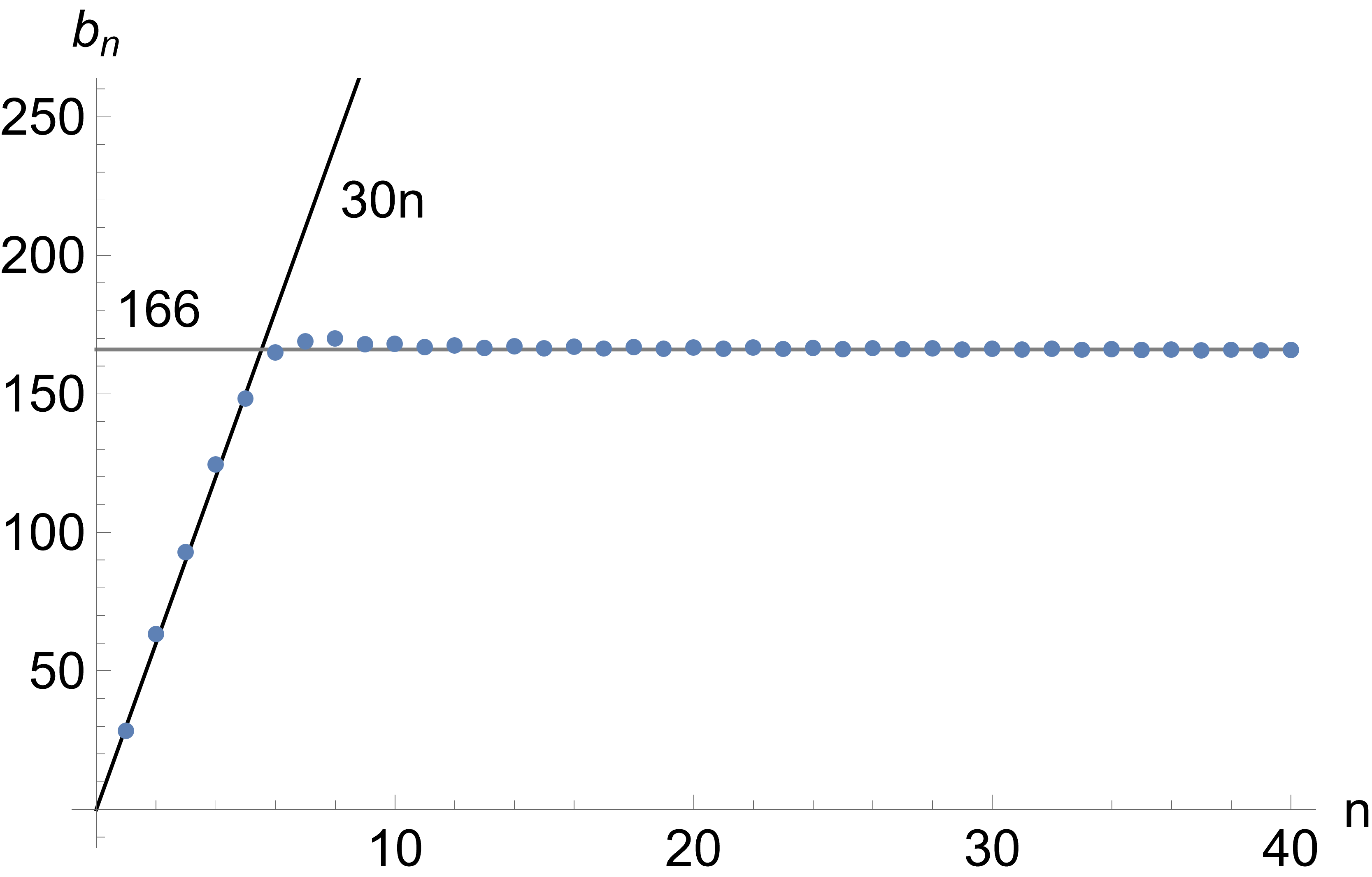}
	\caption{First 140 $b_n$ coefficients obtained from the seed given in Eq. \ref{eq:non gaussian cos} (left) and first 40 $b_n$'s obtained from the seed given in Eq. \ref{eq:non gaussian exp} (right). The corresponding $a_n$'s are all equal to 2001.}
	\label{fig:non gaussian bn}
\end{figure}

\noindent
In both cases, the growth of the $b_n$'s is governed by the decay of the corresponding distribution, as displayed in Fig. \ref{fig:non gaussian bn}. The one that decays as a Gaussian preserves the $b_n\propto\sqrt{\sigma}$ behavior seen in previous cases, while the one with an exponential decay showcases linear growth, as expected from spectral measures with exponential tails. However, these coefficients still reach a maximum around $(B-A)/4=166$, in accordance with the asymptotics dictated by the boundaries of the measure. The $a_n$'s are equal to $2001$ in both cases, also as predicted.

\FloatBarrier
\section{BMN sector: two impurities}

\label{BMNNumerics}
\begin{figure}[h!]
\begin{minipage}{0.49\textwidth}
		\includegraphics[width=1.00\textwidth]{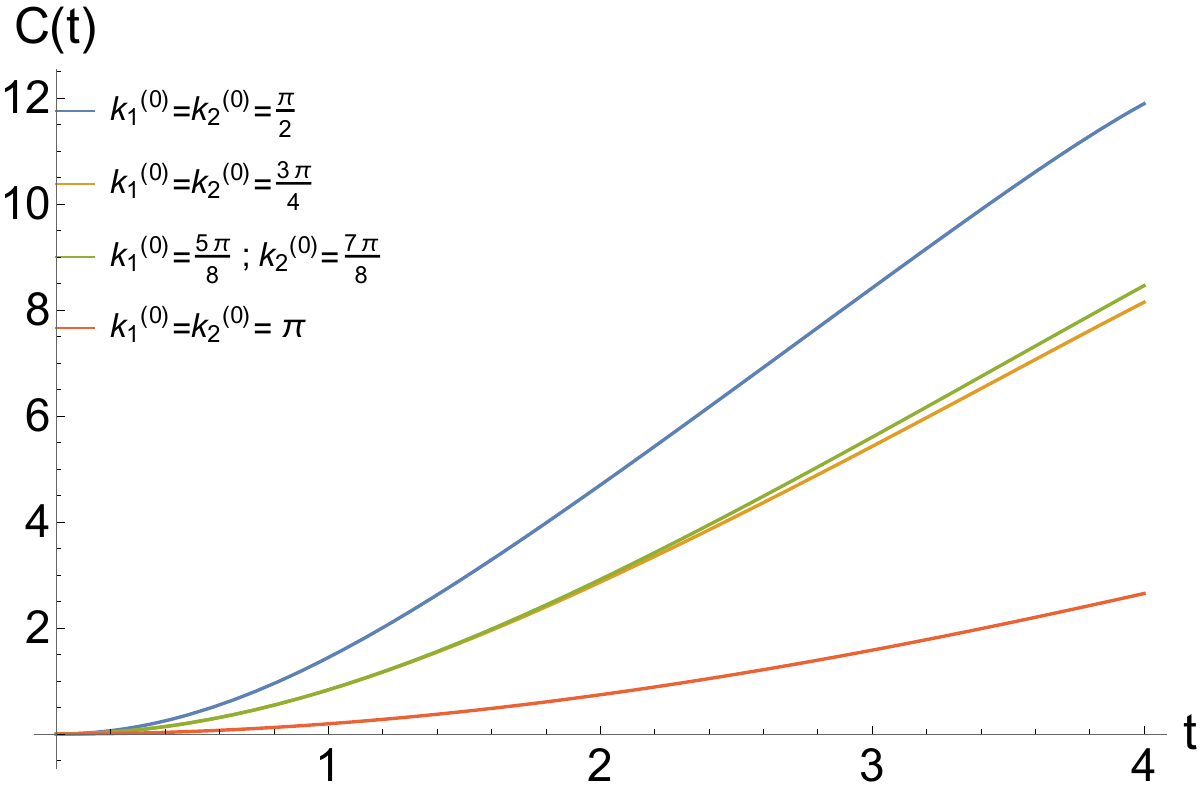}
\end{minipage}
\begin{minipage}{0.49\textwidth}
		\includegraphics[width=1.00\textwidth]{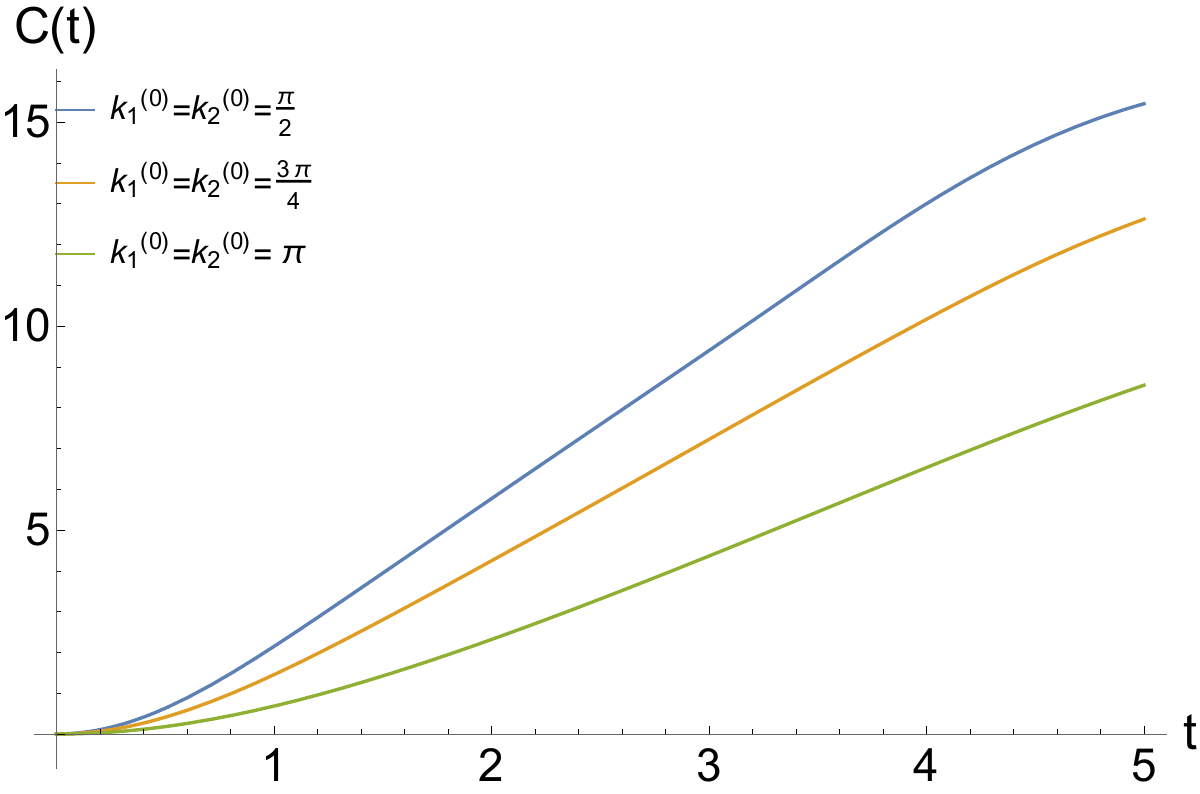}
\end{minipage}
	\caption{The complexity at early times for $L=16$ and $\sigma = \frac{1}{2}$ (left), $\sigma = \frac{3}{4}$ (right).  We find the slowest growth when the packet is near the band edge and fastest growth when the packet is centered near the middle of the band.  }
	\label{fig:TwoMagnonsEarlyTime}
\end{figure}

\begin{figure}[h!]
	\centering
		\includegraphics[width=0.6\textwidth]{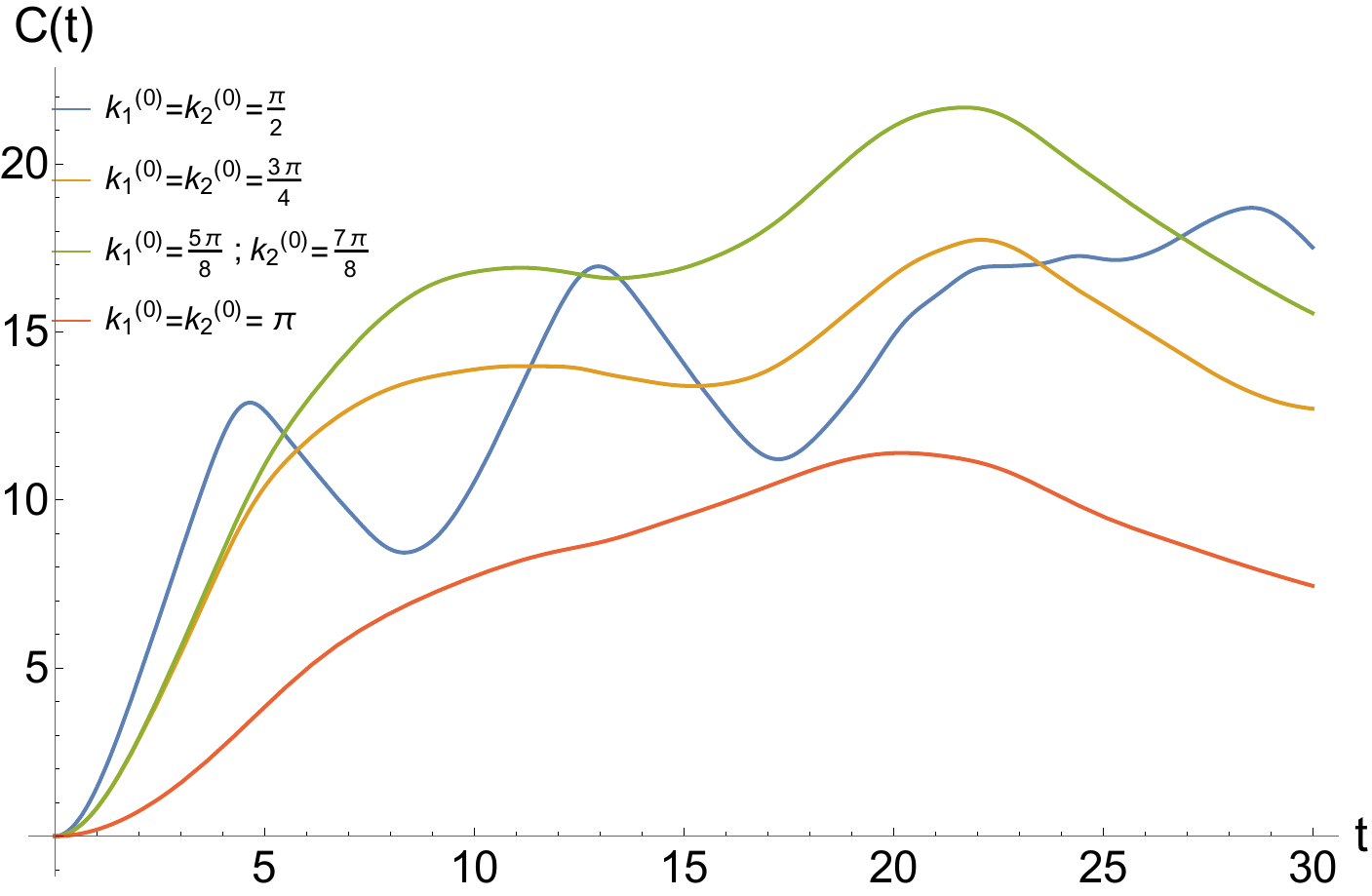}
	\caption{The complexity at later times for $\sigma = \frac{1}{2}$ and $L=16$.  }
	\label{fig:TwoMagnonsInterTime}
\end{figure}

\begin{figure}[h!]
	\centering
		\includegraphics[width=0.6\textwidth]{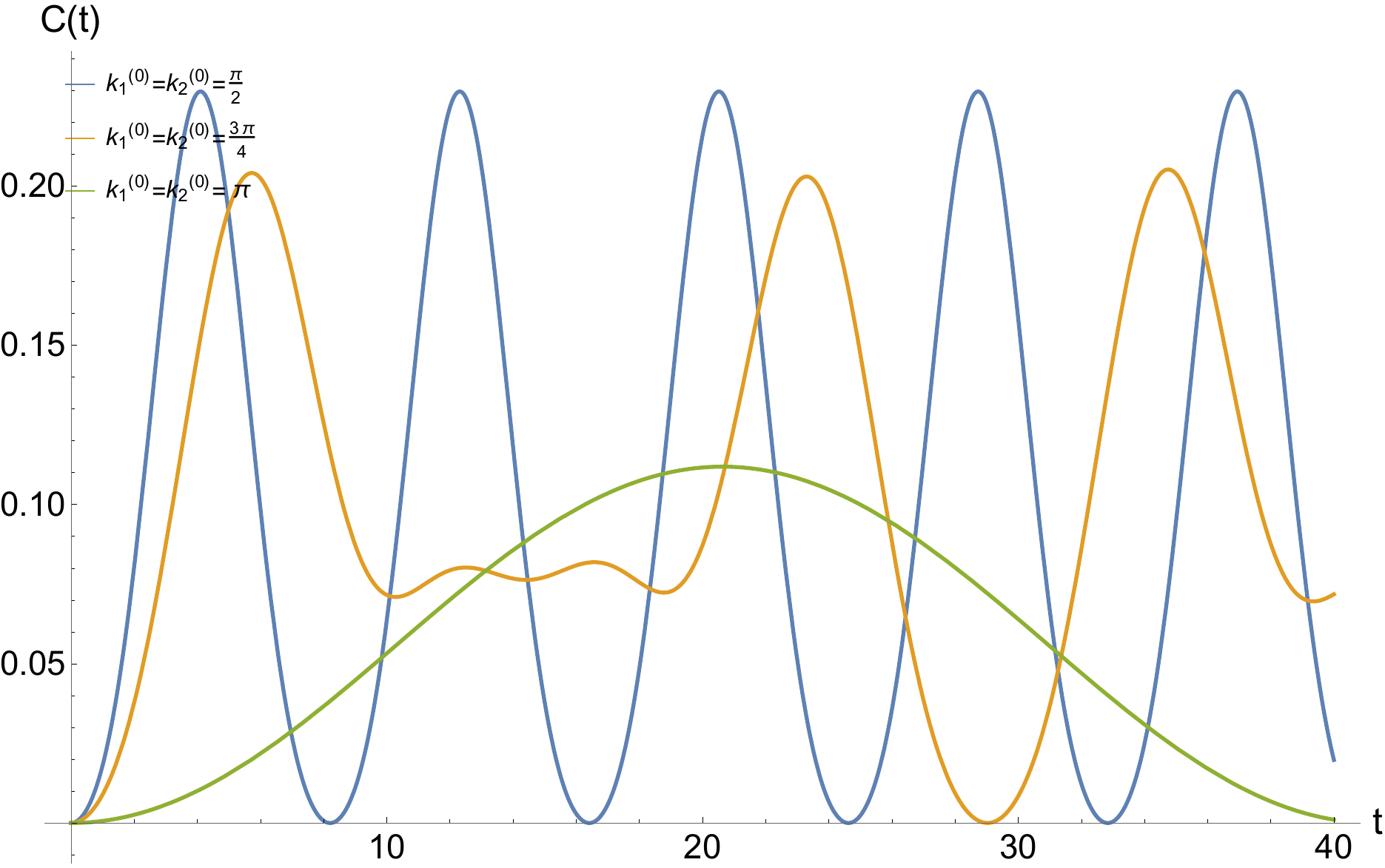}
	\caption{The spread complexity for $L=16$, and $\sigma = \frac{1}{8}$.  The complexity remains small and only a few energies are relevant.   }
	\label{fig:TwoMagnonsSmallSigma}
\end{figure}

The states and spectrum are fully described in the main text, but we restate the key details here for the sake of convenience.  We have a $L^2$ dimensional space of states labeled by pairs of momenta
\begin{equation}
    k = \frac{2 \pi n}{L} \ \ \ n = 0,1,2,\cdots, L,
\end{equation}
which have energies
\begin{equation}
    H|k_1, k_2 \rangle = 2 g (2 - \cos(k_1) - \cos(k_2)) |k_1, k_2 \rangle.
\end{equation}
We will set $g = 1$ in what follows.  As reference state we take a Gaussian superposition of these eigenstates
\begin{equation}
    |K_0\rangle = N \sum_{k_1,k_2} e^{-\frac{(k_1 - k_{1}^{(0)})^2}{4 \sigma_1^2}-\frac{(k_2 - k_{2}^{(0)})^2}{4 \sigma_2^2}}
|k_1,k_2\rangle,
\end{equation}
with $\sigma_1 = \sigma_2 = \sigma$ for the sake of simplicity.  We have plotted the early time complexity in Fig. (\ref{fig:TwoMagnonsEarlyTime}).  As explained in the main text, we find the slowest early growth when the packet is near the band edge and the fastest early growth when the packet is near the middle of the band.  At later times this may no longer hold (see Fig. (\ref{fig:TwoMagnonsInterTime}). Finally, when $\sigma$ is small, we find that the complexity approaches $0$, in line with the expectation that the packet will approach a single eigenstate.  In Fig. (\ref{fig:TwoMagnonsSmallSigma}) we have plotted this situation for $\sigma = \frac{1}{8}$.  It is clear that only a few eigenvalues are relevant to the evolution of the reference state.\\

\FloatBarrier
\noindent
The spectrum studied here is highly degenerate, and the Lanczos algorithm doesn't distinguish degenerate energy eigenstates. As a consequence, the number of dimensions of the Krylov basis obtained from a packet in the space of momenta is much smaller than the number of dimensions of the packet itself, and corresponds to the number of energy eigenvalues accessible to the seed.\\

\begin{figure}[tbh]
	\centering
    \includegraphics[width=0.40\textwidth]{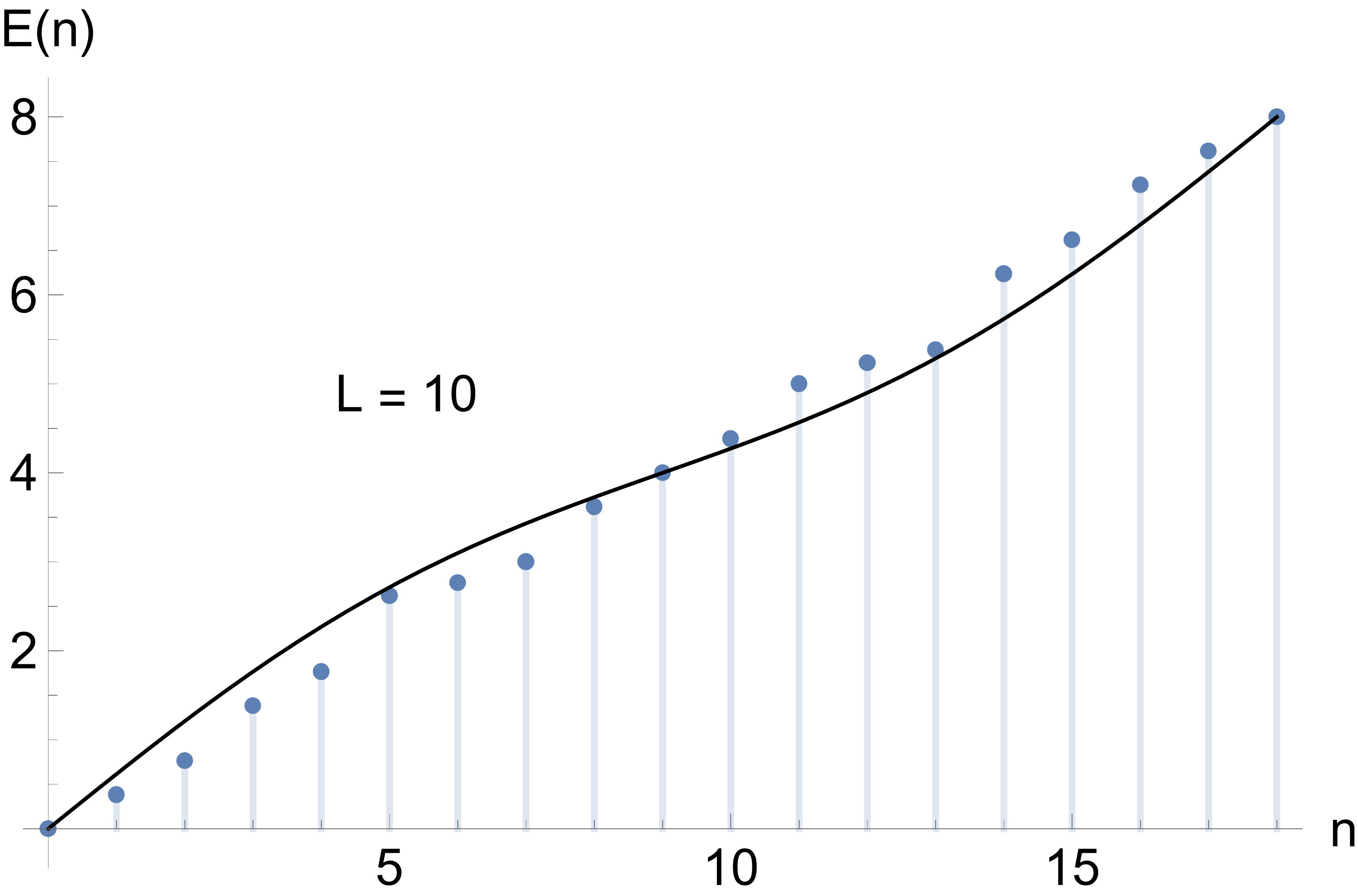}
    \includegraphics[width=0.40\textwidth]{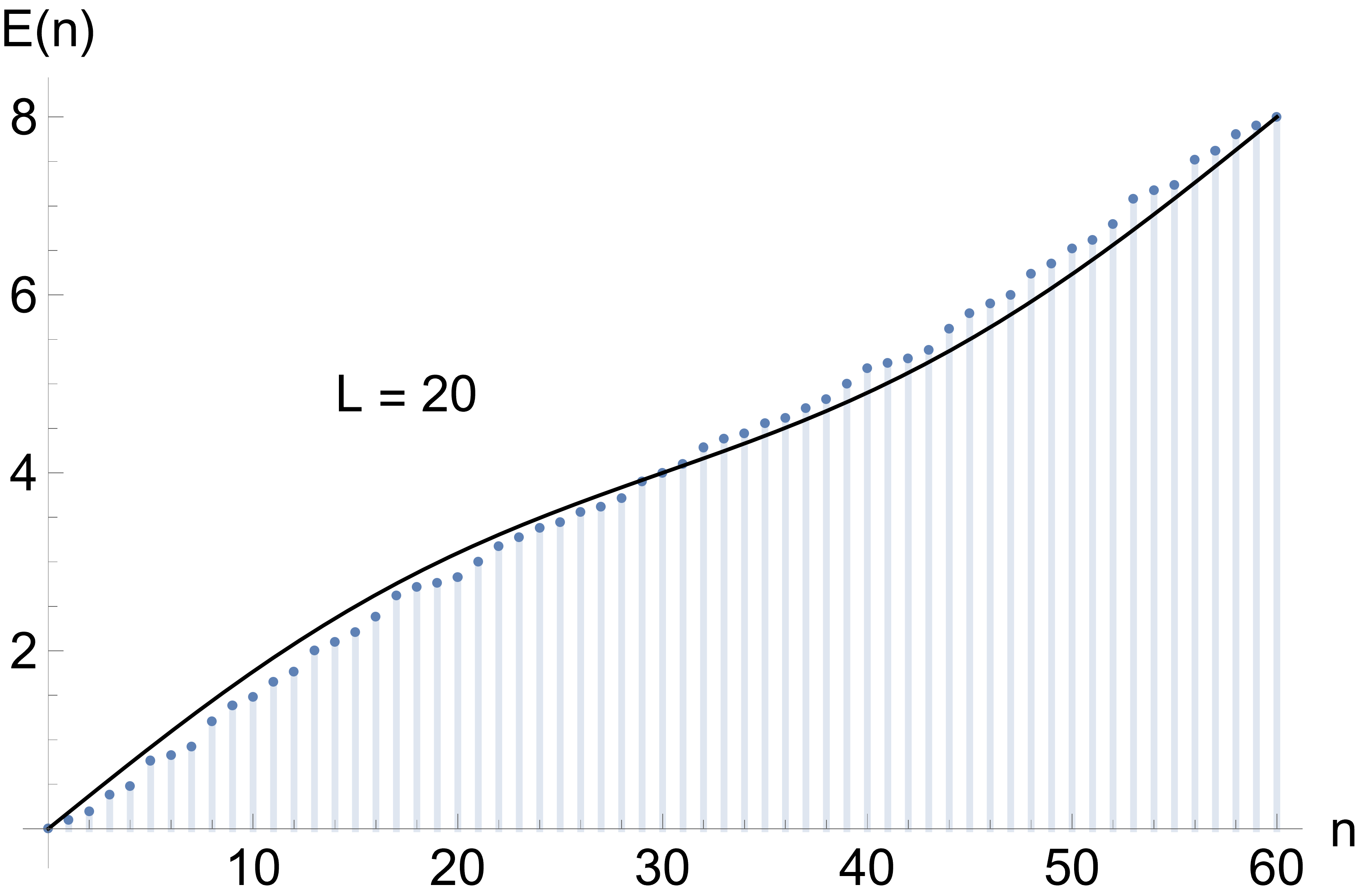}
    \includegraphics[width=0.40\textwidth]{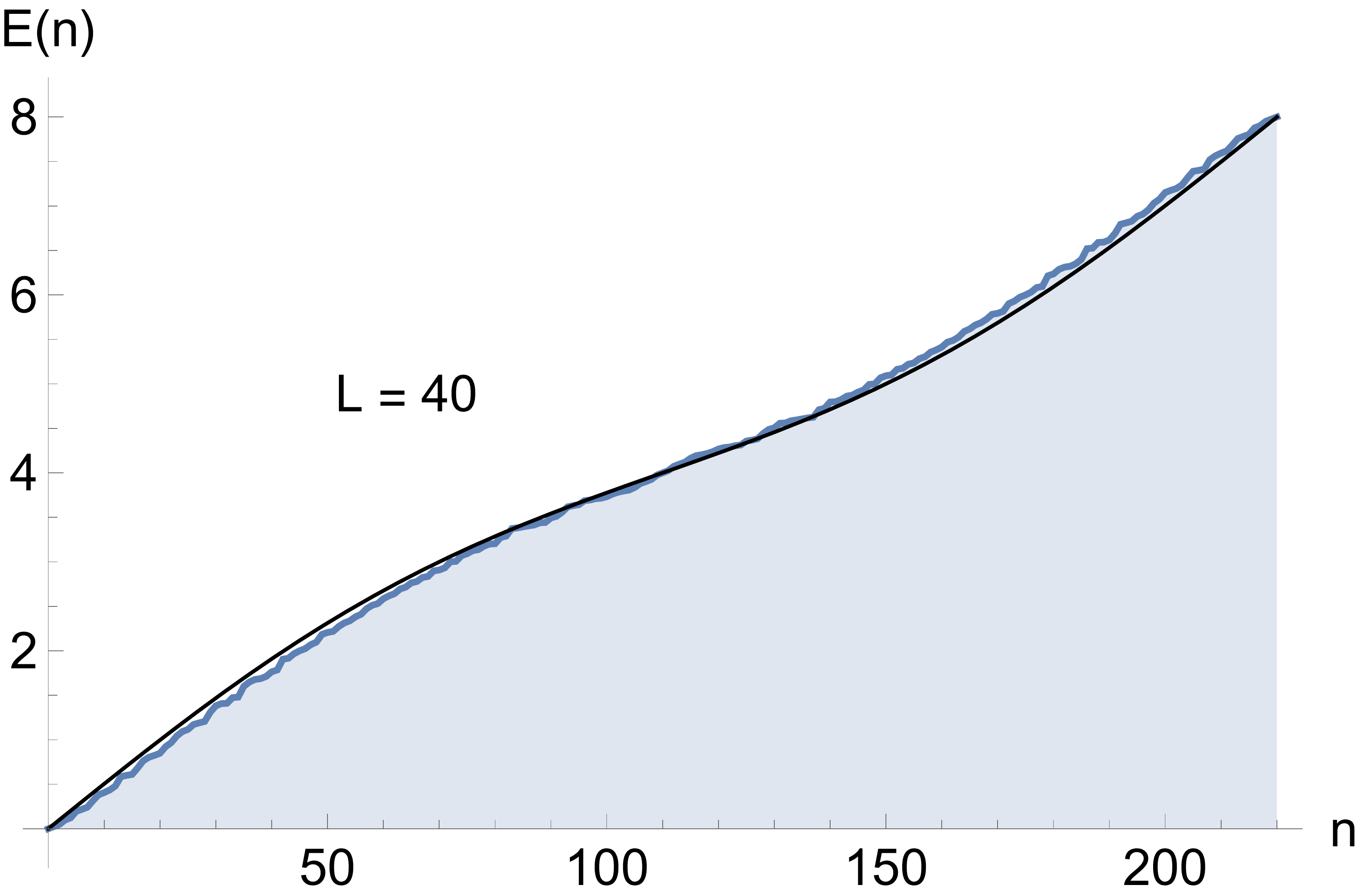}
    \includegraphics[width=0.40\textwidth]{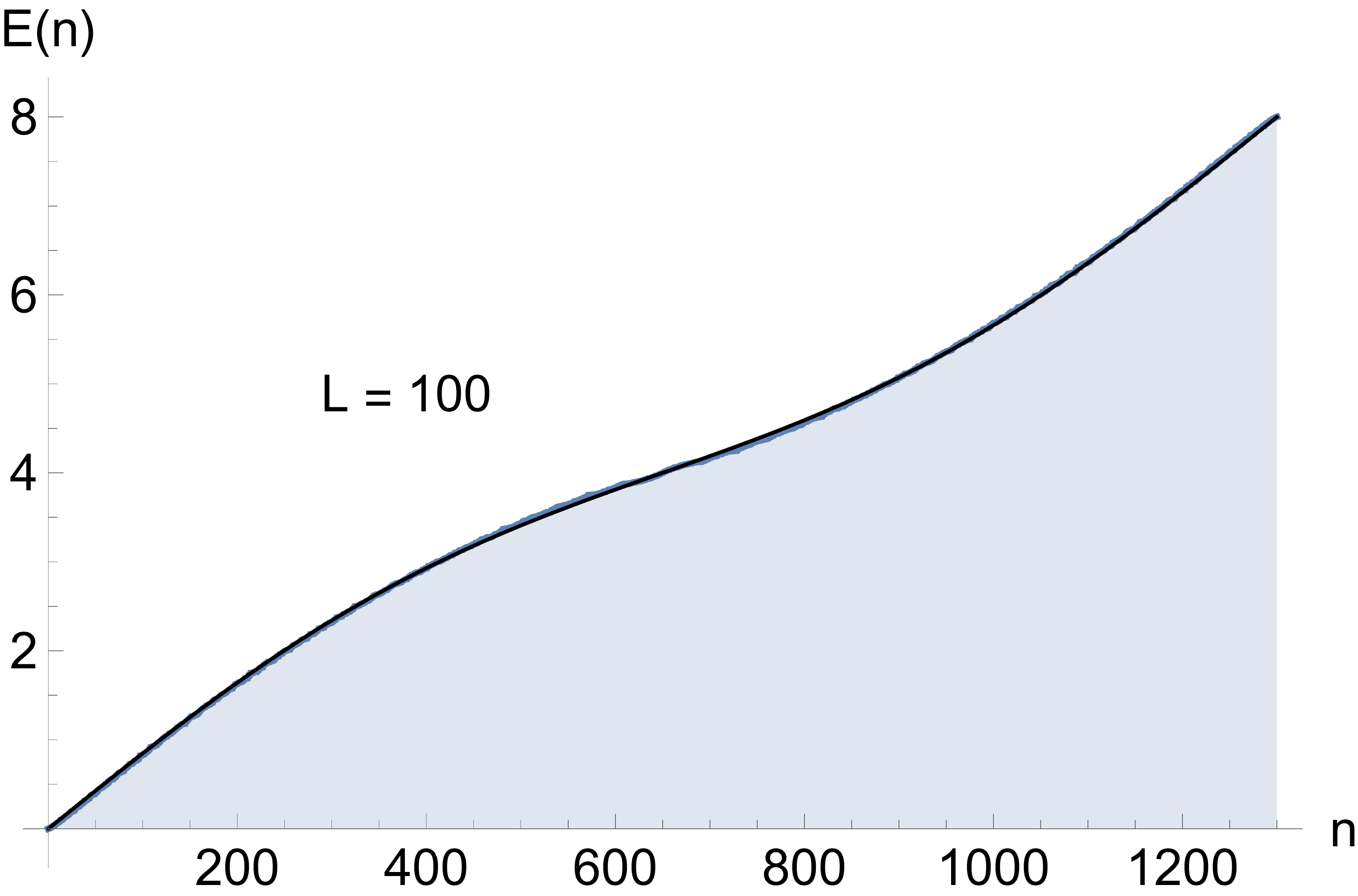}
	\caption{Available energies for superpositions of momentum eigenstates with different values of $L$. The energy eigenvalues are labeled by an integer variable $n$. The function in Eq. \ref{eq:energy distribution magnons}, with $\Lambda=1/2$, is plotted as a black curve.}
	\label{fig:energies g1}
\end{figure}

\begin{figure}[tbh]
	\centering
    \includegraphics[width=0.6\textwidth]{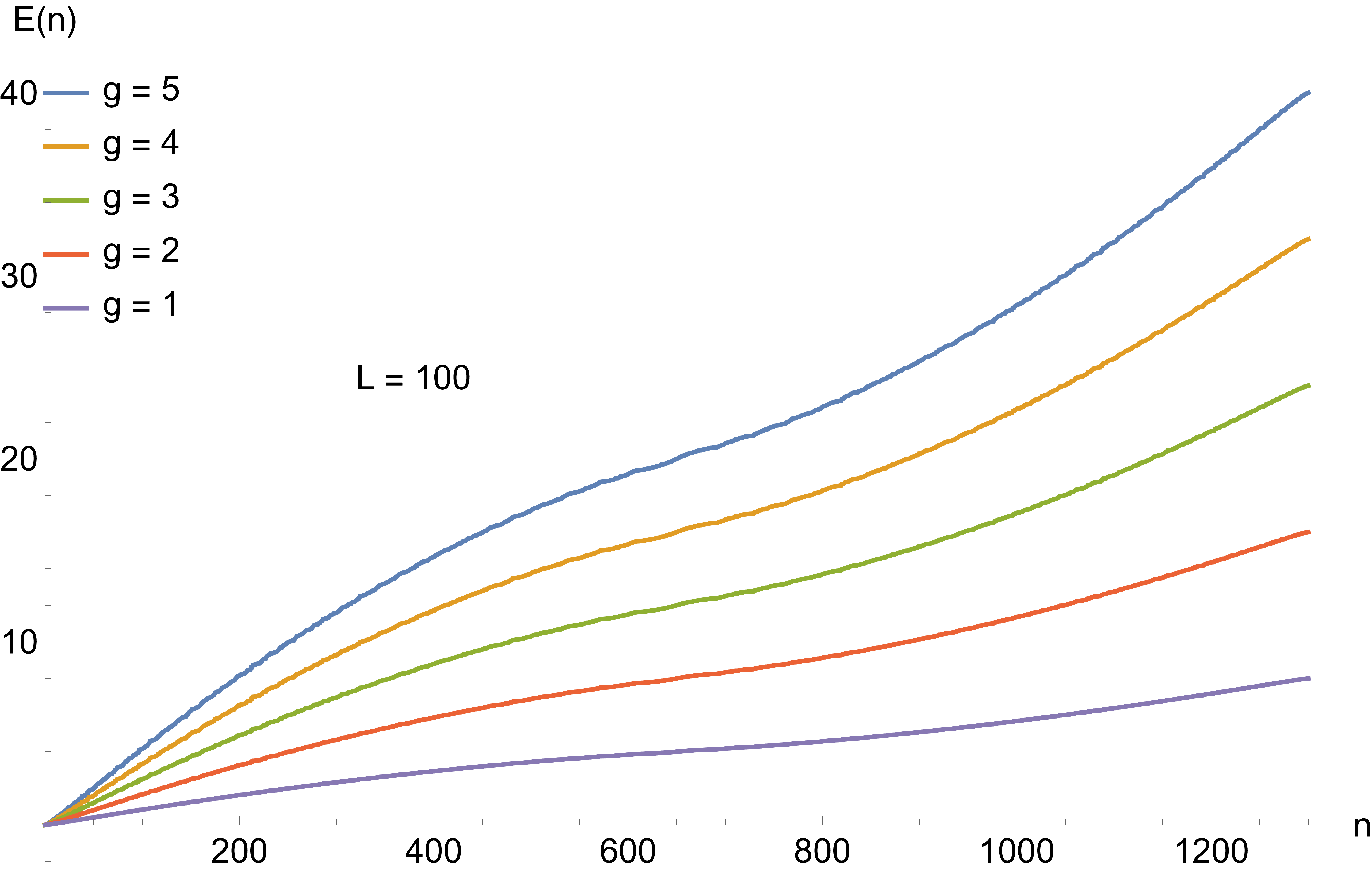}
	\caption{Available energies for superpositions of momentum eigenstates with $L=100$ and different values of $g$.}
	\label{fig:energies many g}
\end{figure}

\noindent
We can find the distribution of allowed energy eigenvalues numerically. Fig. \ref{fig:energies g1} presents the distribution of energies $E(n)$ for different values of $L$, with the energies labeled by an integer variable $n$ ranging from $0$ to $n_{max}(L)$.
In the continuous limit, the distribution of energies seems to approximate a function of the type
\begin{equation}\label{eq:energy distribution magnons}
    E(n) \approx \frac{8gn}{n_{max}} + \Lambda g \sin\left(\frac{2\pi n}{n_{max}}\right),
\end{equation}
for some constant $\Lambda$. Fig. \ref{fig:energies many g} shows how this varies with $g$.
\\

\begin{figure}[tbh]
	\centering
    \includegraphics[width=0.45\textwidth]{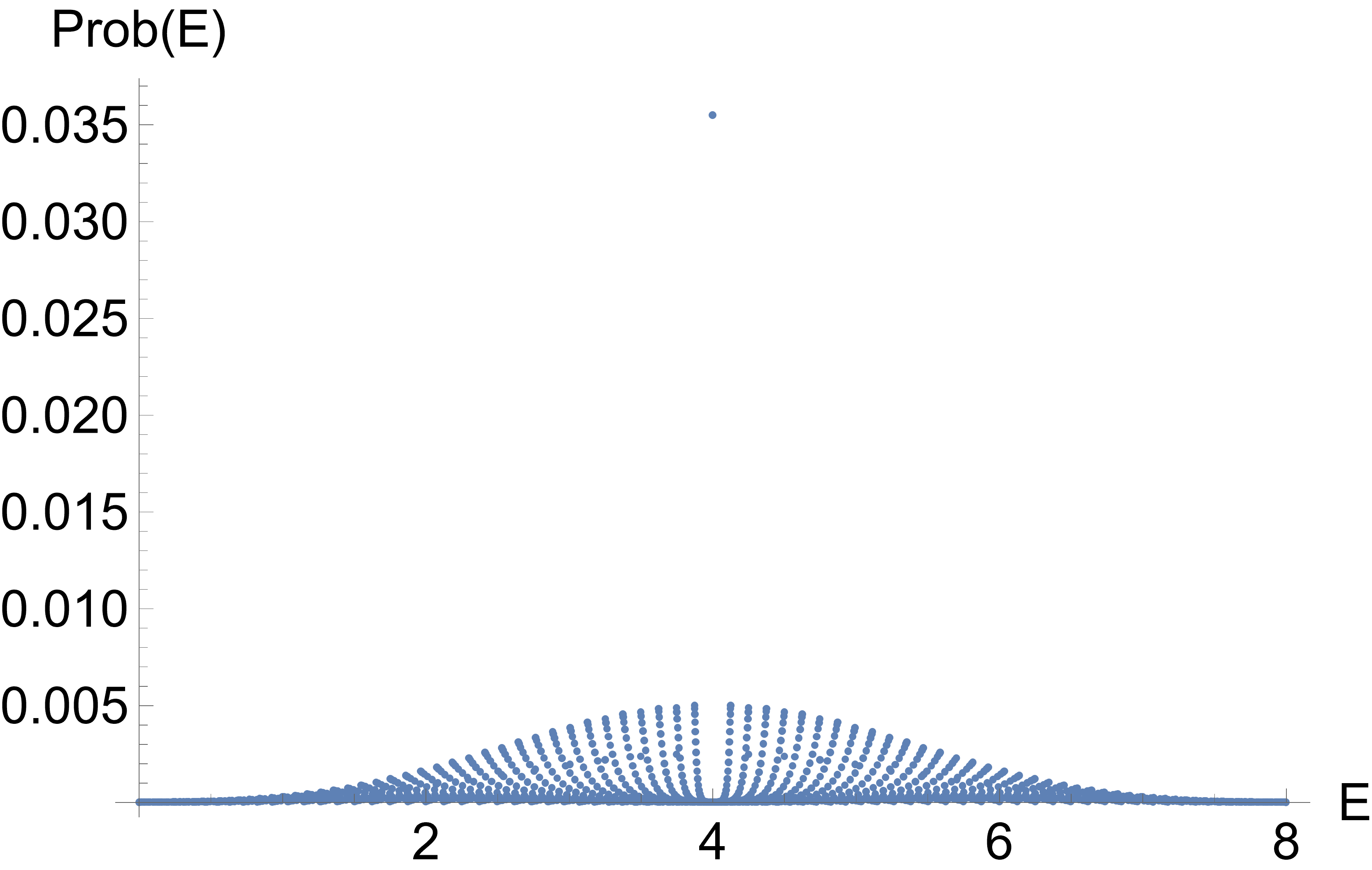}
    \includegraphics[width=0.45\textwidth]{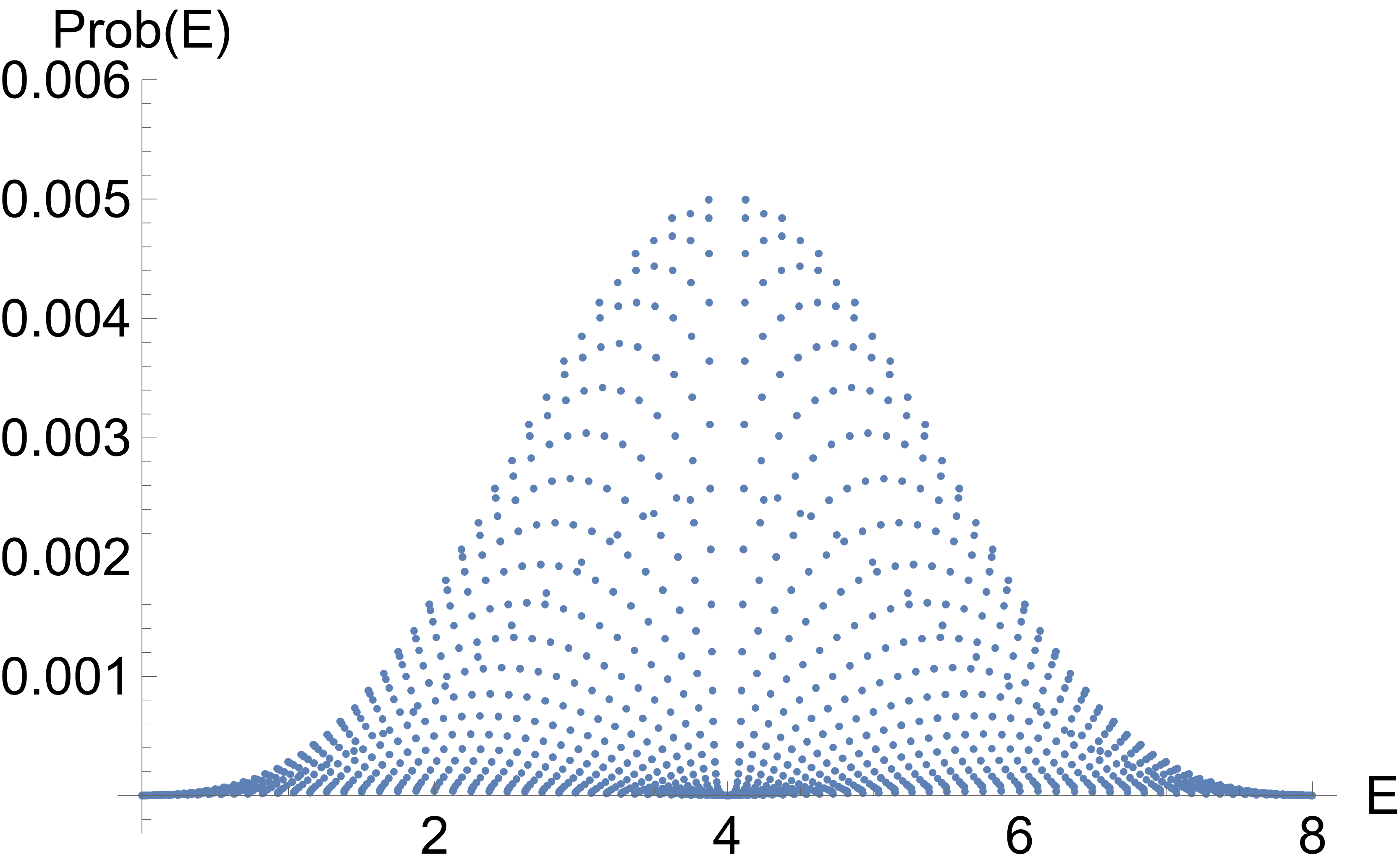}
	\caption{Distribution of probabilities of energy eigenvalues resulting from the Gaussian superposition over $k_1$ and $k_2$ and with parameters $\{g=1,\,L=100,\,\sigma_1=\sigma_2=1/2,\,k_{1}^{(0)}=k_{2}^{(0)}=\pi/2\}$. On the left we display the full set of probabilities; on the right, the eigenvalue with highest probability is excluded in favor of better visualization.}
	\label{fig:dist prob eigenvalues magnons}
\end{figure}

\noindent
Additionally, we can obtain the distribution of probabilities for each energy eigenstate resulting from the gaussian distributions in $k_1$ and $k_2$. An example, corresponding to an arbitrary set of parameters, is given in Fig. \ref{fig:dist prob eigenvalues magnons}.

\FloatBarrier

\bibliographystyle{JHEP}
\bibliography{biblio.bib}

@article{Parker:2018yvk,
  author        = {Parker, Daniel E. and Cao, Xiaoliang and Avdoshkin, Alexander and Scaffidi, Thomas and Altman, Ehud},
  title         = {A Universal Operator Growth Hypothesis},
  journal       = {Phys. Rev. X},
  volume        = {9},
  number        = {4},
  pages         = {041017},
  year          = {2019},
  eprint        = {1812.08657},
  archivePrefix = {arXiv},
  primaryClass  = {cond-mat.stat-mech}
}

@article{Maldacena:1997re,
  author        = {Maldacena, Juan M.},
  title         = {The Large N limit of superconformal field theories and supergravity},
  journal       = {Adv. Theor. Math. Phys.},
  volume        = {2},
  pages         = {231--252},
  year          = {1998},
  eprint        = {hep-th/9711200},
  archivePrefix = {arXiv},
  primaryClass  = {hep-th}
}

@article{Beisert:2010jr,
  author        = {Beisert, Niklas and others},
  title         = {Review of AdS/CFT Integrability: An Overview},
  journal       = {Lett. Math. Phys.},
  volume        = {99},
  pages         = {3--32},
  year          = {2012},
  eprint        = {1012.3982},
  archivePrefix = {arXiv},
  primaryClass  = {hep-th}
}

@article{Minahan:2002ve,
  author        = {Minahan, Joseph A. and Zarembo, Konstantin},
  title         = {The Bethe-ansatz for N=4 super Yang-Mills},
  journal       = {JHEP},
  volume        = {03},
  pages         = {013},
  year          = {2003},
  eprint        = {hep-th/0212208},
  archivePrefix = {arXiv},
  primaryClass  = {hep-th}
}

@article{Balasubramanian:2001nh,
  author        = {Balasubramanian, Vijay and Berkooz, Micha and Naqvi, Asad and Strassler, Matthew J.},
  title         = {Giant gravitons in conformal field theory},
  journal       = {JHEP},
  volume        = {04},
  pages         = {034},
  year          = {2002},
  eprint        = {hep-th/0107119},
  archivePrefix = {arXiv},
  primaryClass  = {hep-th}
}

@article{Corley:2001zk,
  author        = {Corley, Steven and Jevicki, Antal and Ramgoolam, Sanjaye},
  title         = {Exact correlators of giant gravitons from dual N=4 SYM theory},
  journal       = {Adv. Theor. Math. Phys.},
  volume        = {5},
  pages         = {809--839},
  year          = {2002},
  eprint        = {hep-th/0111222},
  archivePrefix = {arXiv},
  primaryClass  = {hep-th}
}

@article{deMelloKoch:2007uu,
  author        = {de Mello Koch, Robert and Ramgoolam, Sanjaye and Wen, C.},
  title         = {On Young diagrams, Schur polynomials and AdS/CFT},
  journal       = {JHEP},
  volume        = {06},
  pages         = {106},
  year          = {2007},
  eprint        = {hep-th/0701066},
  archivePrefix = {arXiv},
  primaryClass  = {hep-th}
}

@article{Nastase:2026lhz,
    author = "Nastase, Horatiu and Nunez, Carlos and Roychowdhury, Dibakar",
    title = "{Holographic Krylov Complexity for Charged, Composite and Extended Probes}",
    eprint = "2604.07432",
    archivePrefix = "arXiv",
    primaryClass = "hep-th",
    month = "4",
    year = "2026"
}

@article{Lin:2004nb,
    author = "Lin, Hai and Lunin, Oleg and Maldacena, Juan Martin",
    title = "{Bubbling AdS space and 1/2 BPS geometries}",
    eprint = "hep-th/0409174",
    archivePrefix = "arXiv",
    reportNumber = "PUPT-2136",
    doi = "10.1088/1126-6708/2004/10/025",
    journal = "JHEP",
    volume = "10",
    pages = "025",
    year = "2004"
}

@article{Chattopadhyay:2023fob,
    author = "Chattopadhyay, Arghya and Mitra, Arpita and van Zyl, Hendrik J. R.",
    title = "{Spread complexity as classical dilaton solutions}",
    eprint = "2302.10489",
    archivePrefix = "arXiv",
    primaryClass = "hep-th",
    doi = "10.1103/PhysRevD.108.025013",
    journal = "Phys. Rev. D",
    volume = "108",
    number = "2",
    pages = "025013",
    year = "2023"
}

@article{deMelloKoch:2007nbd,
    author = "de Mello Koch, Robert and Smolic, Jelena and Smolic, Milena",
    title = "{Giant Gravitons - with Strings Attached (II)}",
    eprint = "hep-th/0701067",
    archivePrefix = "arXiv",
    reportNumber = "WITS-CTP-032",
    doi = "10.1088/1126-6708/2007/09/049",
    journal = "JHEP",
    volume = "09",
    pages = "049",
    year = "2007"
}

@article{deMelloKoch:2007rqf,
    author = "de Mello Koch, Robert and Smolic, Jelena and Smolic, Milena",
    title = "{Giant Gravitons - with Strings Attached (I)}",
    eprint = "hep-th/0701066",
    archivePrefix = "arXiv",
    reportNumber = "WITS-CTP-031",
    doi = "10.1088/1126-6708/2007/06/074",
    journal = "JHEP",
    volume = "06",
    pages = "074",
    year = "2007"
}

@article{Bekker:2007ea,
    author = "Bekker, David and de Mello Koch, Robert and Stephanou, Michael",
    title = "{Giant Gravitons - with Strings Attached. III.}",
    eprint = "0710.5372",
    archivePrefix = "arXiv",
    primaryClass = "hep-th",
    reportNumber = "WITS-CTP-033",
    doi = "10.1088/1126-6708/2008/02/029",
    journal = "JHEP",
    volume = "02",
    pages = "029",
    year = "2008"
}

@article{Gubser:1998bc,
    author = "Gubser, S. S. and Klebanov, Igor R. and Polyakov, Alexander M.",
    title = "{Gauge theory correlators from noncritical string theory}",
    eprint = "hep-th/9802109",
    archivePrefix = "arXiv",
    reportNumber = "PUPT-1767",
    doi = "10.1016/S0370-2693(98)00377-3",
    journal = "Phys. Lett. B",
    volume = "428",
    pages = "105--114",
    year = "1998"
}

@article{Witten:1998qj,
    author = "Witten, Edward",
    title = "{Anti de Sitter space and holography}",
    eprint = "hep-th/9802150",
    archivePrefix = "arXiv",
    reportNumber = "IASSNS-HEP-98-15",
    doi = "10.4310/ATMP.1998.v2.n2.a2",
    journal = "Adv. Theor. Math. Phys.",
    volume = "2",
    pages = "253--291",
    year = "1998"
}

@article{Beisert:2003ys,
    author = "Beisert, Niklas",
    title = "{The su(2|3) dynamic spin chain}",
    eprint = "hep-th/0310252",
    archivePrefix = "arXiv",
    reportNumber = "AEI-2003-087",
    doi = "10.1016/j.nuclphysb.2003.12.032",
    journal = "Nucl. Phys. B",
    volume = "682",
    pages = "487--520",
    year = "2004"
}

@article{Beisert:2005tm,
    author = "Beisert, Niklas",
    title = "{The SU(2|2) dynamic S-matrix}",
    eprint = "hep-th/0511082",
    archivePrefix = "arXiv",
    reportNumber = "PUTP-2181, NSF-KITP-05-92",
    doi = "10.4310/ATMP.2008.v12.n5.a1",
    journal = "Adv. Theor. Math. Phys.",
    volume = "12",
    pages = "945--979",
    year = "2008"
}

@article{deMelloKoch:2016nxq,
    author = "de Mello Koch, Robert and Mathwin, Christopher and van Zyl, Hendrik J. R.",
    title = "{LLM Magnons}",
    eprint = "1601.06914",
    archivePrefix = "arXiv",
    primaryClass = "hep-th",
    doi = "10.1007/JHEP03(2016)110",
    journal = "JHEP",
    volume = "03",
    pages = "110",
    year = "2016"
}

@article{deMelloKoch:2018tlb,
    author = "de Mello Koch, Robert and Kim, Minkyoo and Zyl, Hendrik J. R.",
    title = "{Integrable Subsectors from Holography}",
    eprint = "1802.01367",
    archivePrefix = "arXiv",
    primaryClass = "hep-th",
    doi = "10.1007/JHEP05(2018)198",
    journal = "JHEP",
    volume = "05",
    pages = "198",
    year = "2018"
}

@article{Nandy:2024evd,
    author = "Nandy, Pratik and Matsoukas-Roubeas, Apollonas S. and Mart{\'\i}nez-Azcona, Pablo and Dymarsky, Anatoly and del Campo, Adolfo",
    title = "{Quantum dynamics in Krylov space: Methods and applications}",
    eprint = "2405.09628",
    archivePrefix = "arXiv",
    primaryClass = "quant-ph",
    reportNumber = "RIKEN-iTHEMS-Report-24",
    doi = "10.1016/j.physrep.2025.05.001",
    journal = "Phys. Rept.",
    volume = "1125-1128",
    pages = "1--82",
    year = "2025"
}

@article{Rabinovici:2025otw,
    author = "Rabinovici, Eliezer and S{\'a}nchez-Garrido, Adri{\'a}n and Shir, Ruth and Sonner, Julian",
    title = "{Krylov Complexity}",
    eprint = "2507.06286",
    archivePrefix = "arXiv",
    primaryClass = "hep-th",
    reportNumber = "CERN-TH-2025-128",
    month = "7",
    year = "2025"
}

@article{Baiguera:2025dkc,
    author = "Baiguera, Stefano and Balasubramanian, Vijay and Caputa, Pawel and Chapman, Shira and Haferkamp, Jonas and Heller, Michal P. and Halpern, Nicole Yunger",
    title = "{Quantum complexity in gravity, quantum field theory, and quantum information science}",
    eprint = "2503.10753",
    archivePrefix = "arXiv",
    primaryClass = "hep-th",
    reportNumber = "YITP-25-39",
    doi = "10.1016/j.physrep.2025.11.001",
    journal = "Phys. Rept.",
    volume = "1159",
    pages = "1--77",
    year = "2026"
}

@article{Lanczos1950AnIM,
  title={An iteration method for the solution of the eigenvalue problem of linear differential and integral operators},
  author={Cornelius Lanczos},
  journal={Journal of research of the National Bureau of Standards},
  year={1950},
  volume={45},
  pages={255-282}
}

@book{viswanath2008recursion,
  title={The recursion method: application to many-body dynamics},
  author={Viswanath, VS and M{\"u}ller, Gerhard},
  volume={23},
  year={2008},
  publisher={Springer Science \& Business Media}
}

@article{Muck:2022xfc,
    author = {M{\"u}ck, Wolfgang and Yang, Yi},
    title = "{Krylov complexity and orthogonal polynomials}",
    eprint = "2205.12815",
    archivePrefix = "arXiv",
    primaryClass = "hep-th",
    doi = "10.1016/j.nuclphysb.2022.115948",
    journal = "Nucl. Phys. B",
    volume = "984",
    pages = "115948",
    year = "2022"
}

@article{Adhikari:2025vdl,
    author = "Adhikari, Kiran",
    title = "{Krylov polynomials and quantum query complexity}",
    eprint = "2510.11786",
    archivePrefix = "arXiv",
    primaryClass = "quant-ph",
    doi = "10.1016/j.physleta.2026.131601",
    journal = "Phys. Lett. A",
    volume = "584",
    pages = "131601",
    year = "2026"
}

@inproceedings{Griffel1979AnIT,
  title={An Introduction to Orthogonal Polynomials},
  author={D. H. Griffel and Theodore S. Chihara},
  year={1979},
  url={https://api.semanticscholar.org/CorpusID:120149773}
}

@article{Balasubramanian:2025xkj,
    author = "Balasubramanian, Vijay and Caputa, Pawel and Sim{\'o}n, Joan",
    title = "{Variations on a theme of Krylov}",
    eprint = "2511.03775",
    archivePrefix = "arXiv",
    primaryClass = "hep-th",
    reportNumber = "YITP-25-171",
    doi = "10.1007/JHEP04(2026)172",
    journal = "JHEP",
    volume = "04",
    pages = "172",
    year = "2026"
}

@article{Muck:2024fpb,
    author = {M{\"u}ck, Wolfgang},
    title = "{Black holes and Marchenko-Pastur distribution}",
    eprint = "2403.05241",
    archivePrefix = "arXiv",
    primaryClass = "hep-th",
    doi = "10.1103/PhysRevD.109.126001",
    journal = "Phys. Rev. D",
    volume = "109",
    number = "12",
    pages = "126001",
    year = "2024"
}

@article{Alishahiha:2026fnu,
    author = "Alishahiha, Mohsen and Vasli, Mohammad Javad",
    title = "{Krylov distribution}",
    eprint = "2602.06150",
    archivePrefix = "arXiv",
    primaryClass = "hep-th",
    doi = "10.1103/bdlf-jjtw",
    journal = "Phys. Rev. D",
    volume = "113",
    number = "12",
    pages = "126004",
    year = "2026"
}

@article{Caputa:2021sib,
    author = "Caputa, Pawel and Magan, Javier M. and Patramanis, Dimitrios",
    title = "{Geometry of Krylov complexity}",
    eprint = "2109.03824",
    archivePrefix = "arXiv",
    primaryClass = "hep-th",
    doi = "10.1103/PhysRevResearch.4.013041",
    journal = "Phys. Rev. Res.",
    volume = "4",
    number = "1",
    pages = "013041",
    year = "2022"
}

@article{Balasubramanian:2022tpr,
    author = "Balasubramanian, Vijay and Caputa, Pawel and Magan, Javier M. and Wu, Qingyue",
    title = "{Quantum chaos and the complexity of spread of states}",
    eprint = "2202.06957",
    archivePrefix = "arXiv",
    primaryClass = "hep-th",
    doi = "10.1103/PhysRevD.106.046007",
    journal = "Phys. Rev. D",
    volume = "106",
    number = "4",
    pages = "046007",
    year = "2022"
}

@article{Berenstein:2002jq,
    author = "Berenstein, David Eliecer and Maldacena, Juan Martin and Nastase, Horatiu Stefan",
    title = "{Strings in flat space and pp waves from N=4 superYang-Mills}",
    eprint = "hep-th/0202021",
    archivePrefix = "arXiv",
    doi = "10.1088/1126-6708/2002/04/013",
    journal = "JHEP",
    volume = "04",
    pages = "013",
    year = "2002"
}

@article{McGreevy:2000cw,
    author = "McGreevy, John and Susskind, Leonard and Toumbas, Nicolaos",
    title = "{Invasion of the giant gravitons from Anti-de Sitter space}",
    eprint = "hep-th/0003075",
    archivePrefix = "arXiv",
    reportNumber = "SU-ITP-00-09",
    doi = "10.1088/1126-6708/2000/06/008",
    journal = "JHEP",
    volume = "06",
    pages = "008",
    year = "2000"
}

@article{Perelomov,
author = {A. M. Perelomov},
title = {{Coherent states for arbitrary Lie group}},
volume = {26},
journal = {Communications in Mathematical Physics},
number = {3},
publisher = {Springer},
pages = {222 -- 236},
year = {1972},
doi = {cmp/1103858078},
URL = {https://doi.org/}
}

@article{Levkovich-Maslyuk:2019awk,
    author = "Levkovich-Maslyuk, Fedor",
    title = "{A review of the AdS/CFT Quantum Spectral Curve}",
    eprint = "1911.13065",
    archivePrefix = "arXiv",
    primaryClass = "hep-th",
    doi = "10.1088/1751-8121/ab7137",
    journal = "J. Phys. A",
    volume = "53",
    number = "28",
    pages = "283004",
    year = "2020"
}

@article{Gromov:2013pga,
    author = "Gromov, Nikolay and Kazakov, Vladimir and Leurent, Sebastien and Volin, Dmytro",
    title = "{Quantum Spectral Curve for Planar $\mathcal{N} = 4$ Super-Yang-Mills Theory}",
    eprint = "1305.1939",
    archivePrefix = "arXiv",
    primaryClass = "hep-th",
    reportNumber = "IMPERIAL-TP-13-SL-02",
    doi = "10.1103/PhysRevLett.112.011602",
    journal = "Phys. Rev. Lett.",
    volume = "112",
    number = "1",
    pages = "011602",
    year = "2014"
}

@article{Fatemiabhari:2025cyy,
    author = "Fatemiabhari, Ali and Nastase, Horatiu and Roychowdhury, Dibakar",
    title = "{Holographic Krylov complexity in N=4 SYM theory}",
    eprint = "2511.19286",
    archivePrefix = "arXiv",
    primaryClass = "hep-th",
    doi = "10.1103/6999-p31b",
    journal = "Phys. Rev. D",
    volume = "113",
    number = "10",
    pages = "106033",
    year = "2026"
}

@article{Beisert:2003jj,
    author = "Beisert, Niklas",
    title = "{The complete one loop dilatation operator of N=4 superYang-Mills theory}",
    eprint = "hep-th/0307015",
    archivePrefix = "arXiv",
    reportNumber = "AEI-2003-056",
    doi = "10.1016/j.nuclphysb.2003.10.019",
    journal = "Nucl. Phys. B",
    volume = "676",
    pages = "3--42",
    year = "2004"
}

@article{Caputa:2024sux,
    author = "Caputa, Pawel and Chen, Bowen and McDonald, Ross W. and Sim{\'o}n, Joan and Strittmatter, Benjamin",
    title = "{Spread complexity rate as proper momentum}",
    eprint = "2410.23334",
    archivePrefix = "arXiv",
    primaryClass = "hep-th",
    reportNumber = "YITP-24-137",
    doi = "10.1103/7zs8-9zpg",
    journal = "Phys. Rev. D",
    volume = "113",
    number = "4",
    pages = "L041901",
    year = "2026"
}

\end{document}